\documentclass[12pt]{article} 
\usepackage[utf8]{inputenc}

\newcommand{\mytitle}{Self-Energy of the Lorentzian EPRL-FK Spin Foam Model of Quantum Gravity}
\newcommand{\myauthorname}{Aldo }
\newcommand{\myauthorsurname}{Riello}
\newcommand{\mydate}{\date{February 7, 2013}}
\newcommand{\mylab}{Centre de Physique Théorique, Case 907, Luminy, F-13288 Marseille, France}
\newcommand{\myendfrontpage}{\mylab}

\usepackage{mypackbasic}
\usepackage{mypackadv}

\usepackage[a]{esvect}

\newcommand{\bFIG}{\begin{figure}[!t]\begin{center}\includegraphics}
\newcommand{\eFIG}{\end{center}\end{figure}}
\newcommand{\be}{\begin{equation}}
\newcommand{\ee}{\end{equation}}
\newcommand{\ab}{{ab}}
\newcommand{\ba}{{ba}}
\newcommand{\uD}{{\underline\D}}
\newcommand{\DD}{{\mathcal D}}
\newcommand{\bra}[1]{\langle #1 |}
\newcommand{\ket}[1]{| #1 \rangle}

\newcommand{\sgn}{\text{sgn}}
\newcommand{\ch}{\text{ch}}
\newcommand{\sh}{\text{sh}}

\newcommand{\Tr}{\text{Tr}}
\newcommand{\Y}{{\mathrm{Y}_\gamma}}
\newcommand{\rank}{\text{rank}}
\newcommand{\K}{{\text{K}_\gamma}}
\newcommand{\sldc}{{SL(2,\mathbb{C})}}
\newcommand{\SU}{{SU(2)}}

\allowdisplaybreaks

\title{\bf \mytitle}
\author{\bf \myauthorname \myauthorsurname\\
}
\mydate

\begin{document}

\maketitle\thispagestyle{empty}

\abstract{We calculate the most divergent contribution to non-degenerate sectorof the self-energy (or ``melonic'') graph in the context of the Lorentzian EPRL-FK Spin Foam model of Quantum Gravity. We find that such a contribution is logarithmically divergent in the cut-off over the $\SU$-representation spins when one chooses the face amplitude guaranteeing the face-splitting invariance of the foam. We also find that the dependence on the boundary data is different from that of the bare propagator. This fact has its origin in the non-commutativity of the EPRL-FK $\Y$-map with the projector onto $\sldc$-invariant states. In the course of the paper, we discuss in detail the approximations used during the calculations, its geometrical interpretation as well as the physical consequences of our result.
}
\newpage
\tableofcontents

\newpage
\section{Introduction}
The Spin Foam program is an attempt to quantize General Relativity in a background independent and Lorentz covariant way \cite{Rovelli:Zakopane}. Its strategy is aimed to reproduce in three plus one dimensions the success of the Ponzano-Regge model in providing a viable quantum theory of three dimensional (Euclidean) quantum gravity \cite{PonzanoRegge}.  
The result is the EPRL-FK model \cite{EPRL:EPRL},\footnote{However, see \cite{Perez:SpinfoamApproach} and references therein for a recent review of the spin foam approach to quantum gravity and more details about the EPRL-FK model, and also \cite{BianchiHellmann} for a an interesting comparison between different approaches to its construction.} which is an $\sldc$- (i.e. Lorentz-) covariant \cite{RovelliSpeziale:Covariance} state sum model, defined on two-complexes,\footnote{Sometimes a lower degree of generality is preferred, and the admitted cellular complexes are reduced to those dual to triangulated (pseudo-)manifolds; see for example \cite{Gurau:TopoSingGFT}.\label{cgft}} having (projected) $\SU$ spin networks as boundaries. These spin networks span the same Hilbert space as those obtained in canonical Loop Quantum Gravity.\\

Independence from the cellular complex\footnote{In the three dimensional context of the Ponzano-Regge model, this is not an issue. Indeed, modulo regularization difficulties, the theory is triangulation independent. This reflects the fact that three dimensional gravity is a topological theory, i.e. it has no local degree of freedom.} (and background independence) is recovered in a second moment either via a refining of the complex itself or via a summation over all admitted complexes (e.g. over all the triangulations of a given manifold). The cellular-complex independence of the theory is connected to its continuum limit\footnote{Remark that the continuum limit is \emph{not} the same as its semi-classical limit. See scheme at p.21 of \cite{Rovelli:Zakopane}.} which is intrinsically non-perturbative, and as such it is likely to require some kind of renormalization procedure; in turn, any renormalization procedure needs a thorough understanding of the divergences of the theory.\\

This work aims to perform a first step in this direction for the EPRL-FK model, through the calculation of the most diverging part of the ``self-energy'' (or ``melonic'') cellular complex at fixed boundary data (see \autoref{melone1}). This complex is dual to two four-simplices with four tetrahedral faces identified among them.  The result of the calculation is that the self-energy two complex is found to diverge \emph{logarithmically} in the cut-off on the $\SU$ representations, at least for the specific face amplitude which preserves the face-splitting property of the spin foam ($\mu(j_f)=2j_f+1$). This result is in accordance with what was already known for the Euclidean version of the theory\footnote{Remark that in \cite{KrajewskiMagnen:EPRL-GFT}, another face weight is used (i.e. $\mu(j_f)=(2j_f^+ + 1)(2j_f^- +1)=|\gamma^2-1| j_f^2 + O(j_f)$) and the accordance with this paper is modulo a redefinition of the face weight. Even this looks trivial, it is important to realise that in the Group Field Theory model construction the face and edge weights are strictly related and cannot be modified independently as for Spin Foams. In the case of reference \cite{KrajewskiMagnen:EPRL-GFT}, in particular, the face weight is dictated by the choice of a ``local'' (trivial) vertex and a ``dynamical'' (non-trivial) propagator.} \cite{PeriniRovelliSpeziale:SelfEnergy,KrajewskiMagnen:EPRL-GFT}. However, here one more step is performed: the dependence of the amplitude from the boundary data is worked out in detail, and is found to be different from that of two tetrahedra directly identified among them without the mediation of any dynamical process. Finally, it is interesting to notice that the two spacetime parities of the virtual states which mediate the process play a crucial role in generating the leading divergences.\\

\bFIG[width=8cm]{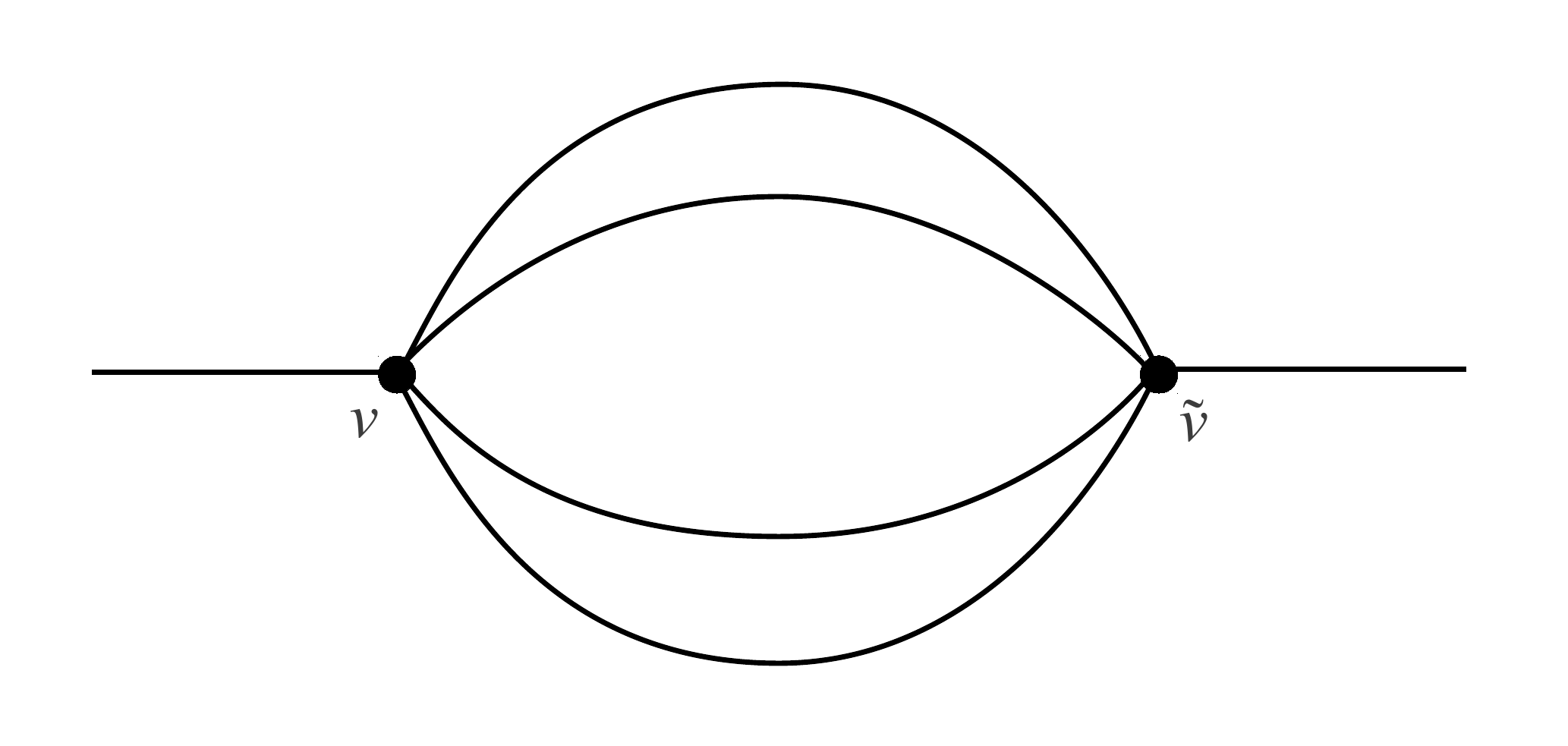}
\caption{The vertex and edge structure of the self-energy cellular complex considered in this paper (see \protect\autoref{melon_simple} for its face structure).}
\label{melone1}
\eFIG

A QFT approach to the Spin Foam program exists and is referred to as Group Field Theory (GFT) \cite{ReisenbergerRovelli:gft}. A GFT is a particular type of a QFT, whose fields live on multiple copies of a group. It can be shown that any spin foam can arise as a particular Feynman diagram of one such GFT,\footnote{The GFT formulations of the Euclidean and Lorentzian EPRL-FK models can be found in \cite{KrajewskiMagnen:EPRL-GFT} and \cite{Krajewski:GFTs}, respectively.} which automatically reproduces its amplitude and at the same time provides a canonical prescription for their summation.\footnote{Furthermore, this summation naturally implements not only a ``sum over geometries'', but a ``sum over topologies'' as well. For more details about GFTs, see \cite{Freidel:GFToverview,Oriti:Book} and references therein.} In a specific class of GFTs (the so called ``coloured'' GFTs and other models derived from them; see  \cite{Gurau:cgft,BonzomGurauRiv:uncoloring}, as well as \cite{GurauRyan:coloredtensors} and references therein), the self-energy graph studied here plays a fundamental role: in simple enough cases it was explicitly shown to drive the non-perturbative behaviour of such models ($1/N$ expansion and critical behaviour \cite{Gurau:1/N,BonzomGurauRielloRiv:CriticalBehav}, renormalization flow \cite{BenGelounRiv:Ren4dTFT,BenGelounSamari:ren3dtft,CarrozzaOriti:RenU1GFT}). This result opened the way to an extension of matrix models and matrix field theories to higher dimensions. In this context, the GFT self-energy graph was first called \emph{melon graph}, a terminology we shall adopt throughout the paper.\\

Spin Foam divergences, at least in the simplest models, can be given two different interpretations, which are essentially dual to one another. The first is mainly geometrical, and associates the divergences to unconstrained sub-simplices which are free to become arbitrarily large, as if some vertices of the triangulation were let free to ``escape to infinity'' (\emph{spikes}). The second is in terms of unconstrained hypersurfaces within the dual complex (\emph{bubbles}), which play a role similar to Feynman diagram loops.\\

In the Ponzano-Regge model, spike divergences are known to be related to the shift symmetry of BF theories and are expected to be associated to the background independence of four dimensional gravity \cite{FreidelLouapre:SFandDiff,BaratinOriti:Diffeos}. Still in the context of the Ponzano-Regge model, in \cite{ChristLangvik:PRspike} it was suggested that spike divergences are also intimately connected with the implicit sum over parities that is present in the theory.\footnote{The presence of a sum over parities can be tracked back to the fact that the Ponzano-Regge model, as any other spin foam model, stems from the Einstein-Cartan action (rather than from the Einstein-Hilbert action) which is parity-change sensitive.} Here we give some arguments which suggest that the same might happen in the EPRL-FK spin foam model.\\

The ERPL-FK model can also be extended to the case of General Relativity with cosmological constant in a non-trivial way, both in its Euclidean \cite{Han:qEPRL,Han:qEPRL_Vassiliev,FairbairnMeusburger:qEPRL} and Lorentzian \cite{FairbairnMeusburger:qLorEPRL,FairbairnMeusburger:qEPRL} versions. Such an extension uses the $q$-deformed Lorentz group, with the $q$-deformation parameter related to the cosmological constant, and turns out to be (perturbatively) finite. The existence of a finite model does not mean that the issue of large radiative corrections can be ignored: it may still happen that some higher order graphs have large amplitudes and therefore drive a renormalization flow, possibly even through phase transitions. Qualitatively, the scale which imposes the infrared\footnote{Here, the term ``infrared'' must be understood as relating to large physical distances; analogously, an ``ultraviolet'' cut-off, in the sense of a short distance cut-off, is naturally present in any spin foam models, via the existence of the \emph{area gap} \cite{RovelliSmolin:discreteness}. It must however be kept in mind, that the roles of the words ``infrared'' and ``ultraviolet'' are interchanged when referring to the momentum space of the gauge group.} finiteness of the theory is given by the cosmological constant, which is of the order of the radius of the Universe; therefore, at our - or smaller - length scales, it can be considered as infinite for practical purposes (but see comments in the conclusions).\\

In this paper, in order to study the simplest EPRL-FK divergence, we introduce a cut-off $\Lambda$ to the $\SU$ representations $j$. The physical meaning of such a cut-off is that of imposing a maximal value for the area operator, which can be thought as the introduction of a finite size for the Universe itself. A bound to the area operator is typical of the $q$-deformed version of the EPRL-FK model. Therefore the introduction of such a cut-off can be hoped to be a simple implementation of the main feature of the $q$-deformed EPRL-FK model within the much more manageable non-deformed version. At the light of this (qualitative) correspondence, the calculation of this paper can be also given a more physical, though possibly naive, interpretation in which the cut-off $\Lambda$ \emph{is} a physical quantity and corresponds  - at least in order of magnitude - to the cosmological constant $\Lambda_{CC}$ expressed in Planck units of area: $\Lambda\approx \Lambda_{CC}/l_P^2\sim10^{120}$.\\

The goal of this work is to calculate the most divergent contribution to the self-energy of the EPRL-FK spin foam model coming from its geometrically non-degenerate sector (the terminology will be clarified later). In order to do this, various types of approximations are introduced which will be explained in detail. The main tool used is an appropriate application of the large spin analysis, first introduced in the context of the EPRL-FK model in \cite{BarrettDowdall:EuclAsympt,BarrettHellmann:LorAsympt}, for a single vertex, and in \cite{HanZhang:EuclAsympt,HanZhang:LorAsympt}, for a general two simplex. Thanks to the geometric content of the equations involved in the analysis, it will be possible to interpret the most diverging contribution to the self-energy as stemming from the constructive contribution of two parity-related ``chunks'' of spacetime (i.e. of a quantum of spacetime with a quantum of \emph{anti}-spacetime, in the terminology of \cite{ChristRielloRov:AntiSpacetime}).\\

The analysis of the degenerate sector is left for future studies. However, it must be remarked at this point that hints coming from the Euclidean version of the model seem to say that this sector  could be the one dominating the melon graph (see the Appendix C of \cite{KrajewskiMagnen:EPRL-GFT}). An intuitive picture of why this may happen, expressed in the language of this paper, shall be given in \autoref{appendix_degenerate}.\\

The paper starts with a brief introduction of the notation in \autoref{section_notations}. Then, in order to help the reader through the multitude of technical details, as well as an argument of plausibility of the approximation schemes, the calculation is first performed in the context of $\SU$ BF theory, where the result is known and can be double-checked. This is the subject of \autoref{section_BF}. The definition of the EPRL-FK model on an arbitrary two complex, and the preliminary study of the EPRL-FK melon graph symmetries are performed in \autorefs{section_EPRL} and \ref{section_EPRL_melon}, respectively. \Autoref{section_solve} introduces the stationary phase approximation for the melon graph amplitude and, consequently, the large spin limit of the EPRL-FK model. The equations which define the stationary points are then solved in \autoref{section_solutionstationaryphase}. This will be first done in complete generality from an algebraic, rather abstract, point of view, and later given a much more transparent geometric interpretation (which however is too weak to capture all the details of the algebraic solution). This analysis shall point out the existence of two non-degenerate sectors of solutions, related to an Euclidean and a Lorentzian geometry respectively, as well as a degenerate one (which deserve a thorough analysis on their own). The role of the two-complex face orientation is discussed in \autoref{section_face}, which is crucial to correctly evaluate the melon amplitude at the stationary point. The radiative corrections related to the three sectors mentioned above are evaluated in \autoref{section_radiativecorrections}, and the total graph amplitude is presented in \autoref{section_totalamplitude}. In \autoref{section_approx}, the approximations  and hypothesis used during the whole calculations are briefly reviewed. The spike interpretation, as much as its limitations, is discussed in \autoref{section_spike}. Finally, we conclude in \autoref{section_conclusions}.

\section{Spinors, Vectors and Bivectors}\label{section_notations}
In this section we briefly introduce some notations and technical tools that will be extensively used in the following.\\

Written explicitly in components, a spinor reads:
\be
\ket{w} \equiv w = \left(\begin{array}{c} w^0 \\ w^1\end{array}\right)\in\mathbb C^2 \quad\text{and}\quad \bra{w}= \left(\begin{array}{c} w^0 \\ w^1\end{array}\right)^\dag= \left( \bar{w}^0,\;\bar{w}^1\right)\;,
\ee
where the bar stands for complex conjugation. In the first of these equalities we introduced the simplified notation for ``ket'' spinors. The hermitian scalar product between spinors, is
\be
\langle{w}|{ z}\rangle =\bar{w}^0z^0 + \bar{w}^1 z^1\;,
\ee
hence the squared norm $||w||^2:=\langle w|w\rangle\geq0$.\\

We define the antilinear operator $J$:
\be
J:\mathbb C^2\rightarrow \mathbb C^2\;,\quad\;J \left(\begin{array}{c} w^0 \\ w^1\end{array}\right) :=  \left(\begin{array}{c} -\bar{w}^1 \\ \bar{w}^0\end{array}\right).
\ee
It is easy to verify that $\bra{w}J\ket{w}=0$ for all $w\in\mathbb C^2$, and that that $JgJ^{-1} = g^\dag{}^{-1}$ for all $g\in\sldc$.\\

To a spinor one can associate a four-vector $\iota(w)$:
\be
w= \left(\begin{array}{c} w^0\\w^1 \end{array}\right)
\quad \longrightarrow\quad 
\iota(w)^I := \bra{w}\sigma^I\ket{w}= \left(\begin{array}{c} ||w||^2 \\ 2\Re\left(w^0\bar{w}^1\right)\\ 2\Im\left(w^0\bar{w}^1\right) \\ |w^0|^2-|w^1|^2 \end{array}\right), \quad\text{where}\quad \sigma^I = \left(\begin{array}{c} \mathbb I\\\sigma^i \end{array}\right)\;
\label{vector}
\ee
and $\sigma^i$ are the Pauli matrices.\footnote{With the usual conventions: $$
\sigma^1=\left(\begin{array}{cc} 0 & 1 \\ 1 & 0\end{array}\right),\quad
\sigma^2=\left(\begin{array}{cc} 0 & -\I \\ \I & 0\end{array}\right)\quad\text{and}\quad
\sigma^3=\left(\begin{array}{cc} 1 & 0 \\ 0 & -1\end{array}\right).
$$}
Equivalently, in a notation more adapted to normalized spinors:
\be
w = ||w|| \E^{\I\phi}\left(\begin{array}{c} \E^{\I\psi}\cos\theta \\ \E^{-\I\psi}\sin\theta \end{array}\right)
\quad \longrightarrow\quad 
\iota(w)^I = ||w||^2\left(\begin{array}{c} 1 \\ \sin2\theta\; \cos2\psi\\ \sin2\theta\;\sin2\psi \\ \cos2\theta \end{array}\right),
\ee
From this expression, it is readily verified that $w^I$ is light-like. Remark that the four-vector $\iota(w)$ looses the information about the global phase of the spinor. The spatial part of $\iota(w)$ shall be noted $\hat w$, and its components $w^i$.\\

The spinor $Jw$ is associated to the four-vector
\be
\iota(Jw)^I=\left(\begin{array}{c} ||w||^2 \\ -2\Re\left(w^0\bar{w}^1\right)\\ -2\Im\left(w^0\bar{w}^1\right)\\ -\left(|w^0|^2-|w^1|^2\right) \end{array}\right)=\mathcal P\iota(w)^I,
\ee
where $\mathcal P$ is the parity reversal operator.\\

The four-vectors $\iota(w)$ do actually carry a vectorial representation of $\sldc$:
\be
g\rhd\iota(w) =\iota(gw),
\ee
where the $g$ in the l.h.s. is understood to be in the vectorial representation of $\sldc$, i.e. to be an element of $SO^+(1,3)$, the proper orthochronus Lorentz group. Similarly, the vectors $\hat w$ carry a vectorial representation of $\SU$:
\be
h\rhd \hat w = \widehat{hw},
\ee
where the $h$ in the l.h.s. is understood to be in the vectorial representation of $\SU$, i.e. to be an element of $SO(3)$. Remark that this is the $\SU$ subgroup of $\sldc$ which stabilizes the time normal $t^I:=(1,\vec 0)^T$. \\

Let $m\in\mathbb C^2$ be a normalized spinor and $\hat m$ the space component of $\iota(m)$. Define the $2\times2$ complex matrix $D(m):=(m,Jm)$ having $m$ and $Jm$ as columns.\footnote{
\label{footnoteD(w)}
In vectorial representation $D(m)$ becomes an $SO(3)$ rotation matrix, written in terms of the Euler angles $(2\phi,2\theta,2\psi)$. Then, $\hat m$ is just the image of the unit vector $\hat z$ via $D(m)$: $$\hat m = D(m)\rhd \hat z.$$
 This construction is natural and simple. Indeed, in spin $\frac{1}{2}$ representation, the spinor associated to the $\hat z$-axis is just $(1,0)^T$ (see \autoref{spinor_z}); therefore for it to be sent into $m$ by a matrix $D$, the first column of $D$ must be $m$ itself. Requiring $D\in\SU$, gives $D=D(m)$.
} It is an element of $\SU$. Using the previous notation, it can be written as
\be
\SU\ni D(m):=(m,Jm)=\E^{\I\psi\sigma_z}\E^{-\I\theta\sigma_y}\E^{\I\phi\sigma_z}.
\ee\\

The Livine-Speziale (LS) coherent states are defined as \cite{LivineSpeziale:Coherent}
\be
\ket{m}_j:=D^{(j)}(m)\ket{j,j}\equiv D(m)\rhd \ket{j,j},
\ee
where $\ket{j,j}$ is the usual maximal magnetic number state in the standard $j$-th $\SU$ irreducible representation $\mathcal H^{(j)}$, and $D^{(j)}(m)$ is the representation of the matrix $D(m)$ acting on $\mathcal H^{(j)}$.\footnote{Explicitly: 
$$D^{(j)}(m)=\E^{2\I\psi\mathcal J^{(j)}_z}\E^{-2\I\theta\mathcal J^{(j)}_y}\E^{2\I\phi\mathcal J^{(j)}_z},$$
$\{\mathcal J^{(j)}_i\}_{i=x,y,z}$ being the three generators of $\SU$ in representation $j$.}These states are coherent in the sense that their expectation value of the angular momentum operator $\vv{\mathcal{J}}$ is peaked around the vector $j\hat m\in\mathbb R^3$ with a relative uncertainty which goes to zero as $\sim1/\sqrt{j}$ in the semi-classical large-spin limit. In formulas:
\be
\bra{m}\vv{\mathcal J}\ket{m}_j = j\hat m\quad\text{and}\quad \bra{m}{\vv{\mathcal J}}^{\;2}\ket{m}_j - \big|\bra{m}\vv{\mathcal J}\ket{m}_j\big|^2 = j,
\ee
where $\bra{m}m'\rangle_j$ is the Hermitian scalar product in $\mathcal H^{(j)}$. The LS states can therefore be used to described semi-classical three-vectors  of length $j$ and direction $\hat m\in S_2$. \\

Observe that in the previous notation:
\be
m\equiv\ket{m}\equiv\ket{m}_\frac{1}{2} = D(m)\rhd\ket{\frac{1}{2},\frac{1}{2}},
\ee
which holds thanks to $\ket{\frac{1}{2},\frac{1}{2}}=(1,0)^T$.\footnote{Indeed, $\ket{\frac{1}{2},\frac{1}{2}}$ is defined by $\mathcal J^{(1/2)}_z\ket{\frac{1}{2},\frac{1}{2}}=\frac{1}{2}\ket{\frac{1}{2},\frac{1}{2}}$. Being $\mathcal J^{(1/2)}_z=\frac{1}{2}\sigma_z$, one finds  $\ket{\frac{1}{2},\frac{1}{2}}=(1,0)^T$.\label{spinor_z}}\\

LS states are an (over-)complete basis of the vector space $\mathcal H^{(j)}$ supporting the representation $j$ of $\SU$. They allow to write the following resolution of the identity
\be
\mathbb I_j = \frac{(2j+1)}{2\pi} \int_{\SU} \D D(m) \;\;\ket{m}_j\bra{m}_j,
\ee
where the measure appearing in the integral is the normalized Haar measure on $\SU$.\\

The phase of $\ket{m}_j$ plays no role both in the resolution of the identity and in the formula associating to $\ket{m}_j$ a vector $j\hat m$. Indeed, its phase has no physical meaning and can be fixed to any value. Since the phase $\phi$ of the spinor $m$ translates into a phase $2j\phi$ for $\ket{m}_j$, for what concerns the definition of the LS states it can be fixed to zero.\\
Therefore, we modify the previous notation into:
\be
\ket{m}_j := D(\hat m)\rhd  \ket{j,j}\quad\text{with}\quad m\in S_2
\ee
and why the resolution of the identity shall rather be written as
\be
\mathbb I_j = (2j+1) \int_{S_2} \D\hat m \;\ket{m}_j\bra{m}_j=:\int_{S_2}\uD_j\hat m\;\ket{m}_j\bra{m}_j,
\label{id_resolution}
\ee
where $\D \hat m$ is the normalized measure on the two-sphere $S_2$, while $\underline\D_j \hat m:=(2j+1)\D \hat m$. When no risk of confusion arises, $\underline\D_j\hat m$ is simply noted $\underline\D \hat m$.


\newpage
\section{BF Theory\label{section_BF}}

In this section the main steps of this article are briefly reproduced in the simplified context of BF theory. This is done, despite their high redundancy, with a twofold purpose: this section aims to clarify the ideas lying behind the rather intricate algebra of the calculations in the EPRL-FK model, and to justify (or at least strengthen) the reliability of the approximation and simplification schemes used. Indeed, such schemes are shown to lead to the correct result within BF theory.

\subsection{BF Graph Amplitudes in the Livine-Speziale representation\label{subsect3.1}}
The $\SU$-BF amplitude of a closed 2-complex $\mathcal C$, given an (arbitrary) orientation of its edges $\{e\}$ and faces $\{f\}$, is formally defined by
\be
W_\mathcal{C}^{\text{BF}} :=  \left[\prod_{e}\int_\SU\D h_e\right] \prod_{f} \delta\left( h_f  \right),\quad\text{where}\quad h_f:=\overleftarrow{\prod_{e\in\partial f}}  h_e^{\epsilon_{ef}}
\label{eq_BF_deltas}
\ee
Here $\overleftarrow{\prod}_{e\in\partial f}$ stands for the ordered (non-commutative) product over the edge index $e$ along the boundary of the face $f$. Also, $\epsilon_{ef}$ are the entries of the edge-face incidence matrix: it is non-zero only if $e\in\partial f$, and it is 1 if the face and edge orientations agree and $-1$ if they do not.  Finally, $\delta$ is the delta distribution on the group $\SU$ for its Haar measure $\D h$, i.e. $\int_\SU \D h \; \delta(h_0^{-1}h) f(h) = \int_\SU \D h \; \delta(h h_0^{-1}) f(h) =f(h_0)$.\\

The previous expression can be readily generalized in order to include open complexes \cite{Rovelli:Zakopane}. In this case the boundary of the two complex intersects the external (half-)edges at points (nodes $\{n\}$) and the open faces at segments (links $\{l\}$). The link associated to a certain open face $f$ connects the nodes associated to the (half-)edges bounding $f$. At each link is associated an $\SU$ group element $h_l$. Therefore, one defines the open complex amplitude by extending the previous definition of $h_f$ to open faces. This is done just by formally treating, in the formula for $h_f$, the nodes as vertices and the link as edges. Despite of the similarity the interpretation of the complex amplitude is different: the $h_l$ constitute the boundary data and are \emph{not} integrated over:
\be
W_\mathcal{C}^{\text{BF}}(h_l) :=  \left[\prod_{e}\int_\SU\D h_e\right] \prod_{f} \delta(h_f)\;,
\label{BF_delta}
\ee
where the label $e$ applies only to the internal edges, while the label $f$ to both closed and open faces.\\

Despite of its appealing simplicity, this can be only a formal definition of $W_\mathcal{C}^{\text{BF}}$, since for a general complex $\mathcal C$ it contains \emph{a priori} multiple products of $\delta(\mathbb{I})$. One way to regularize these divergences is through the Fourier-transform of the $\SU$-BF amplitude \ref{BF_delta} via the Peter-Weyl theorem. In this way it is possible to introduce a sharp cut off at spin $j=\Lambda$. Indeed, the $\SU$ delta function can be decomposed onto the $\SU$ representations labelled by $j$ as
\be
\delta(h) = \sum_{j\in\frac{1}{2}\mathbb N} (2j+1)\Tr_j(h)\quad\forall h\in\SU,
\ee
where $\Tr_j(h)$ stands for the trace of the matrix $h$ in representation $j$, i.e. the $j$-th character of $h$. By using the fact that each of the addends in the previous formula is finite, one can define the regularized version of the $\SU$-BF amplitude of a complex:\footnote{In the context of BF theory, it is often more effective to use a different, possibly smooth, regularization (e.g. the heat kernel regularization, see among others \cite{BenGelounBonzom:BFren,CarrozzaOriti:RenU1GFT}). Nevertheless, here we shall focus on this one, because it is more easily implemented in the EPRL-FK model.}
\be
W_\mathcal{C}^{\text{BF},\Lambda}(h_l):=  \sum_{\{j_f<\Lambda\}}\left\{\left[\prod_{e}\int_\SU\D h_e \right]\prod_{f} (2j_f+1) \;\Tr_{j_f}\left(h_f \right)\right\},
\ee
where the sum and the product $\prod_f$ are over all faces of the graph, either close or open. For future convenience we introduce the amplitudes
\be
W_\mathcal{C}^{\text{BF},\Lambda}(j_l,h_l):=  {\sum_{\{j_f<\Lambda\}}}' \left\{\left[\prod_{e}\int_\SU\D h_e \right]{\prod_{f}}' (2j_f+1) \;\Tr_{j_f}\left(h_f \right)\right\},
\label{BF_spin}
\ee
where the primed summation indicates that the sum is taken only over the close faces, in such a way that the $j_l$'s are external data for these amplitudes, while the primed product means that the weights $(2j_l+1)$ are omitted for the external faces:
\be
{\prod_f}' (2j_f+1) \;\Tr_{j_f}\left(h_f \right) \equiv \left[\prod_{f\;\text{open}}\Tr_{j_l}\left(h_f \right)\right]\left[\prod_{f\;\text{close}}(2j_f+1) \;\Tr_{j_f}\left(h_f \right)\right].
\ee

In passing, notice also that the cut-off delta function $\delta_\Lambda(h):=\sum_{j<\Lambda}(2j+1)\Tr_j(h)$ when calculated at the identity diverges as $\Lambda^3$, since $\Tr_j(\mathbb I)=\dim\mathcal H^{(j)}=2j+1$.\\

In order to deal with a formalism closer to the one used in the following sections, it is necessary to further manipulate \autoref{BF_spin}. First, every group element associated to one edge is decomposed into two group elements associated to each half edge:
\be
h_e= h_{s(e)e}^{-1}h_{t(e)e},
\ee 
the functions $s(e)$ and $t(e)$ denoting the source and the target vertices of the oriented edge $e$, respectively. In this way, there is one group element $h_{ve}$ associated to each vertex-edge couple $(ve)$. By construction, $h_{ve}$ always ``enters'' the vertex.\\

Then, for every face, the trace in \autoref{BF_spin} is decomposed by inserting between each $h_{s(e)e}$ and $h_{t(e)e}$ the resolution of the identity in $\mathcal H^{(j)}$ onto the LS states $\ket{m}_j$
\be
\mathbb{I}_j = \int_{S_2} \underline\D_j \hat m \; \ket{m}_j\bra{m}_j.
\ee %
Furthermore, the information about each boundary group element $h_l$ is translated into the (redundant) information about two LS states $m_{nl}$ and $m_{n'l}$, being $n=s(l)$ and $n'=t(l)$:
\be
h_l\equiv\ket{m_{n'l}}\bra{m_{nl}}+\ket{Jm_{n'l}}\bra{Jm_{nl}}.
\ee

Hence, the  $\SU$-BF amplitude can finally be written in the LS representation:
\begin{align}
& W_\mathcal{C}^{\text{BF},\Lambda}(j_l, m_{nl})={\sum_{\{j_f<\Lambda\}}}'w^{\text{BF}}_\mathcal{C}(j_l,m_{nl};j_f)
\label{BF_m}
\\
&w^{\text{BF}}_\mathcal{C}(j_l,m_{nl};j_f):=\\
&\quad:=\left[\prod_f \prod_{e\in\partial f}\int_{S_2}\uD_{j_f} m_{ef}\right]
\left[ \prod_v\prod_{e: v\in\partial e}\int_\SU\D h_{ve}    \right]
{\prod_{f}}' (2j_f+1)  \prod_{v\in\partial f}\bra{m_{e'f}} h_{ve'}^{-1}h_{ve}\ket{m_{ef}}_{j_f}\;,\notag
\end{align}
where the \emph{partial amplitudes} $w^{\text{BF}}_\mathcal{C}$ have been introduced.
Here the labels $(ve)$ and $(ve')$ denote the two half edges belonging to the face $f$ ending in the vertex $v$ and such that $e'$ is the edge  following $e$ according to the orientation of $f$. Furthermore $\bra{m}h\ket{n}_j$ is a shorthand notation for the Hermitian scalar product in $\mathcal H^{(j)}$. Remark that in this representation the boundary data are more conveniently chosen to be the spins associated to the open faces $j_l$ and the unit vectors $\hat m_{nl}$ associated to each couple $(n,l)$ composed of an external \emph{half} edge $n$ and an open face $l$, to which $n$ belongs ($n\in\partial l$). These boundary data enter the previous expression thanks to the following two conventions: (1) if $f$ is the open face such that $l\in\partial f$, then $j_f=j_l$; (2) if $f$ is the open face such that $l\in\partial f$ and $(ve)$ (or $(ve')$) is an external half edge ending on the node $n\in\partial l$, then  $\bra{m_{e'f}} h_{ve'}^{-1}h_{ve}\ket{m_{ef}}_{j_f}=\bra{m_{e'f}} h_{ve'}^{-1}h_{ve}\ket{m_{nl}}_{j_l}$ (or $\bra{m_{e'f}} h_{ve'}^{-1}h_{ve}\ket{m_{ef}}_{j_f}=\bra{m_{nl}} h_{ve'}^{-1}h_{ve}\ket{m_{nl}}_{j_l}$, respectively).\\

In the following, the spins associated to the closed faces of a graph shall be often referred to as \emph{internal} spins. This is because they are associated to faces living in the interior of the graph.\\

Finally, remark that in \autoref{BF_m} the elements $\{h_{ve}\}$ appear always multiplied two by two, in such a way that one $h_{ve}$ multiplies only other group elements at the same vertex. Thanks to Haar measure's translation invariance, this fact implies that at each vertex $v$ one of the $h_{ve}$'s, say $h_{ve_0}$ can always be reabsorbed into the integrations over others $h_{ve}$. Therefore the integrations over the various $h_{ve_0}$ are redundant. This has no consequence for a compact group like $\SU$. However, it is judicious to gauge fix one group element per vertex in order to avoid meaningless divergences in the amplitudes when generalizations to non-compact gauge group are envisaged. This will be our case, since the gauge group of the Lorentzian EPRL-FK model is the Lorentz group double covering group $\sldc$, which is non-compact.

\subsection{The BF Melon Graph}
The aim of this subsection is to sketch the calculation of the dominant contribution to the $\SU$-BF melon graph in the $\Lambda\rightarrow\infty$ regime, while keeping the boundary data fixed. To do this, we shall focus on the limit in which all the internal spins $j_f$ are scaled to infinity via a parameter $\lambda$. The idea is to understand the asymptotic behaviour at large spins of the partial amplitudes $w_\mathcal{M}^\text{BF}$, i.e. of the graph amplitudes at fixed spins. This is the relevant regime for studying the convergence properties of the full amplitude: the full amplitude is obtained via a sum over six variables, therefore it converges only if the partial amplitudes $w_\mathcal{M}^\text{BF}$ scale to zero faster  than $\lambda^{-6}$, and it diverges otherwise. In particular the divergence is logarithmic in the cut-off if $w_\mathcal{M}^\text{BF}\sim\lambda^{-6}$.\\

It is however important to remark that this procedure is not justifiable on a strict mathematical basis, since divergences could be hidden in those sectors where one of the internal spins is kept small. However, there are at least two independent reasons why to be interested in the uniformly large spin regime. First, it is a regime with a sensible geometrical meaning, as it shall be shown in the course of this paper. Then, the restriction to this regime is \emph{a posteriori} justified by the fact that it is enough to account for the asymptotic behaviour of the melon graph in the context of the $\SU$-BF theory.
\subsubsection{Notation}\label{Section_BF_notation}
The melon graph $\mathcal M$ (see \autoref{melon_simple}) has two external half edges, two vertices ($v$ and $\tilde v$), four internal edges (labelled by indices $a,b,\dots\in\{1,\dots,4\}$), six internal faces (labelled by unordered couples $ab$), and four external faces (labelled by indices $a$). Each internal edge belongs to a unique external face, whose spin is thus noted $j_a$. In the same way the boundary unit vectors are noted $n_a$ and $\tilde n_a$, depending on the vertex belonging to their boundary. Since the internal faces have two edges for boundary, their spins are noted $j_\ab=j_\ba$. The unit vectors used to resolve the identity within the internal edge $a$, are noted $\hat m_\ab$ if they are inserted in the trace relative to the face $ab$ (remark that $\hat m_\ab\neq \hat m_\ba$), and just $\hat m_a$ if they are inserted in the trace relative to the external face $a$. Finally the group elements associated to the internal half edges $h_{ve}$ and $h_{\tilde v e}$ are noted $h_a$ or $\tilde h_a$, following the obvious identification. The last two group elements associated to the external half edges are noted simply $h$ and $\tilde h$.\\

\bFIG[width=8.5cm]{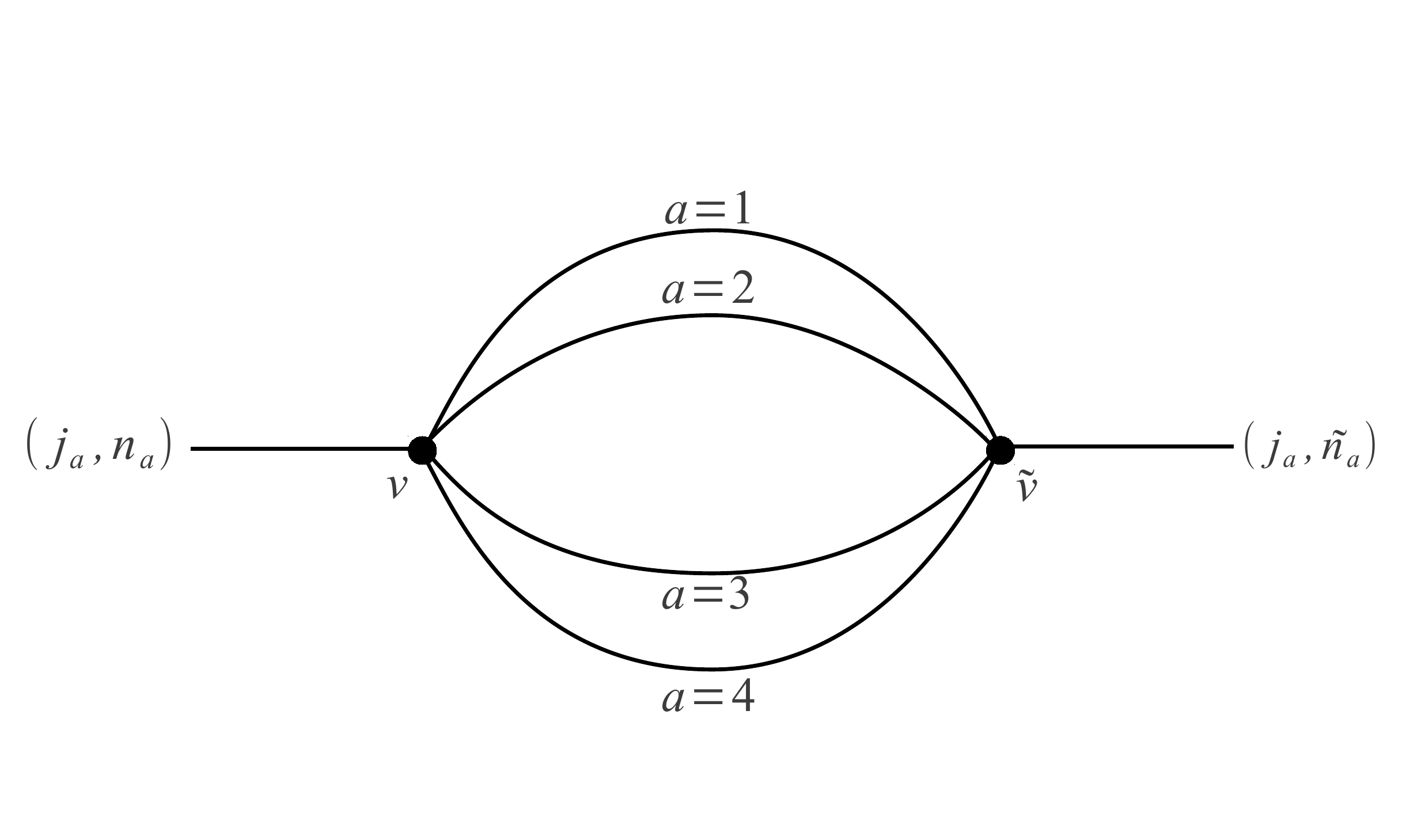}
\includegraphics[width=8.5cm]{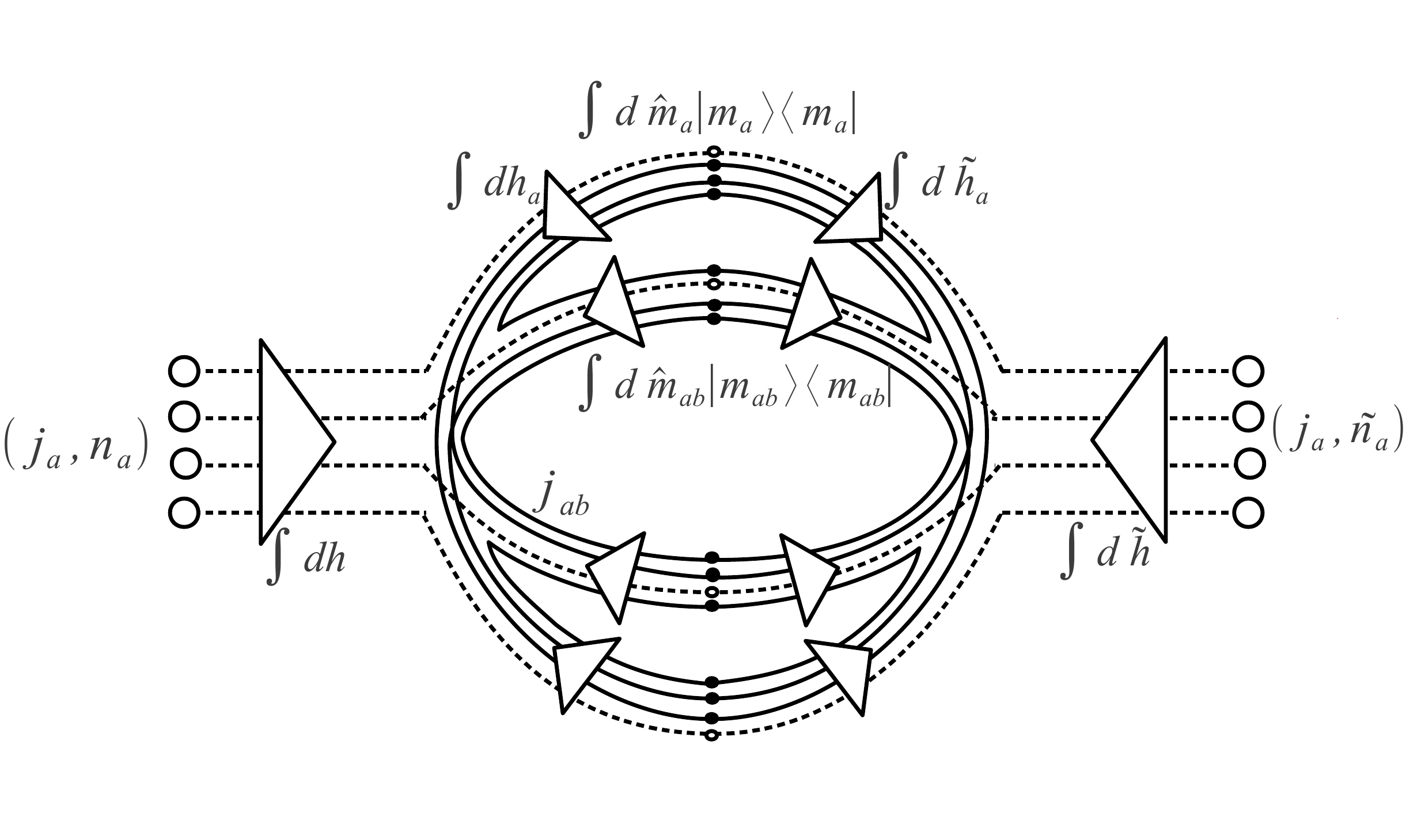}
\caption{The melon graph. On the right, its faces and extra structure entering the LS representation are put into evidence. Triangles point in the direction of the action of the $\SU$ elements $\{h, h_a,\tilde h, \tilde h_a\}$. Dots represent insertions of the resolution of the identity. External faces are drawn in a dashed line.}
\label{melon_simple}
\eFIG

With this notation the melon graph amplitude reads
\begin{align}
W^{\text{BF},\Lambda}_{\mathcal M}(j_a,n_a,\tilde n_a) &=\sum_{\{j_\ab<\Lambda\}}w^\text{BF}_\mathcal{M}(j_a,n_a,\tilde n_a;j_\ab)\\
w^\text{BF}_\mathcal{M}(j_a,n_a,\tilde n_a;j_\ab)&:=\\
&\hspace{-1cm}:=\left[\int_{\SU^{\otimes2}} \D h \D\tilde h\; \prod_a \int_{\SU^{\otimes2}} \D h_a\D\tilde h_a \right]\left[\prod_a\int_{S_2}\uD\hat m_a\right]\left[\prod_{\ab:a<b}\int_{S_2{}^{\otimes2}}\uD\hat m_\ab\uD\hat m_\ba\right]\notag\\
&\hspace{0cm}\delta(h)\delta(\tilde h) \left[\prod_a\bra{n_a}h^{-1}h_a\ket{m_a}_{j_a}\bra{m_a}\tilde h_a^{-1}\tilde h\ket{\tilde n_a}_{j_a} \right]\times\notag\\
&\hspace{1.7cm}\times\left[\prod_{ab:a<b}(2j_\ab+1)\;\bra{m_\ab}\tilde h_a^{-1}\tilde h_b\ket{ m_\ba}_{j_\ab} \bra{m_\ba}h_b^{-1}h_a\ket{m_\ab}_{j_\ab}\right],\notag
\end{align}
where the $\delta$ functions were introduced in order to implement the gauge fixing conditions discussed at the end of the previous subsection (\autoref{subsect3.1}). Integrating out $h$ and $\tilde h$, and introducing the short-hand notation
\be
\int\DD h\DD\hat m := \left[ \prod_a \int_{\SU^{\otimes2}} \D h_a\D\tilde h_a \right]\left[\prod_{\ab:a<b}\int_{S_2{}^{\otimes2}}\uD\hat m_\ab\uD\hat m_\ba\right]
\ee
and performing the integrals over the $\hat m_a$'s, one gets immediately:
\begin{align}
w^{\text{BF}}_{\mathcal M}(j_a,n_a,\tilde n_a;j_\ab)=&
\int\DD h\DD\hat m\;\left[\prod_a\bra{n_a}h_a\tilde h_a^{-1}\ket{\tilde n_a}_{j_a}\right] \times\notag\\
&\hspace{1.7cm}\times\left[\prod_{ab,a<b}(2j_\ab+1)\bra{m_\ab}\tilde h_a^{-1}\tilde h_b\ket{ m_\ba}_{j_\ab} \bra{m_\ba}h_b^{-1}h_a\ket{m_\ab}_{j_\ab}\right]\phantom{\int}\hspace{-3mm}.
\label{BF_melon_0}
\end{align}

\subsubsection{The BF Large Spin Limit}
In order to give to the right hand side of \autoref{BF_melon_0} a form adapt to the large spin analysis, we need to write it in such a way that an exponentiated action proportional to the face-spin appears. Such a form will very closely remind that of a path integral. This can be easily done via the following observation. Since $\ket{j,j}=\bigotimes_{2j}\ket{\frac{1}{2},\frac{1}{2}}$, it is immediate to show that $\langle{m'}|{m}\rangle_j=\langle{m'}|{m}\rangle^{2j}=\exp(2j\langle{m'}|{m}\rangle)$, where $\ket{m}\equiv\ket{m}_\frac{1}{2}\equiv m \in \mathbb C^2$ (see \autoref{section_notations}). Therefore, \autoref{BF_melon_0} can be given the sought form:
\begin{align}
W^{\text{BF},\Lambda}_{\mathcal M}(j_a,n_a,\tilde n_a)&=\sum_{\{j_\ab<\Lambda\}} w^{\text{BF}}_{\mathcal M}(j_a,n_a,\tilde n_a;j_\ab) \label{BF_W}\\
w^{\text{BF}}_{\mathcal M}(j_a,n_a,\tilde n_a;j_\ab)&:=\int\DD h\DD\hat m\; \left[\prod_a \bra{n_a}h_a\tilde h_a^{-1}\ket{\tilde n_a}_{j_a}\right]\left[ \prod_{\ab, a<b} (2j_\ab+1)\;\E^{S^{\text{BF}}_\ab}\right]\;,
\label{BF_melon}
\end{align}
where the face action $S^{\text{BF}}_\ab$ was defined
\be
S^{\text{BF}}_\ab:=2 j_\ab \ln \bra{m_\ba}h_b^{-1}h_a\ket{m_\ab} + 2j_\ab \ln \bra{m_\ab}\tilde h_a^{-1} \tilde h_b\ket{m_\ba}
\label{BF_face_action}
\ee
which is proportional to face-spin $j_\ab$.\\

We also define a total (internal) action $S^{\text{BF}}$ by adding the various internal face actions:
\be
S^{\text{BF}} := \sum_{\ab,\; a<b} S^{\text{BF}}_\ab \;.
\ee
Remark that due to Cauchy-Schwarz's inequality, the real part of each face action is null or negative, and hence also that of the total action
\be
\Re(S^{\text{BF}})\leq0.
\ee

Before delving into the analysis of the semi-classical (large spin) limit of the melon-graph partial amplitudes, we shall analyse the symmetries of $S^\text{BF}$. First of all, it is clear that at each vertex one can left-rotate all the $h_a$ (resp. $\tilde h_a$) by an arbitrary $\SU$ element $k$ (resp. $\tilde k$), without affecting \autoref{BF_face_action}:
\be
h_a\mapsto k h_a\quad \text{and}\quad \tilde h_a \mapsto \tilde k \tilde h_a\;.
\label{BF_v_invariance}
\ee
Furthermore, one can as well rotate the spinors on a given edge $\{m_\ab\}_{b,b\neq a}$ by a $k_a\in\SU$, provided one simultaneously right-rotates the sets $\{h_a\}$ and $\{\tilde h_a\}$ by $k_a^{-1}$:
\be
\left\{\begin{array}{l}
{m_\ab}\mapsto k_a{m_\ab}\quad\forall b,\; b\neq a\\
(h_a, \tilde h_a) \mapsto (h_a k_a^{-1}, \tilde h_a k_a^{-1})
\end{array}\right.\;.
\label{BF_e_invariance}
\ee

The identification of these symmetries will be fundamental in correctly evaluating the degree of divergence of the graph. For reasons explained in more detail in \autoref{Sect_sym}, these symmetries shall be often referred to as \emph{gauge} symmetries. \\

In order to study the large spin limit of the partial amplitudes $w^{\text{BF}}_{\mathcal M}$, in which all the internal spins uniformly scale to infinity, we formally rescale them by a common factor $\lambda$. In the limit $\lambda\rightarrow\infty$, it is clear that the integrals are dominated by the stationary points of the total action $S^{\text{BF}}$.\footnote{Rigorously speaking, these are not stationary \emph{points}, but rather stationary \emph{hypersurfaces} on which $S^{\text{BF}}$ is constant. However, the contributions of the external faces to the amplitude is not constant on such hypersurfaces (not in every direction, at least). Therefore, it will turn out that the partial amplitude will be given at the dominant order by the average of the external face contributions on these hypersurfaces. This ``averaging'' procedure would not have been necessary if another gauge fixing were used instead of $h=\tilde h=\mathbb I$. More on this topic at the end of \autoref{Sect_sym}. }
Following the same steps of \cite{HanZhang:LorAsympt}, adapted at the present case, it is easy to find the following stationary point equations:
\be
\Re(S^{BF})= 0 \quad\text{iff} \quad
\left\{\begin{array}{l}
h_b {m_\ba}=\E^{-\I\varphi_\ab} h_a {m_\ab}\\
\tilde h_b {m_\ba}=\E^{-\I\tilde\varphi_\ab}\tilde h_a{ m_\ab}
\end{array}\right.
\label{crit_eq_BF1}
\ee
\be
\delta_{h_a}S^{BF}=0=\delta_{\tilde h_a}S^{BF}  \quad\text{iff} \quad \sum_{b, b\neq a} \epsilon_\ab j_\ab \hat m_\ab = \vec 0,
\label{crit_eq_BF2}
\ee
where $\varphi_\ab$ and $\tilde\varphi_\ab$ are arbitrary phases, and $\epsilon_\ab=-\epsilon_\ba=-\tilde\epsilon_\ab\in\{\pm1\}$ accounts for the reciprocal orientation of edges and faces.\footnote{More precisely $\epsilon_\ab$, $\tilde\epsilon_\ab$ are a rewriting (adapted to the present notation) of the edge-face incidence matrix ($\epsilon_{ef}$), up to an overall undetermined sign $\mu$. See \autoref{section_face}. } Finally, the variation of the action with respect to the unit vectors $\hat m_\ab$ does not lead to any further equation.


\subsubsection{The Geometrical Interpretation}
The stationary point equations (\autorefs{crit_eq_BF1} and \ref{crit_eq_BF2}) can be shown to define (see \autoref{section_....}) at each vertex a three dimensional tetrahedron in $\mathbb R^3$.\\

Each (internal) edge $a$ corresponds to a face of the tetrahedron. Hence one can associate to the vertex opposite to this face the same label $a$. Then the edge $\vec\ell_\ab$ going from the vertex $a$ to the vertex $b$ is given by a solution\footnote{Not \emph{any} set of spins $\{j_\ab\}$ admits a solution to the stationary point equations. Explicit conditions for the existence of a solution shall be given further. Also, the role of the degenerate sectors of such geometries is not yet completely clear. However, see \autoref{appendix_degenerate}.} of the stationary point equations via the formula $\vec\ell_\ab:=h_a\rhd\epsilon_\ab j_\ab \hat m_\ab$. Remark that by \autoref{crit_eq_BF2}  the edges relative to the same face ``close'' forming a triangle, as they should in a geometrical tetrahedron. Moreover, the dihedral angle $\Theta_\ab$ between faces $a$ and $b$ (``around'' the edge $\vec\ell_\ab$) is, essentially, given by $2\epsilon_\ab\varphi_\ab$.\\

Furthermore, the combinatorics of the melon graph, tells how the faces of the two tetrahedra have to be identified to form a (closed) three-manifold. This is done along the graph edges. In other term the face $a$ of a tetrahedron is identified with the face $a$ of the other one. Topologically, the so obtained manifold is a three-sphere.\footnote{In order to visualize this fact, it is useful to go one dimension down: glueing two copies of a triangle by their congruent sides, and ``inflating'' the so obtained ``sandwich'', one clearly obtains a two-sphere.} However, subtleties can arise from orientation issues. 
For the time being, it is enough to observe that, since the two tetrahedra have the same edge-lengths,  they must be congruent or one the parity-reversed of the other. Mathematically this choice is reflected at each vertex in the choice of an overall sign ($\nu_v$) in the relation between the angles $\Theta_\ab$ and the phases $2\epsilon_\ab\varphi_\ab$.

\bFIG[height=7cm]{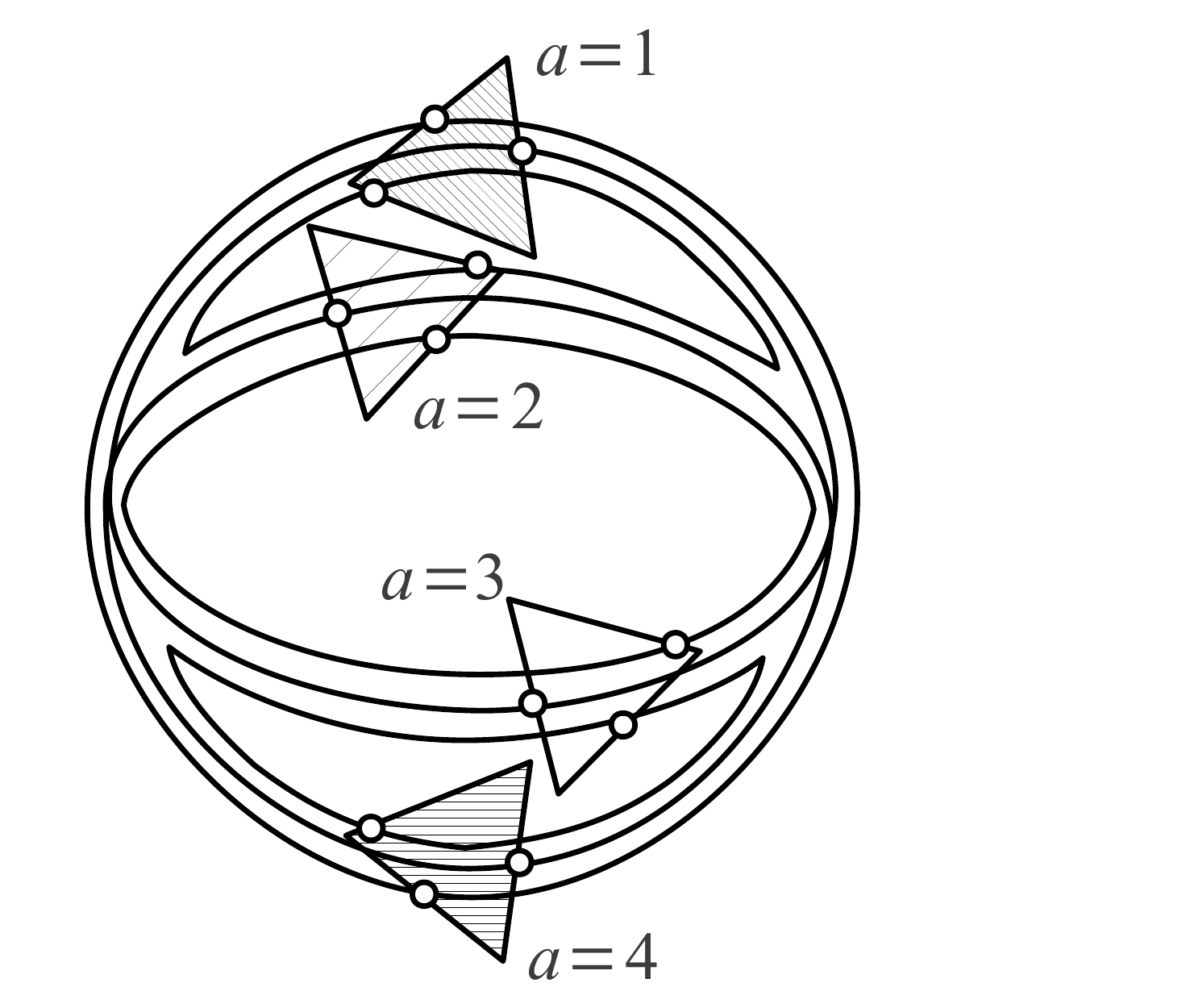}
\includegraphics[height=7cm]{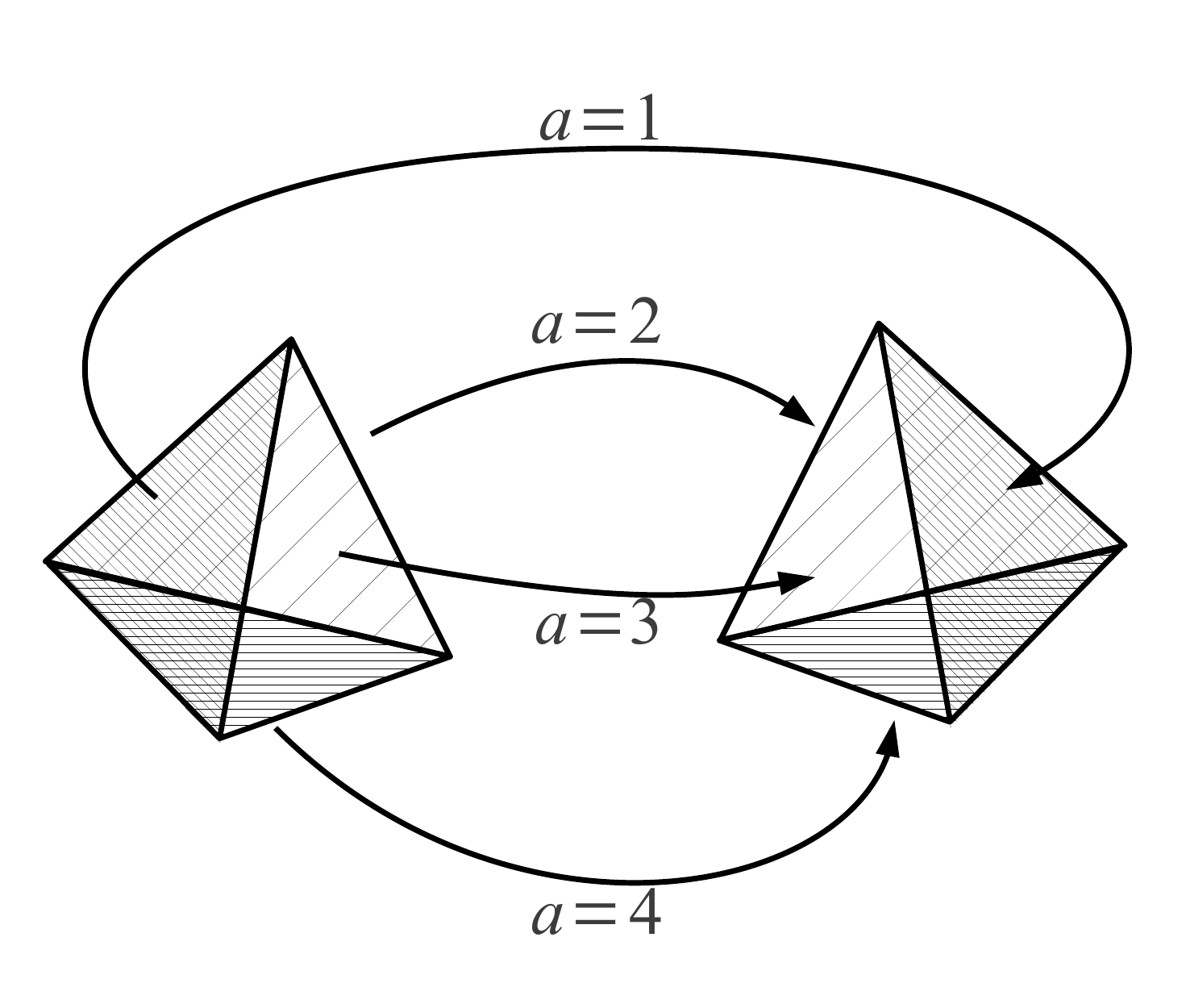}
\caption{The figure shows the combinatoric structure of the internal faces of the melon graph, and how it corresponds to two tetrahedra with faces identified.}
\label{tetrahedra_id}
\eFIG

\subsubsection[The BF Amplitude]{The BF Amplitude\label{section_BFevaluation}\protect\footnote{This section makes use of some results demonstrated only later, in the context of the EPRL-FK model. Most of these results apply unaltered to the BF case, just by restricting them to elements of \textit{SU}(2). However, this shall not always be the case. Whenever this happens it will be explicitly remarked in the text.}}

As discussed in the previous section, the stationary point equations \ref{crit_eq_BF1} and \ref{crit_eq_BF2} define two geometrical tetrahedra (one per each graph vertex) of edge-lengths given by the internal spins $\{j_\ab\}$. These tetrahedra can be either the same or be related by a parity transformation. In particular it will be shown that the spins $\{j_\ab\}$ basically determine the ${m_\ab}$ up to the transformation in \autoref{BF_e_invariance} and, once the ${m_\ab}$'s are fixed, they also determine up to a parity transformation the group elements
\be
H_\ab:=h_a^{-1}h_b\in\SU.
\ee
This is everything\footnote{This is almost true. More complete formula shall be given in the next sections. Our intention now is to skip over unimportant technicalities.} is needed to evaluate the melon graph amplitude in the stationary phase approximation.\\

As a result, the action at the stationary point is (essentially) given by
\be
S^\text{BF}_\text{stat} = \I\;(\nu_v+\nu_{\tilde v})\sum_{ab,a<b} j_\ab\;\Theta_\ab,
\ee
where the $\Theta_\ab\in[0,\pi]$ are the external dihedral angles associated to the geometrical tetrahedron of sides $\{j_\ab\}$, and $\nu_v$ and $\nu_{\tilde v}$ are sign factors ($\pm1$) assigning an explicit orientation to the tetrahedra associated to each vertex of the melon graph. Hence, it is natural to introduce the Regge action\footnote{
This definition is justified by the following fact. Given a closed triangulated 3-manifold $M$, define for each tetrahedron $\tau$ of edges $\{l_i\}$ the action $S^\tau_\text{Regge}[l_i]:=\sum_{i=1}^6 l_i\Theta^\tau_i[l]$, and hence the total action $S^M=\sum_\tau S^\tau_\text{Regge}$. On the other hand, the Regge action for $M$ is defined by $S^M_\text{Regge}=\sum_i l_i \delta_i$, where  $\delta_i = \left(\sum_\tau\Theta^\tau_i\right)-2\pi$. Therefore, $S^M=S^M_\text{Regge}\mod2\pi$ provided that the edges $\{l_i\}$ have integer (positive) length. In our case this is \textit{almost} the case, since $\{j_\ab\}\subset\frac{1}{2}\mathbb N$.
} for one tetrahedron
\be
S_\text{Regge}[j]:=\sum_{ab,a<b} j_\ab\Theta_\ab.
\ee

\paragraph{Equal Tetrahedra} In this case $\nu_v=\nu_{\tilde v}=\pm 1$, and the action at the stationary points is given by
\be
S^\text{BF}_\text{stat, eq} = \pm2\I S_\text{Regge}[j],
\label{equal_BF}
\ee
Moreover, at these stationary points, $h^0_a\neq \tilde h^0_a$ in any gauge (even if they are simply related), where the $0$ emphasizes the fact that this group elements are functions of the spins. This can be easily shown via the observation that $h_a^0=\tilde h_a^0$ implies $H_\ab=\tilde H_\ab$ and therefore (\autoref{BF_face_action}) $\left.S^\text{BF}_\ab\right|_\text{stat} = 0$, which is not consistent with \autoref{equal_BF}. \\

With these elements at hand, it is possible to find a formula for tsymmetrieshe asymptotic (large spin) behaviour of the partial amplitudes $w^\text{BF}_{\mathcal M,\text{eq}}$ at the equal-tetrahedra stationary point. Consider \autoref{BF_melon}, that we report here:
\be
w^{\text{BF}}_{\mathcal M}(j_a,n_a,\tilde n_a;j_\ab)=\int\DD h\DD\hat m\; \left[\prod_a\;\bra{n_a}h_a\tilde h_a^{-1}\ket{\tilde n_a}_{j_a}\right]\left[ \prod_{\ab, a<b} (2j_\ab+1)\;\E^{S^{\text{BF}}_\ab}\right]\;.
\ee
In the limit where all the internal spin scale like $\lambda\rightarrow\infty$, this integral (which is on a compact domain $\SU^{\times8}\times(S_2)^{\times12}$) can be approximated by stationary phase techniques with
\be
w^{\text{BF}}_{\mathcal M,\text{eq}\pm}\sim \left[\prod_{ab,a<b}(2j_\ab+1)^3\right] \lambda^{-\frac{1}{2}\rank(\delta^2S^\text{BF}_\text{eq})} F(j_a,n_a,\tilde n_a;j_\ab),
\ee
where the explicit form of the dependence on the external data is left for the moment unspecified in the function $F$, but shall be clarified very soon. Meanwhile, the rest of the formula deserves an explanation.\\
The first term is a product over the internal faces $(ab)$ of the face weight times the measure factors entering the integrals over $S_2$, contributing with one and two copies per face of the  $(2j_\ab+1)$, respectively. Indeed, recall that $\uD_{j_\ab}\hat m_\ab=(2j_\ab+1)\D \hat m_\ab$, and that $m_\ab\neq m_\ba$. In $\lambda$, this factor scales like $\lambda^{18}$.\\
The second term is the core of the stationary point approximation: for each direction around the stationary point in the integration space modulo gauge symmetries, the integral contributes by a factor\footnote{The $2\pi$'s are omitted in our order-of-magnitude treatment.} $\sqrt{2\pi/\lambda}$, provided that the Hessian of the action $\delta^2S^\text{BF}$ is not degenerate along that direction.\\

Now, in order to calculate $F$ in the previous formula, first suppose that the integrand in the original expression for $w^\text{BF}_{\mathcal M}$ were invariant under the action gauge transformations of \autorefs{BF_v_invariance} and \ref{BF_e_invariance}. In this case the conclusion would just be that $F$ is the evaluation on the stationary phase gauge orbit of the integrand we just mentioned. However, it is immediate to verify that the actual integrand is not invariant under the gauge transformation at the vertices (\autoref{BF_v_invariance}), because of the external face contribution. The way out of this impasse is to consider the average of the integrand on the gauge orbit. At a practical level, this procedure ``undoes'' the previous gauge fixing. Indeed, it result in
\be
F(j_a,n_a,\tilde n_a;j_\ab)=\E^{\pm2\I S_\text{Regge}[j]}\int_{\SU^{\otimes2}}\D k\D\tilde k\; \prod_a\;\bra{n_a}h_a\tilde h_a^{-1}\ket{\tilde n_a}_{j_a}.
\label{F_eq}
\ee
Actually, remark that the integrand would have been invariant, hadn't it been gauge fixed with $\delta(h)\delta(\tilde h)$. Moreover, in the context of $\SU$-BF theory, such gauge fixing was superfluous and, if it hadn't been performed, \autoref{F_eq} would have been obtained straightforwardly.\footnote{This conclusion could have been reached by another way, i.e. by performing in the first place a different gauge fixing (e.g. via $\delta(h_1)\delta(\tilde h_4)$), which would  break the vertex symmetry of \autoref{BF_v_invariance}, and obtain the integrals over $\SU$ in the previous formulas from the integrals (no more gauge-fixed) over the group elements $h$ and $\tilde h$ associated to the external half-edges.}\\

Before giving the explicit formula for $w^\text{BF}_{\mathcal{M},\text{eq}\pm}$, we first estimate the rank of the action Hessian. As already said, to estimate $\rank (\delta^2S^\text{BF}_\text{eq})$, one has to first calculate the dimension of the integration space in \autoref{BF_melon}, and then subtract to it the dimension of the gauge orbit related to each solution of the stationary point equations. This must be done, because the action is constant on each gauge orbit; hence, the directions spanned by the gauge transformations do not contribute to the Gaussian integrals. Therefore:
\be
\rank (\delta^2S^\text{BF}_\text{eq}) = 8\times \dim(\SU) + 12\times \dim(S_2)  - 4\times\dim(\SU) - 2\times\dim(\SU)  = 30,
\ee
where 8 is the number of $h_a$ and $\tilde h_a$, and 12 that of $\hat m_\ab$ over which we are integrating. 
Meanwhile, the 4 copies of $\SU$ we are subtracting correspond to the freedom of redefining the $\hat m_\ab$ at each edge following \autoref{BF_e_invariance}, and the two more copies are due to the freedom \ref{BF_v_invariance}. Remark also that this truly is an upper bound to the rank of the Hessian, which we supposed to attain its maximal value compatible with the symmetries. Therefore what we find is a lower bound on the degree of divergence. It shall be the comparison of our final result with the well-known solution to the problem, to assure \emph{a posteriori} that such lower bound  is attained in the $\SU$-BF case. \\

Finally, one gets for the partial amplitude:
\be
w^{\text{BF}}_{\mathcal M, \text{eq}} \sim \lambda^3 \E^{\pm2\I S_\text{Regge}[j]}\int_{\SU^{\otimes2}} \D k \D \tilde k\;\prod_a \;  \bra{n_a} k h_a^0 \tilde (h_a^0)^{-1} \tilde k^{-1} \ket{\tilde n_a}_{j_a}\;.
\ee
Unluckily, evaluating the total amplitude $W^{\text{BF},\Lambda}_{\mathcal M, \text{eq}}=\sum_{\{j_\ab<\Lambda\}}w^{\text{BF}}_{\mathcal M, \text{eq}}$ is extremely involved, since both the global scale and the dependence on the boundary data is depends on the internal spins $\{j_\ab\}$ in a highly non-trivial way. However, because of the highly oscillatory character of $\exp\pm2\I S_\text{Regge}[j]$ for large spins, one can argue that this term gets suppressed. We shall come back on this point in \autoref{section_eucl_st_pts}.

\paragraph{Parity Related Tetrahedra} In this case $\nu_v=-\nu_{\tilde v}=\pm1$ and the action evaluated at the stationary point is independent on the internal spins and is essentially zero:
\be
S^\text{BF}_\text{stat,p}=0.
\ee
Moreover, it is also possible to show, that in this case
\be
H_\ab=\tilde H_\ab\quad\Longrightarrow\quad h_a^0 = k \tilde h_a^0,
\ee
$k$ being an arbitrary $\SU$ element. Therefore, for these stationary points the partial amplitudes read
\be
w^\text{BF}_{\mathcal{M},\text{p}}\sim \lambda^3\int_\SU \D k\; \prod_a\; \bra{n_a}k\ket{\tilde n_a}_{j_a},
\ee
where the fact was used that the estimation of the rank of the action is the same as for two equal tetrahedra. This is a very simple expression, and the total amplitude can be easily estimated. Indeed, the dependence over the internal spins is confined to the global scale of the partial amplitude, while has completely dropped out of the terms involving the boundary data. Hence:
\be
W^{\text{BF},\Lambda}_{\mathcal M, \text{p}}=\sum_{\{j_\ab<\Lambda\}}w^{\text{BF}}_{\mathcal M, \text{p}}\sim \Lambda^9\int_\SU \D k\;\prod_a\; \bra{n_a}k\ket{\tilde n_a}_{j_a},
\label{div_BF}
\ee
as a consequence of the summation over the six internal faces.\\

At this point, we can put together the two contributions from the equal and parity-reversed tetrahedra solutions to the stationary point equations. Accepting the argument according to which the imaginary exponential of the Regge action appearing in the equal tetrahedra sector causes its suppression, one is lead to write at the dominant order:
\be
W^{\text{BF},\Lambda}_\mathcal{M}\sim\Lambda^9\int_\SU \D k\; \prod_a\;\bra{n_a}k\ket{\tilde n_a}_{j_a}.
\label{BFamplitude}
\ee

This is the result of our analysis of the $\SU$ BF self-energy graph. Notice it is polynomially divergent in the cut-off.

\begin{center}
* * *
\end{center}
This result can be verified through a simple calculation which is available in the BF case. Starting from the formal expression in terms of $\SU$ delta functions of the $\SU$-BF amplitude (\autoref{BF_delta}), it is indeed possible to first perform all the non-trivial integrals, and then to count how many redundant delta functions are left. Only at this point the delta function regularization is introduced (i.e. $\delta(\mathbb I)\mapsto\delta_\Lambda(\mathbb I)\sim\Lambda^3$) and the divergence degree readily calculated.\\

In formulas,it is found%
\footnote{
Explicitly:
$$
W^\text{BF}_\mathcal{M}=\left[\prod_{a}\int_\SU \D H_a\right] \prod_{ab,a<b}\delta(H_a^{\phantom{-}} H_b^{-1})\prod_a \bra{n_a}H_a\ket{\tilde n_a}.
$$
It is clear that the delta functions impose $H_a=H_b\equiv k\;\forall a,b$. However, this is already implied by the delta functions with $a=1$ and $b$ free to vary from 2 to 4.  The remaining delta functions contribute each with $\delta(\mathbb I)$. There are three of these terms, i.e. $(ab)\in\{(23),(24),(34)\}$.
}
\be
W^\text{BF}_\mathcal{M}=\Big[\delta(\mathbb I)\Big]^3\int_\SU \D k\; \prod_a\;\bra{n_a}k\ket{\tilde n_a}_{j_a}\sim\Lambda^9\int_\SU \D k\; \prod_a\;\bra{n_a}k\ket{\tilde n_a}_{j_a}.
\ee
Therefore, \autoref{BFamplitude} correctly captures the dominant behaviour of $W^{\text{BF},\Lambda}_\mathcal{M}$.\\

This is an \emph{a posteriori} check of our procedure in general and of some of its details in particular. E.g. the fact that the rank of the Hessian attains the maximal value allowed by the symmetries of the action; or the fact that the equal-tetrahedra stationary point is actually suppressed; or finally the fact that it is the \emph{uniform} large spin limit the relevant regime to study the most diverging contributions to the melon-graph amplitude.\footnote{It could also be mentioned the fact that this formula shows that the degenerate sector, which we did not mention in the context of BF theory to reduce its dissertation to the minimum, does not contribute at this order. See \autoref{appendix_degenerate}.} We shall use all these facts as a guide in the more complex gravitational case, we shall introduce in the following section about the EPRL-FK Spin Foam model.\\

%

To conclude, we add one last comment about $\SU$-BF theory. It is clear that the methodology employed above is by far not the most economic nor the most effective in this context. Despite this, it has the merit of being directly generalizable to the EPRL-FK model, while most of the other techniques one can find in the literature make substantial use of the peculiar properties of the particularly favourable BF-theory structure. The interested reader can find a much more thorough analysis of the melon graph in $\SU$-BF theory, and in particular of the sub-leading terms in its amplitude, in \cite{BenGelounBonzom:BFren}.

\newpage
\section{Lorentzian EPRL-FK on a 2-Complex\label{section_EPRL}}

Starting from this section, we shall treat the physically interesting case, i.e. the Lorentzian EPRL-FK amplitudes. In particular this section is dedicated to the general definition of the model.\\

The cut-off Lorentzian EPRL-FK amplitude $W_{\mathcal C}$ of an open two-complex $\mathcal C$, in the LS representation reads (e.g. \cite{Rovelli:Zakopane})
\begin{align}
W_\mathcal{C}^\Lambda(j_l, m_{nl}):=&{\sum_{\{j_f<\Lambda\}}}'\left\{
\left[\prod_f \prod_{e\in\partial f}\int_{S_2}\uD_{j_f}\hat m_{ef}\right]
\left[ \prod_v\prod_{e: v\in\partial e}\int_\sldc\D g_{ve} \right]\right.\notag
\\
&\hspace{0cm}\left.\phantom{\left[\prod_f \prod_{e\in\partial f}\int_{S_2}\uD_{j_f}\hat m_{ef}\right]}
{\prod_{f}}' \mu(j_f) \prod_{v\in\partial f}\bra{m_{e'f}} \Y^\dag g_{ve'}^{-1}g_{ve}\Y\ket{m_{ef}}_{j_f}
\right\}\;.\label{EPRL_spin}
\end{align}
where the same notations are used as in the $\SU$-BF case (\autoref{BF_m}). The new ingredients with respect to the BF amplitude are the use of $\sldc$ as the gauge group, and the insertion of the EPRL-FK $\Y$-map. This map provides an embedding of the representations $\mathcal H^{(j)}$ of $\SU$ into the unitary representation $\mathcal H^{(k,p)}=\mathcal H^{(\gamma j,j)}$ of $\sldc$:
\be\begin{array}{rcccc}
\Y &:&\mathcal H^{(j)} &\rightarrow& \mathcal H^{(\gamma j,j)}\\
&&\ket{j,n} &\mapsto & \ket{(\gamma j, j), j, n}
\end{array}\;,
\ee
where $n\in\{-j,\dots,j\}$ is the spin magnetic number. The parameter $\gamma\in\mathbb R$ appearing in the definition of the $\Y$-map, is the Barbero-Immirzi parameter. 
The role of the $\Y$-map is crucial, since it is meant to implement the simplicity constraints which, in a Plebanski formulation of gravity, transform an $\sldc$-BF theory into general relativity \cite{EPRL:EPRL}.\\

Remark that in \autoref{EPRL_spin} the face weight is left unspecified. This is because there is no general consensus thereabout. Mainly two schools of thought exist: $\mu(j)=2j+1$ \cite{BianchiRegoli:EPRLfaceampl,Rovelli:Zakopane}, as in $\SU$-BF theory, or $\mu(j)=(1+\gamma^2)j^2$ \cite{EPRL:EPRL,Perez:SpinfoamApproach}, as in a $\sldc$-BF theory constrained onto the image of the $\Y$-map. We shall keep these two options in mind, while generally noting the asymptotic scaling of $\mu(j)$ by
\be
\mu(\lambda)\sim\lambda^\mu\quad\text{as}\quad \lambda\rightarrow \infty.
\label{mu}
\ee
Nevertheless, the choice $\mu(j)=2j+1$ seems to us more natural, since it makes the face amplitude invariant under splitting \cite{BianchiRegoli:EPRLfaceampl}. 

\newpage
\section{The EPRL-FK Melon Graph: Notation\label{section_EPRL_melon}}
In the particular case of the melon graph $\mathcal M$ the notation used here, parallel those introduced in \autoref{Section_BF_notation}:
\begin{eqnarray}
W^{\Lambda}_{\mathcal M}(j_a,n_a,\tilde n_a) &= &\sum_{\{j_\ab<\Lambda\}} w_\mathcal M (j_a,n_a,\tilde n_a; j_\ab)\\
w_\mathcal M (j_a,n_a,\tilde n_a; j_\ab)&:=&\left[\int_{\sldc^{\otimes2}} \D g \D\tilde g\; \prod_a \int_{\sldc^{\otimes2}} \D g_a\D\tilde g_a \right]\left[\prod_{\ab:a<b}\int_{S_2{}^{\otimes2}}\uD\hat m_\ab\uD\hat m_\ba\right]\notag\\
&&\hspace{.5cm}\phantom{\int}\delta(g)\delta(\tilde g)\left[\prod_a\int_{S_2}\uD\hat m_a\bra{n_a}g^{-1}g_a\ket{m_a}_{\Y,j_a}\bra{m_a}\tilde g_a^{-1}\tilde h\ket{\tilde n_a}_{\Y,j_a} \right]\times\notag\\
&&\hspace{.5cm}\phantom{\int}\times\left[\prod_{ab:a<b}\mu(j_\ab)\;\bra{m_\ab}\tilde g_a^{-1}\tilde g_b\ket{ m_\ba}_{\Y,j_\ab} \bra{m_\ba}g_b^{-1}g_a\ket{m_\ab}_{\Y,j_\ab}\right]\notag\\
&&
\label{Mel_Gr_ampl}
\end{eqnarray}
where we used the shorthand notation $\bra{m}g\ket{\tilde m}_{\Y,j}:=\bra{m}\Y^\dag g \Y\ket{m}_j$.\\

\bFIG[width=11cm]{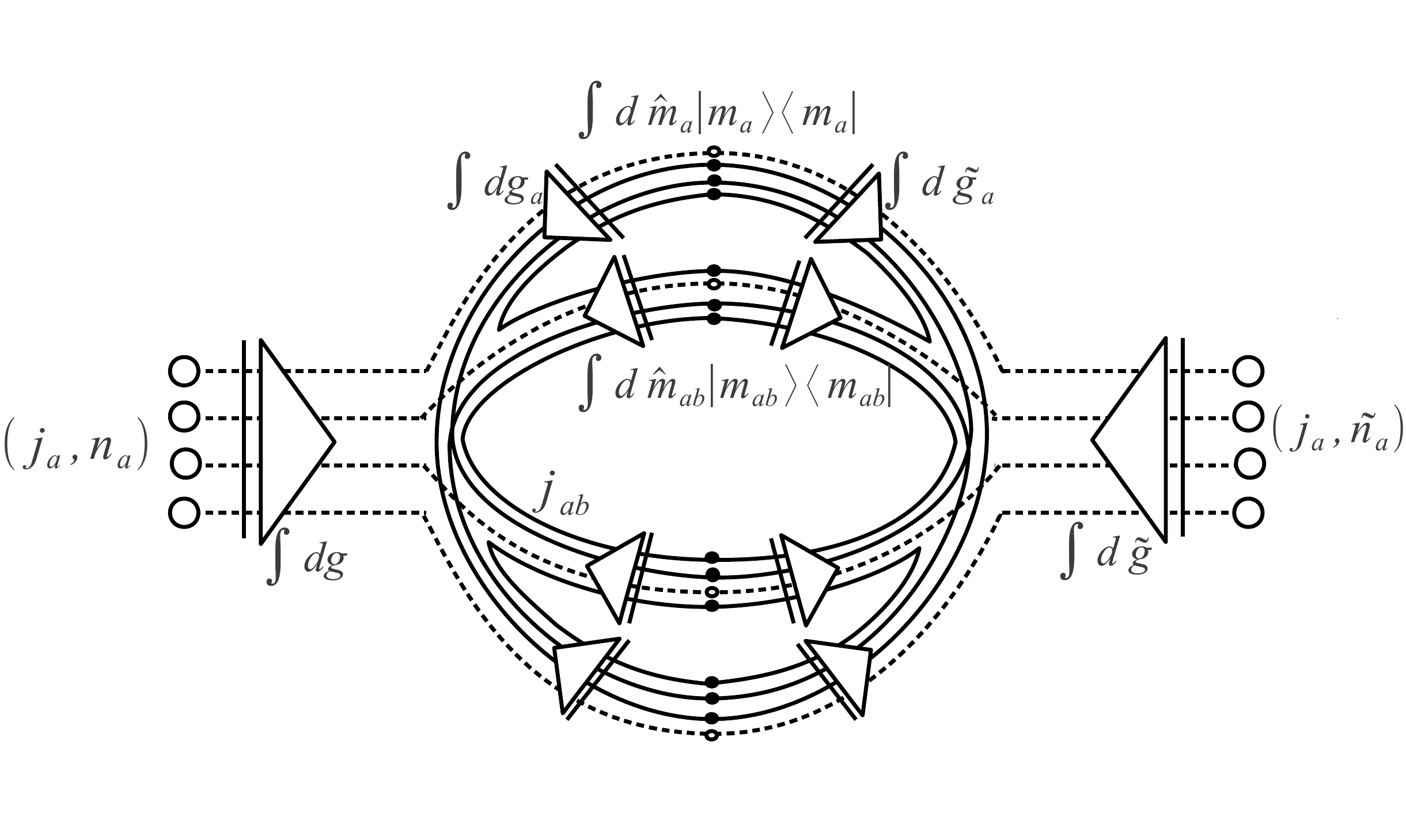}
\caption{The EPRL-FK melon graph. With respect to \protect\autoref{melon_simple} lines are added close to the triangles indicating group averaging: these lines represent the EPRL-FK $\Y$-map. Integrations over the $\mathbb{CP}^1$ spinors $\{z_\ab,\tilde z_\ab\}$ are not shown in the figure.}
\label{fig_eprl_melon}
\eFIG

This amplitude can be set in the form of a path integral. The details of the procedure leading to the path-integral formula can be found in \cite{HanZhang:LorAsympt} (and also \cite{BarrettHellmann:LorAsympt}, where different - but ultimately equivalent - conventions for the splitting of the traces onto coherent states are used). Here, we just give the final result:
\be
w_{\mathcal M}=\int\DD g \DD\hat m\DD z \left[\prod_a\int_{S_2}\uD\hat m_a\bra{n_a}g_a\ket{m_a}_{\Y,j_a}\bra{m_a}\tilde g_a^{-1}\ket{\tilde n_a}_{\Y,j_a}\right]\left[\prod_{ab,a<b}\mu(j_\ab)\;\E^{S_\ab}\right],
%
%
\label{EPRL_pathintegral}
\ee
where a shorthand notation for the integrals has been introduced:
\be
\int\DD g \DD\hat m\DD z:=\left[ \prod_a\int_{\sldc^{\otimes2}}\D g_a \D\tilde g_a\right]\left[\prod_{ab,a<b} \int_{S_2{}^{\otimes2}}\uD \hat m_\ab \uD\hat m_\ba \int_{\mathbb{CP}^1{}^{\otimes2}} \left(\frac{2j_\ab+1}{\pi} \right)^2 \Omega'_\ab \tilde\Omega'_\ab\right].
\ee

Before defining the new piece of this integration measure, we shall give the new form of the face action $S_\ab$:
\begin{eqnarray}
S_\ab&:=&j_\ab \ln \frac{\langle m_\ba| Z_\ba\rangle^2 \langle Z_\ab|m_\ab\rangle^2}{\langle Z_\ba|Z_\ba\rangle \langle Z_\ab|Z_\ab\rangle } + i \gamma j_\ab \ln \frac{\langle Z_\ab|Z_\ab\rangle}{\langle Z_\ba|Z_\ba\rangle} + \notag\\
&&\hspace{3cm}+j_\ab \ln \frac{\langle m_\ab|\tilde Z_\ab\rangle^2 \langle \tilde Z_\ba|m_\ba\rangle^2}{\langle \tilde Z_\ab|\tilde Z_\ab\rangle \langle \tilde Z_\ba|\tilde Z_\ba\rangle } + i \gamma j_\ab \ln \frac{\langle\tilde  Z_\ba|\tilde Z_\ba\rangle}{\langle \tilde Z_\ab|\tilde Z_\ab\rangle}\;.
\label{face_action}
\end{eqnarray}
Remark it is still proportional to $j_\ab$.\\

%
%
%
The $Z_\ab$'s are spinors defined by:
\be
\begin{array}{lcr}
Z_\ab:=g_a^\dag z_\ab& \text{and} & Z_\ba:=g_b^\dag z_\ab
\end{array}\;,\label{Z}
\ee
where ${z_\ab}={z_\ba}\in\mathbb{CP}^1$ (i.e. a spinor defined up to complex rescaling). Remark that $Z_\ab\neq Z_\ba$.\\

Finally, $\Omega'_\ab$ is a measure on $\mathbb{CP}^1$ (i.e. a scaling invariant measure on $\mathbb C^2$), defined by
\be
\Omega'_\ab:=\frac{\Omega_\ab}{\langle Z_\ab| Z_\ab\rangle \langle  Z_\ba| Z_\ba\rangle}\;,
\label{omega'}
\ee
where
\be\Omega_\ab:=\frac{i}{2}(z_\ab^0dz_\ab^1-z_\ab^1dz_\ab^0)\wedge(\bar z_\ab^0d\bar z_\ab^1-\bar z_\ab^1d\bar z_\ab^0)\;.
\label{omega}
\ee

It is understood that equations analogous to \autorefs{Z}, \ref{omega'}, and \ref{omega}, hold at the other vertex of $\mathcal M$ (tilded quantities).\\

The need for the introduction of the $\mathbb{CP}^1$ variables $\{z_\ab,\tilde z_\ab\}$, which are absent in the $\SU$-BF case, has to be traced to the necessity of explicitly writing the scalar products in the $\mathcal H^{(\gamma j, j)}=\Y \mathcal H^{(j)}$ representation of $\sldc$. Indeed, the canonical representation of $\mathcal H^{(k,p)}$ is in terms of homogeneous function of two complex variable \cite{BarrettHellmann:LorAsympt,HanZhang:LorAsympt} (see also \cite{Ruhl:sl2c} for the general approach).

\subsection{About the Symmetries of the Internal Action\label{Sect_sym}}

It is now important to analyse the symmetries of the total ``internal'' action:\footnote{This action is said to be ``internal'' because it is related to the internal faces of the graph only.}
\be
S:=\sum_{ab,a<b}S_\ab.
\label{eq_action}
\ee

Such symmetries will be found to be an obvious generalization of those found in the $\SU$-BF case. First of all, at each vertex one can left-rotate all the $g_a$ (resp. $\tilde g_a$) by an arbitrary $K\in\sldc$ (resp. $\tilde K$), provided one simultaneously transforms the $z_\ab$ (resp. $\tilde z_\ab$) appropriately:
\be
\left\{\begin{array}{l}
g_a \mapsto K g_a \\
Jz_\ab\mapsto KJz_\ab
\end{array}\right.\;\quad\text{and}\quad 
\left\{\begin{array}{l}\tilde g_a \mapsto \tilde K \tilde g_a\\
J\tilde z_\ab\mapsto \tilde KJ\tilde z_\ab
\end{array}\right.\;.
\label{v_invariance}
\ee
Furthermore, one can rotate the spinors on a given edge $\{m_\ab\}_{b,b\neq a}$ by a $k_a\in\SU$, provided one also right-rotates the sets $\{g_a\}$ and $\{\tilde g_a\}$ by the inverse of the same $k_a$:
\be
\left\{\begin{array}{l}
m_\ab\mapsto k_a m_\ab\quad\forall b,\; b\neq a\\
(g_a, \tilde g_a) \mapsto (g_a k_a^{-1}, \tilde g_a k_a^{-1})
\end{array}\right.\;.
\label{e_invariance}
\ee

Once more, we stress on how important the identification of these symmetries will be for a correct evaluation of the degree of divergence of the graph.\\

Since these symmetries are symmetries of the total \emph{internal} action and of the ``internal path integral'' integration measure
\be
\DD g\DD \hat m\DD z = \left[\prod_a \D g_a\D\tilde g_a\right]\left[ \prod_{ab,a< b}\uD\hat m_\ab\uD\hat m_\ba\right]\left[ \prod_{ab,a< b}\left(\frac{2j_\ab+1}{\pi}\right)^2\Omega'_\ab\tilde\Omega'_\ab \right] \;,
\ee
they shall be often referred to as \emph{gauge} symmetries. However, this term is inaccurate. Indeed, the amplitude of the external faces of the melon graph is invariant only under the transformations at the edges\footnote{Provided also $m_a\mapsto k_a m_a$.} (\autoref{e_invariance}), while it is \emph{not} invariant under the transformations at the vertices (\autoref{v_invariance}). This is essentially because of the gauge fixing performed on the external (half) edges (see end of \autoref{subsect3.1}). Therefore, as argued in \autoref{section_BFevaluation} just before \autoref{F_eq}, the evaluation of $w_\mathcal M(j_\ab)$ at the stationary ``points'' of the action is rather an average on the \emph{internal} ``gauge'' orbits of the solutions to the stationary point equations.\\

If this may sound odd, there is an equivalent way of performing this calculation that does not need this ``averaging''. It consists in gauge fixing to the identity a couple of internal holonomies (e.g. $g_1$ and $\tilde g_4$). In this way, there will be no invariance at the vertices for the internal action, and consequently no averaging; nevertheless, the integration over the two $\sldc$ group elements associated to the external edges would play the same role as the averaging in the previous setting. Also, the degree of divergence of the graph would stay obviously unaltered: there would be two less integration on $\sldc$ elements, but also two less $\sldc$ symmetries to take into account. Eventually, the two settings are equivalent.\\

The reason why we did not choose the latter option, is to keep an higher degree of symmetry in the stationary point equations.

\newpage
\section{Stationary Phase Approximation}\label{section_solve}
The basic assumption of this section, and of the entire article, is that it is meaningful to apply to the EPRL-FK melon graph the same techniques introduced in the previous sections in the context of the $\SU$-BF theory. Therefore, we shall suppose that the uniformly large internal spin regime is the pertinent one to be studied in order to understand the dominant term of the EPRL-FK self energy. However, one more assumption is needed here. The stationary phase approximation can indeed be confidently applied only to functions defined on a \emph{compact} domain. On the contrary, the EPRL-FK amplitude requires multiple integrals on $\sldc$ which is non-compact. Therefore special assumption have to be made on the behaviour of the $\sldc$ integrand for large group elements, in order to ensure the applicability of the techniques here proposed.\footnote{We thank M. Han for having pointed out this fact. } Supposing all the underlying hypothesis are satisfied, we go through the same analysis we sketched in the $\SU$-BF case.\\
%

By formally rescaling all the \emph{internal} spins $\{j_\ab\}$ by a common factor $\lambda\rightarrow\infty$, we are lead to consider the stationary points of the total action $S$:
\be
S:=\sum_{ab,a<b}S_\ab.\label{S}
\ee

Maximizing\footnote{In our conventions there is no minus sign in front of the action within the exponential inside the path integral (\autoref{EPRL_pathintegral}).} the real part of the total action (which can be easily shown to be always negative or null via Cauchy-Schwartz inequalities) leads to
\be
\Re(S)=0 \quad\text{iff} \quad
\left\{\begin{array}{l}
m_\ab=\E^{-\I\varphi^a_b}\frac{Z_\ab}{||Z_\ab||}\\
m_\ab=\E^{-\I\tilde\varphi^a_b}\frac{\tilde Z_\ab}{||\tilde Z_\ab||}
\end{array}\right. ,\label{z_stationary}
\ee
for some phases $\varphi^a_b, \tilde\varphi^a_b \in [0,2\pi]$. Using \autoref{Z} and the definition of $J$, this implies:
\be
\left\{\begin{array}{l}
 g_b J m_\ba= \frac{|| Z_\ab||}{||Z_\ba||}\E^{\I\varphi_\ab} g_a J m_\ab \\ 
\tilde g_b J m_\ba= \frac{|| \tilde Z_\ab||}{||\tilde Z_\ba||}\E^{\I\tilde\varphi_\ab} \tilde g_a J m_\ab\end{array}\right.    ,\label{transpJ}
\ee
where $\varphi_\ab:=\varphi^a_b-\varphi^b_a=-\varphi_\ba$, and similarly for their tilded counterparts. The vanishing of the variation of the action with respect to the variables $z_\ab,\;\tilde z_\ab,\;g_a$ and $\tilde g_a$ gives
\begin{eqnarray}
 \delta_{z_\ab}S=0=\delta_{\tilde z_\ab}S &\text{iff} &
\left\{\begin{array}{l}
 g_b m_\ba= \frac{|| Z_\ba||}{||Z_\ab||}\E^{-\I\varphi_\ab}g_a m_\ab  \\ 
\tilde g_b m_\ba= \frac{|| \tilde Z_\ba||}{||\tilde Z_\ab||}\E^{-\I\tilde\varphi_\ab}\tilde g_a m_\ab \end{array}\right.   , \label{transp}\\
\delta_{g_a}S=0=\delta_{\tilde g_a}S  &\text{iff} & \sum_{b, b\neq a} \epsilon_\ab j_\ab \hat m_\ab = \vec 0 \;.   \label{closure}
\end{eqnarray}
Here, as before, $\hat m_\ab:=\langle m_\ab|\vec\sigma |m_\ab\rangle\in S_2$ and $\epsilon_\ab=-\epsilon_\ba=-\tilde\epsilon_\ab\in\{\pm1\}$. The particular choice of face orientations we implicitly chose in writing \autoref{EPRL_pathintegral} gives (\emph{up to a global sign}) $\epsilon_\ab=1$ for $a<b$. Finally, the equations issued by the variations of the $m_\ab$'s do not lead to any further condition. \\

The stationary point equations \autorefs{z_stationary}, \ref{transp} and \ref{closure} are slightly laborious to work out. For this reason, and because of the fact that it is straightforward to apply to our case the techniques used in \cite{BarrettHellmann:LorAsympt,HanZhang:LorAsympt}, we shall not detail the steps which lead to the stationary point equations. The main thing one has to take into account is the fact we are rescaling only the \emph{internal} spins, and not all of them as in the previous references. Concretely, the action to be extremized is that of \autoref{S}, which only involves the data along internal faces of the melon graph. In this sense the graph external faces drop out of the ``closure'' equations \ref{closure}, and no ``parallel transport'' equations\footnote{As it shall be made clear later, ``closure'' equations are equations like \autoref{closure}, while ``parallel transport'' equations are equations like \autorefs{transpJ} and \ref{transp}.} analogous to \autorefs{transpJ} and \ref{transp} has to be considered along such faces. At a practical level, it is as if we were treating the large $j$ limit of a closed melon graph with four strands only. This is graphically represented in \autoref{fig_eprl_melon} by the use of different styles for the internal- and external-face strands.\\

Characterizing the solutions to these equations will be the goal of the following sections. In particular, equations \ref{z_stationary}, \ref{transp} and \ref{closure} shall be explicitly solved for the variables $\{\hat m_\ab\}$, $\{g_a, \tilde g_a\}$ and $\{z_\ab, \tilde z_\ab\}$ as functions of the spins $\{j_\ab\}$ only. These solutions shall be finally used to calculate the phase of the multiple integral of \autoref{EPRL_pathintegral} which defines the melon graph EPRL-FK amplitude at its stationary points.


\section{Solving the Stationary Point Equations\label{section_solutionstationaryphase}}

The aim of this section is to explicitly solve  the stationary point equations relative to the inner faces of the EPRL-FK melon graph (\autorefs{z_stationary}, \ref{transp} and \ref{closure}) in terms of the spins $\{j_\ab\}$ only. Remark that the quantities relative to the two vertices must satisfy the same equations. As it shall be made clear, the sought solution has a neat geometrical interpretation in terms of two glued tetrahedra. Each tetrahedron corresponds to a vertex of the melon graph. It is characterized by its edge lengths, which are given by the spins $\{j_\ab\}$ themselves. For the sake of clarity, we anticipate here the result at one vertex, leaving the explicit calculations at each vertex to the rest of \autoref{section_solutionstationaryphase}, and the glueing of the solutions at the two vertices to \autoref{section_face}.\\

It is useful to start with the following geometrical remark, whose demonstration can be found in \autoref{appendix_euclideanorlorentzian}. Any set of six edge lengths $\{j_\ab\}$ (strictly) satisfying triangular inequalities for any subset $\{j_\ab\}_{b,b\neq a}$ defines up to global rotations and boosts, and up to space inversion, a (non-degenerate) tetrahdron in $\mathbb R^{1,3}$. The tetrahedron can be either Euclidean or Lorentzian. In the first case the edge lengths satisfy also the condition of positivity of the so-called Caley-Menger determinant.\\


At each vertex of the melon graph, the solutions to the stationary point equations can then be classified into four cases, depending on the particular values of the spins $\{j_\ab\}$:

\paragraph{No Solution} If the $\{j_\ab\}_{b,b\neq a}$ do not satisfy the triangular inequalities at each edge $a$, then there is no solution to the stationary point equations (in particular to the closure equations). The corresponding graph amplitude results exponentially suppressed with the spin scale $\lambda$.

\paragraph{Euclidean Sector} If the triangular inequalities are strictly satisfied at every edge, and the Caley-Menger determinant of the spins $\{j_\ab\}$ is positive, then there are (at each vertex) exactly four solutions to the stationary point equations equations, which correspond to two different geometries characterised by a sign choice. These two geometries can be interpreted as the two different, parity-related\footnote{This can be seen as the change in sign of the $\hat z$-axis, corresponding to $\eta\mapsto-\eta$.} Eculidean tetrahedra one can build out of the six side lengths $\{j_\ab\}$. The two parities are characterised by a sign $\nu_v=\pm1$ (\autoref{7.21}). 

\paragraph{Lorentzian Sector} Similarly, if the triangular inequalities are strictly satisfied at every edge, but the Caley-Menger determinant of the spins $\{j_\ab\}$ is negative, then there are (at each vertex) exactly four solutions to the stationary point equations equations, which still correspond to two different geometries characterised by a sign choice.  However, these two geometries can now be interpreted as the two different, time-reversal-related ``Lorentzian tetrahedra'' one can build out of the six (Minkowskian) side lengths $\{j_\ab\}$. Remark that $j_\ab^2>0$, hence the reconstructed tetrahedron has space-like sides and faces. Once again the two possible geometries are characterised by a sign $\nu_v=\pm1$ (\autoref{7.21}).

\paragraph{Degenerate Sector} If the triangular inequalities are satisfied at each edge by the $\{j_\ab\}_{b,b\neq a}$, but they are saturated at least at one edge, then the graph is said to belong to the (geometrically) \emph{degenerate sector}. It shall as well be called degenerate, the sector spanned by the $\{j_\ab\}$ strictly satisfying the triangular inequalities, whose Caley-Menger determinant is null (zero volume tetrahedra). The study of this sector is left for future work, however see \autoref{appendix_degenerate}.

\begin{center}
* * *
\end{center}
This section is organized as follows. In the first part we shall completely solve the full set of
stationary point equations in a purely algebraic way. In this way the existence of the two different families of solutions is shown (Euclidean, Lorentzian). In the second part, we shall provide a geometrical interpretation to the stationary point equations for the non-degenerate Euclidean and Lorentzian solutions. Such an interpretation shall allow us to give an explicit meaning to the different quantities involved in the solutions to the stationary point equations at the end of this section.\\

All along the calculation we shall take advantage of the possibility of solving all the equations explicitly, trying to add a flavour of concreteness to the more abstract works of \cite{BarrettHellmann:LorAsympt,HanZhang:LorAsympt}.

\subsection{Algebraic Solution\label{section7.1}}
At an edge $a$, \autoref{closure},  hereafter called the ``closure equation'', implies that the three vectors $\{\hat m_\ab\}_{b,b\neq a}$ are coplanar. Therefore, we can use the freedom of $\SU$ rotating the spinors at each edge (see \autoref{e_invariance}), in order to gauge fix the plane the vectors lie on, to be the $xz$-plane. In the notation of \autoref{section_notations}, this means that the $m_\ab$'s take the form\footnote{Remark that this is the case for any choice of $\psi_\ab=\frac{\pi}{2}k$, $k\in\{0,1,2,3\}$, possibly at the cost of redefining the phase $\phi_\ab$ and the angle $\theta_\ab$. Explicitly: $$\phi_\ab \mapsto \phi_\ab+\frac{\pi}{2}k,\quad\text{and}\quad \theta_\ab\mapsto (-1)^k\theta_\ab.$$}
\be
m_\ab = \E^{\I\phi_\ab}\left(\begin{array}{c} \cos\theta_\ab \\\sin\theta_\ab \end{array}\right)\;.
\label{eq8}
\ee
Remark that the $m_\ab$'s are normalized spinors, according to what is needed for the resolution of the identity $\mathbb{I}_j = \int_{S_2}\uD_j \hat m \ket{m}_j\bra{m}_j$ (see \autoref{id_resolution}).\\

Via a  further rotation of the $\{\hat m_\ab\}_{b,b\neq a}$ within the $xz$-plane, it is possible to set $\theta_\ab=0$ for a particular $b$. Then, it is straightforward to solve the closure equation for the differences $\theta^a_{bc}:=(\theta_\ab-\theta_{ac})$, provided that the $\{j_\ab\}_{b,b\neq a}$ respect the triangular inequalities (otherwise \emph{no} solution to this equation exists). This gives
\be
\cos2\theta^a_{bc} = -\epsilon_\ab\epsilon_{ac} \frac{j_\ab^2+j_{ac}^2-j_{ad}^2}{2 j_\ab j_{ac}}\;;
\label{coseno}
\ee
furthermore, the closure equation fixes the relative signs between the previous differences in the following way:
\be
\sgn[\epsilon_\ab\sin2\theta^a_{bc}) ] = \sgn[\epsilon_{ad}\sin2\theta^a_{cd} ].
\label{signs}
\ee
where in the last two equations the convention is used, that different letters stand for different values of the indices. This convention should be used throughout the rest of this section.

 It is now convenient to define a new set of angles $\{\vartheta_\ab\}$, in such a way to reabsorb the dependence from the $\{\epsilon_\ab\}$:
\be
\vartheta_\ab : = \theta_\ab + \frac{\epsilon_\ab-1}{4}\pi \quad \Longrightarrow \quad \cos2\vartheta_{bc}^a =\epsilon_\ab\epsilon_{ac}\cos2\theta^a_{bc},\;\;\sin2\vartheta_{bc}^a =\epsilon_\ab\epsilon_{ac}\sin2\theta^a_{bc}
\label{vartheta}
\ee
This new definition allows to write the previous two equations in the cleaner form:
\be
\cos2\vartheta_{bc}^a = -\frac{j_\ab^2+j_{ac}^2-j_{ad}^2}{2 j_\ab j_{ac}}\;
\ee
\be
\sgn[\sin2\vartheta^a_{bc}] = \sgn[\sin2\vartheta^a_{cd} ].
\label{sinesign}
\ee
The first equation has two solutions for $2\vartheta_{bc}^a$ (modulo $2\pi$) related by a sign change.\footnote{While it has four solutions for $\vartheta_{bc}^a$.} The second equation fixes the relative signs of these angles at a given edge ($a$, in this case). Therefore, at fixed $\{\epsilon_\ab\}$, exactly two solutions per edge exist.\\

The attentive reader has surely noticed that the previous equation simply follows from the interpretation of the vectors $\{\epsilon_\ab j_\ab \hat m_\ab\}_{b,b\neq a}$ in terms of the sides of a triangle, as dictated by the closure equation. Nevertheless, we shall postpone the geometrical discussion.\\

Now that the closure equations have been solved by assigning a particular value to the $m_\ab$'s, we can focus on the other two stationary point equations. These shall allow us to solve for the $\sldc$ group elements $\{g_a\}$. To do this, we rewrite equations \autorefs{transpJ} and \ref{transp} as a unique matrix equation;
\be
g_b D(m_\ba) = g_a D(m_\ab)\left(\begin{array}{cc} \frac{|| Z_\ba||}{||Z_\ab||}\E^{-\I\varphi_\ab} & 0 \\ 0 & \frac{|| Z_\ab||}{||Z_\ba||}\E^{\I\varphi_\ab}  \end{array}\right),
\label{transp_mJm}
\ee
where $D(m):=(m,Jm)\equiv\E^{\I\psi\sigma_z}\E^{-\I\theta\sigma_y}\E^{\I\phi\sigma_z}$ is the $\SU$ matrix with the two (normalized) spinors $m$ and $Jm$ as columns.\footnote{Cf. \autoref{footnoteD(w)}. 
} Remark that in this formula the $m_\ab$ are gauge fixed in such a way that $\psi_\ab=0$. It is then convenient to define
\be
\eta_\ab\equiv-\eta_\ba:=\I(\varphi_\ab-\phi_\ab+\phi_\ba) + \ln\frac{||Z_\ab||}{||Z_\ba||} \;\in \mathbb C\;,
\label{eta}
\ee
and
\be
G_\ba\equiv G_\ab^{-1} := g_b^{-1} g_a\;\in\sldc\;.
\ee
In the next section, it shall be shown that these two quantities have a neat geometrical interpretation.\\

In terms of the new variables, \autoref{transp_mJm} simply fixes the form of the $G_\ab$'s:
\be
G_\ba = \E^{-\I\theta_\ba\sigma_y}\E^{\eta_\ab\sigma_z}\E^{\I\theta_\ab\sigma_y}\;.
\label{transp_G}
\ee

However, most of the physical information lies now in the fact that the $G_\ab=g_a^{-1}g_b$ are not all independent from one another. Indeed they must satisfy three consistency equations:
\be
\left\{\begin{array}{l}
G_{ab}G_{bc}G_{ca}=\mathbb{I}\\
G_{ac}G_{cd}G_{da}=\mathbb{I}\\
G_{ab}G_{bd}G_{da}=\mathbb{I}
\end{array}\right. .\label{consistency}
\ee
A fourth equation follows from the previous three: $G_{bc}G_{cd}G_{db}=\mathbb I$.\\

An important feature of these equations is that they involve only the differences between the angles $\theta$ at a same edge of the graph. E.g., the first one involves the differences $\theta^b_{ac}\equiv(\theta_{ba}-\theta_{bc})$, $\theta^c_{ba}$ and $\theta^a_{cb}$. As already observed, such differences are essentially determined by the spins $j_\ab$'s via the closure equation. This means that \autorefs{consistency} can be considered as equations for the complex variables $\eta_\ab$.\\

Also, notice that since the $m_\ab$'s are in common between the equations relative to the two vertices, $\Im(\tilde\eta_\ab)=\tilde\varphi_\ab-\phi_\ab+\phi_\ab$, for the \emph{same} $\phi_\ab$ and $\phi_\ba$ appearing in $\Im(\eta_\ab)$. This fact shall be of importance later, where it will let the unphysical phases $\phi_\ab$ drop out of the final formula.\\

Now, in order to solve this system in \autoref{consistency}, we start by analysing one single equation, e.g. the first one. Using \autoref{transp_G}, it can be cast into the form:
\be
\E^{\eta_\ba\sigma_z}\E^{\I\theta^b_{ac}\sigma_y}\E^{\eta_{cb}\sigma_z}=\E^{-\I\theta^a_{cb}\sigma_y}\E^{-\eta_{ac}\sigma_z}\E^{-\I\theta^c_{ba}\sigma_y},
\ee 
where the differences between the angles $\theta^b_{ac}\equiv(\theta_{ba}-\theta_{bc})$ are put into evidence.\\

One way to solve this equation is to explicitly write the two matrices appearing on its left and right hand side
\begin{eqnarray}
&\hspace{0cm}\left(
\begin{array}{cc}
\E^{\eta_\ba+\eta_{cb}}\cos(-\theta^b_{ac})&-\E^{\eta_\ba-\eta_{cb}}\sin(-\theta^b_{ac})\\
\E^{-\eta_\ba+\eta_{cb}}\sin(-\theta^b_{ac}) & \E^{-(\eta_\ba+\eta_{cb})}\cos(-\theta^b_{ac})
\end{array}
\right)&=\\&&\hspace{-7.5cm}=
\left(
\begin{array}{cc}
\E^{-\eta_{ac}}\cos\theta^a_{cb}\cos\theta^c_{ba}-\E^{\eta_{ac}}\sin\theta^a_{cb}\sin\theta^c_{ba} & -\E^{\eta_{ac}} \sin\theta^a_{cb}\cos\theta^c_{ba}-\E^{-\eta_{ac}}\cos\theta^a_{cb}\sin\theta^c_{ba}\\
\E^{\eta_{ac}} \cos\theta^a_{cb}\sin\theta^c_{ba} + \E^{-\eta_{ac}}\sin\theta^a_{cb}\cos\theta^c_{ba} & \E^{\eta_{ac}} \cos\theta^a_{cb}\cos\theta^c_{ba}-\E^{-\eta_{ac}}\sin\theta^a_{cb}\sin\theta^c_{ba} 
\end{array}
\right).\notag
\label{app_bigmatrix}
\end{eqnarray}
and taking the sums and the differences of the elements along the two diagonals:
\be
\left\{\begin{array}{l}
\ch(-\eta_\ba-\eta_{cb})\cos(-\theta^b_{ac})=\ch\eta_{ac}\cos(\theta^c_{ba}+\theta^a_{cb})\\
\ch(-\eta_\ba+\eta_{cb})\sin(-\theta^b_{ac})=\ch\eta_{ac}\sin(\theta^c_{ba}+\theta^a_{cb})\\
\sh(-\eta_\ba-\eta_{cb})\cos(-\theta^b_{ac})=\sh\eta_{ac}\cos(\theta^c_{ba}-\theta^a_{cb})\\
\sh(-\eta_\ba+\eta_{cb})\sin(-\theta^b_{ac})=\sh\eta_{ac}\sin(\theta^c_{ba}-\theta^a_{cb})
\end{array}\right. .
\label{app_sys}
\ee

This system of equations can now be easily solved for 
 $\ch 2 \eta_{ac}$, yielding
\be
\ch2\eta_{ac} = \frac{\cos2\theta^c_{ba}\cos2\theta^a_{cb}-\cos2\theta^b_{ac}}{\sin2\theta^c_{ba}\sin2\theta^a_{cb}}.
\ee
And using \autoref{vartheta}:
\be
\ch2\eta_{ac} = \frac{\cos2\vartheta^c_{ba}\cos2\vartheta^a_{cb}+\cos2\vartheta^b_{ac}}{\sin2\vartheta^c_{ba}\sin2\vartheta^a_{cb}}=:\Delta_{ac}[j]\;.
\label{ch2eta}
\ee
Here we defined the function $\Delta_{ac}[j]\in\mathbb R$, which depends only on (all) the spins $\{j_\ab\}$ (as well as on a choice of sign at the two edges $a$ and $c$. See the discussion after \autoref{vartheta}). In particular it does \emph{not} depend on the choice of the $\{\epsilon_\ab\}$.\\

Since the function $\Delta[j]$ is real, $2\eta$ can take only particular values (modulo $2\I\pi$):
\begin{itemize}
\hypertarget{delta_1}{\item[{[Eucl]}]} if $\Delta\in[-1;1]$, then $2\eta=\pm\I\;\text{acos}\Delta$,
\hypertarget{delta_2}{\item[{[Lor1]}]} if $\Delta\in[1;\infty]$, then $2\eta=\pm\text{ach}{\Delta}$,
\hypertarget{delta_3}{\item[{[Lor2]}]} if $\Delta\in[-\infty;-1]$, then 
$2\eta=\pm\left(\text{ach}{\Delta}+
\I\pi\right)$,\label{delta_3}
\end{itemize}
where we neglected cases in which particular numerical coincidences among the values of the $\{j_\ab\}$ are such that $\Delta\in\{\pm1,\pm\infty\}$.\footnote{These values correspond to the degenerate geometrical sector. As it shall become clear later, $\Delta=\pm1$ implies a null volume of the tetrahedron  with no degenerate faces, while $\Delta=\pm\infty$ implies a null volume of the tetrahedron because of the degeneracy of some of its faces.} Remark that the number of solutions for $\eta$ (modulo $2\I\pi$) itself doubles with respect to the number of solutions for $2\eta$.\\

Solutions of type [Eucl] ([Lor]) are said ``Euclidean'' (``Lorentzian'') and will be shown to correspond to the Euclidean (Lorentzian) geometrical sector. Indeed, the parameters $2\eta$ will be shown to be related to the dihedral angles among faces of a tetrahedron: real and imaginary values of this angles correspond to pure rotations and boosts, respectively.\\

Now that \autoref{ch2eta} gives the norm of every $2\eta_{ab}$ as a function of the $\{j_\ab\}$, one would like to control also its sign. First of all, notice that if $\{2\eta_\ab\}$ solves \autorefs{app_sys}, then so does $\{-2\eta_\ab\}$. Therefore, there is no way to extract from these equations the absolute sign of the $2\eta_\ab$'s. Nevertheless, it is possible to extract their relative signs.\\

A preliminary step in this direction is to show that if one of the $\eta_\ab$'s is in the Euclidean sector, then so are all the others. To do this, it is enough to add among them the products of the first by the fourth and the second by the third of \autorefs{app_sys}, to obtain
\be
\frac{\sh2\eta_\ba}{ \sin 2\theta^c_{ba}} =\frac{ \sh 2\eta_{ac}}{ \sin 2\theta^b_{ac}}\;,
\ee
or, equivalently:
\be
\epsilon_{ab}\frac{\sh2\eta_\ab}{ \sin 2\vartheta^c_{ab}} = \epsilon_{ca}\frac{ \sh 2\eta_{ca}}{ \sin 2\vartheta^b_{ca}}.
\label{sh2eta}
\ee
As an immediate consequence of this equation, it is found that if $2\eta_\ab\in\I\mathbb R$ at one face $(ab)$ (case \hyperlink{delta_1}{[Eucl]} in the list above), then so will be at any other face. In this case $G_\ab\in\SU \;\forall ab$ (\autoref{transp_G}). Otherwise, the $G_\ab$'s are genuine elements of $\sldc$ (even if not the most general, \autoref{transp_G} and cases \hyperlink{delta_2}{[Lor1]} and \hyperlink{delta_3}{[Lor2]} of the list above).\\

To proceed with the analysis of the signs, it is first necessary to perform another gauge fixing on the $\{m_\ab\}$. The will is to fix the signs of the $\sin2\vartheta^a_{cd}$'s, which appear in \autorefs{sinesign}, \ref{ch2eta} and \ref{sh2eta}, in some \emph{global} way. This shall be geometrically interpreted as the fixing of a coherent orientation throughout the tetrahedron.\\
We start by noticing that at \emph{one} edge $a$, the sign of all the $\vartheta^a_{cd}$ (see \autoref{sinesign}) can be easily changed by acting on the $\{m_\ab\}_{b, b\neq a}$ with $k_a=\E^{\I\frac{\pi}{2}\sigma_z}\in\SU$:\footnote{Remark that this gauge fixing is compatible with the previous one, done at the very beginning of this section.}
\be
m_\ab=\E^{\I\phi_\ab}\left(\begin{array}{c}\cos\theta_\ab\\\sin\theta_\ab\end{array}\right) \xrightarrow{\E^{\I\frac{\pi}{2}\sigma_z}} \E^{\I\phi_\ab+\I\frac{\pi}{2}}\left(\begin{array}{c}\cos(-\theta_\ab)\\\sin(-\theta_\ab)\end{array}\right)\;\qquad \forall b\neq a.
\label{gauge_transf_pimezzi}
\ee
Indeed, this implies (modulo $2\pi$):
\be
2\vartheta_\ab \xrightarrow{\E^{\I\frac{\pi}{2}\sigma_z}} -2\vartheta_\ab  \;\;\forall b \neq a \quad\text{and therefore} \quad 2\vartheta^a_{bc} \xrightarrow{\E^{\I\frac{\pi}{2}\sigma_z}} -2\vartheta^{a}_{bc} \;\;\forall b,c\neq a.
\ee
The gauge fixing then goes as follows. Using the completely antisymmetric tensor $\epsilon_{abcd}$ such that $\epsilon_{1234}=+1$, and an assignement of a (different) number from 1 to 4 to each edge, the gauge transformation \autoref{gauge_transf_pimezzi}  is performed at the edge $a$ if and only if $\sgn[\sin2\vartheta^{a}_{bc}]\neq\epsilon_{abcd}$, where $d$ is the only element of the set $\{1,2,3,4\}\setminus\{a,b,c\}$. Remark that this prescription is consistent with \autoref{sinesign} (which was in turn implied by the closure condition) since $\epsilon_{abcd}=\epsilon_{acdb}$.\\

Within this gauge choice, \autoref{sh2eta} acquires a very simple meaning when combined with the explicit solution for $\ch2\eta_\ab$ (\autoref{ch2eta}):
\be
2\epsilon_\ab\eta_\ab = \nu_v \Theta_\ab(j)\mod 2\I\pi\;,\quad\text{where}\;\Theta_\ab(j)\in\left\{
\begin{array}{lc}
\I[0,\pi]&\text{[Eucl]}\\
\mathbb R^+ + \{0,\I\pi\}&\text{[Lor]}
\end{array}
\right.
\label{7.21}
\ee
i.e. where $\Theta_\ab(j)\in\I[0,\pi]$ in the Euclidean sector (\hyperlink{delta_1}{[Eucl]}), while $\Theta_\ab(j)\in\big(\mathbb R^+ + \{0,\I\pi\}\big)$ in the Lorentzian one (\hyperlink{delta_2}{[Lor1]} or \hyperlink{delta_3}{[Lor2]}). Moreover, $\Theta_\ab(j)$ is a function of the spins only, and in particular it is \emph{independent} of the $\epsilon_\ab$'s. Finally $\nu_v\in\{\pm1\}$ is an overall sign that can be independently chosen at each vertex $\{v,\tilde v\}$. \\

The last step in the solution of the stationary point equations, is to find the solutions for the $\eta_\ab$'s, instead of the $2\eta_\ab$ as in \autoref{7.21}. This step is not a completely trivial, since the $\eta_\ab$'s are defined modulo $2\I\pi$. Therefore, the possible solutions for $\eta_\ab$ are both $\frac{1}{2}\times(2\eta_\ab)$ and $\frac{1}{2}\times(2\eta_\ab)+\I\pi$. However, \autorefs{consistency} put some constraints on these choices, which are not all independent from one another. In other words, the choice at one face $(ab)$ influences the choice at the other faces. Unluckily, there is no geometrical interpretation of this choice. Indeed, as it shall be clear from the discussion in the next section, the geometrical interpretation stems from the spin $1$ representation of the parallel transport equations, which need on the contrary be solved in their fundamental (spin $\frac{1}{2}$) representation.\\

Unfortunately, the way the transformation $\eta_{ab}\mapsto\eta_{ab}+\I\pi$ influence the other $\{\eta_{cd}\}_{(cd)\neq(ab)}$ is not unique, and various solutions are possible. Indeed, it is not hard to convince oneself via \autoref{app_sys}, that if e.g. $\eta_{12}\mapsto\eta_{12}+\I\pi$ then one of the following cases is given:
\begin{itemize}
\item also $\eta_{23},\;\eta_{34},\;\eta_{14}$ acquire an $\I\pi$;
\item also $\eta_{23},\;\eta_{24}$ acquire an $\I\pi$;
\item also $\eta_{13},\;\eta_{14}$ acquire an $\I\pi$;
\item also $\eta_{13},\;\eta_{24},\;\eta_{34}$ acquire an $\I\pi$.
\end{itemize}

We shall not go into the details of these solutions, which may be cumbersome to write down explicitly. However, we shall generally note the fact that ``some'' $\I\pi$'s may be needed ``here and there'' with the following symbolical expression
\be
\eta_\ab = \frac{1}{2}\nu_v\epsilon_\ab\Theta_\ab(j)+[\I\pi]_\ab\; \mod 2\I\pi\;, \quad\text{where}\;\Theta_\ab(j)\in\left\{
\begin{array}{lc}
\I[0,\pi]&\text{[Eucl]}\\
\mathbb R^+ + \{0,\I\pi\}&\text{[Lor]}
\end{array}
\right.
\label{7.22}
\ee
Luckily, in \autoref{section_face} it will be shown that the detailed knowledge of these terms is not necessary. This is because of some global properties of the graph amplitude, which will allow us to bypass this issue.


\subsection{Geometric Interpretation}
Since the work of Ponzano and Regge \cite{PonzanoRegge}, the standard idea behind the large spin limit of spin foams is that of recovering a classical discrete geometry. We shall show that also in the case of the EPRL-FK melon graph with fixed external spins, such an interpretation is possible. Since our analysis practically concerns only three out of four strands per edge, it shall be related only to a three dimensional geometry, rather than to a four dimensional one as expected for more complex graphs \cite{HanZhang:LorAsympt}.\\

At the beginning of the previous section, it was already mentioned how the closure equations (\autoref{closure}) define at each edge a geometrical triangle $a$ embedded in $\mathbb R ^3$, and how the latter is determined by its side lengths $\{j_\ab\}_{b,b\neq a}$ up to rotations. It was then algebraically shown that it was possible to gauge fix such triangles to lie in the $xz$-plane. In the following, we shall make explicit how the other stationary point equation can be also interpreted in terms of these triangles: they will be seen to encode their parallel transports from the ``edge reference frame'' to the ``vertex reference frame'', in such a way to constitute in the latter frame the faces of a geometrical tetrahedron. Such a tetrahedron shall be Euclidean (embeddable in $\mathbb R^3$) or Lorentzian (embeddable in $\mathbb R^{2,1}$), according to the values of their edge lengths (or, equivalently, to the values of the $\Delta[j]$'s in \autoref{ch2eta}).\\

On a concrete level, the goal of this section is to give a neat geometrical meaning to the rather abstract functions $\Theta_\ab(j)$ of the previous section, which will be found to be anything but the dihedral angle (respectively boost parameters, in the Lorentzian sector) among the faces of the tetrahedron associated to each graph vertex. During this quest, the following geometrical meaning will be found for the $G_\ab$'s: their spin 1 representation will be identified with the rotations (respectively Lorentz transformations) taking from one face of the tetrahedron to another.\\

Despite their algebra and interpretation are very similar, the Euclidean and the Lorentzian sectors shall be treated separately since the beginning, because of technical reasons. \\



\subsection{The Euclidean Sector}

\bFIG[width=11cm]{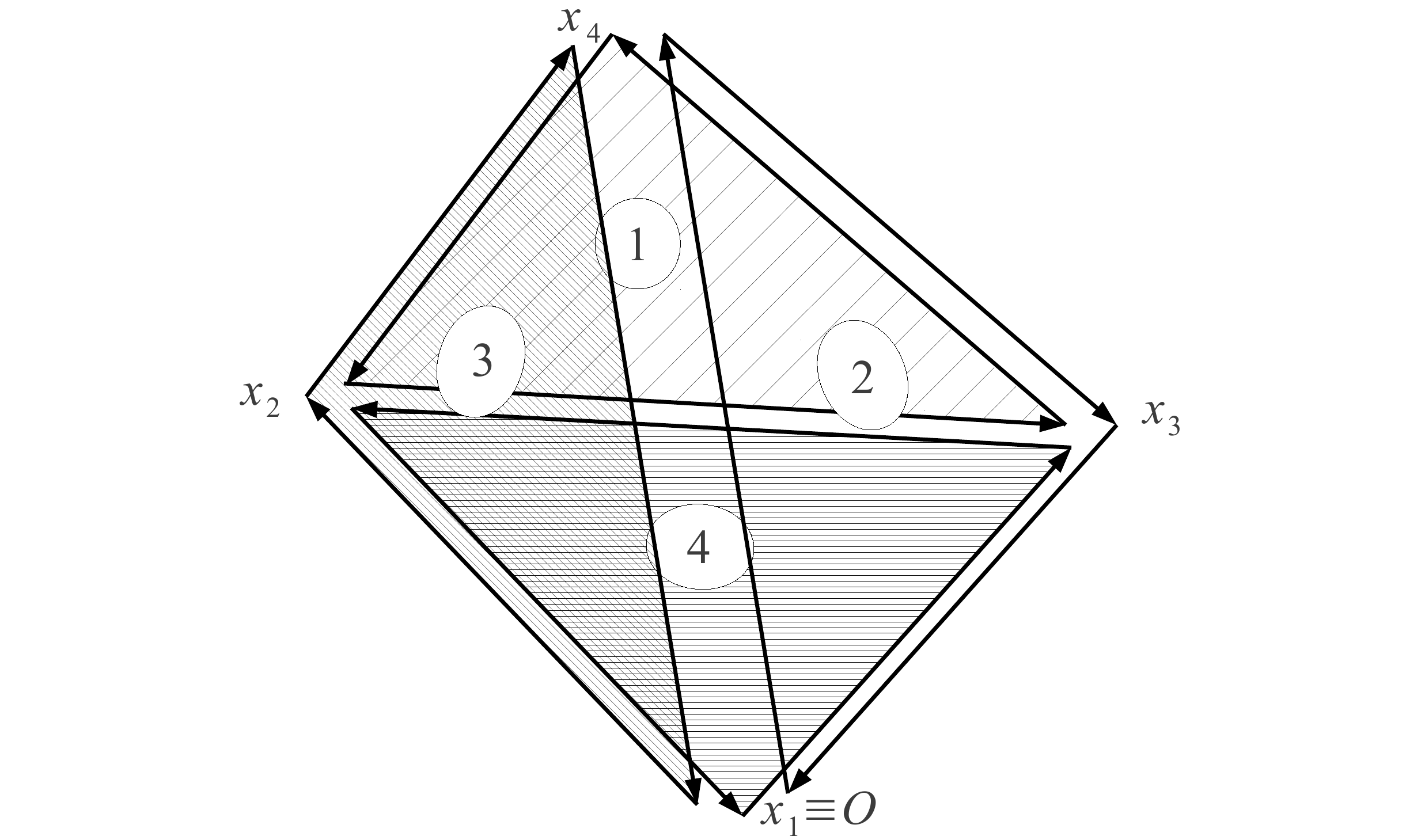}
\caption{A graphical representation of the vectors $\{h_a\rhd\vec\ell_\ab\}$ in the Euclidean sector. Remark how they close into triangular faces (\autoref{closure_h}) and into a tetrahedron (\autoref{transp_h}).}
\eFIG

In the Euclidean sector, $\Re(\eta_\ab)=0$, and therefore $G_\ab\equiv g_a^{-1} g_b\in\SU$. Hence, the two $\sldc$ parallel transport equations (\autorefs{transpJ} and \ref{transp}) become equivalent to one another. Moreover, such a solution for the $G_\ab$'s implies the following form for the $g_a$'s:
\be
g_a= g h^0_a \forall a,\;\text{with}\; h^0_a\in\SU,\;\text{and any}\;g\in\sldc,
\label{7.23}
\ee
where the $h_a^0$ must be understood to be determined by the $j_\ab$'s.\\

Define now
\be
\vec \ell_\ab := \epsilon_\ab j_\ab \hat m_\ab.
\ee
Remark that $j_\ab = j_\ba$, $\epsilon_\ab=-\epsilon_\ba$, while $m_\ab$ has \emph{a priori} no relation with $m_\ba$ other than \autorefs{transpJ} and \ref{transp}. Also, the symbol $\epsilon_\ab=-\epsilon_\ba$ is the same which appears in the closure equation (\autoref{closure}).\\

The closure equation (\autoref{closure}) being written at each edge, can be acted upon with $h_a\in\SU$:
\be
\sum_{b,b\neq a} h^0_a\rhd \vec \ell_\ab = \vec 0\;.
\label{closure_h}
\ee
Also, in the Euclidean case,  \autorefs{transpJ} and \ref{transp} when written in spin 1 representation, read:
\be
h^0_a \rhd \vec \ell_\ab = - h^0_b\rhd \vec \ell_\ba\;.
\label{transp_h}
\ee
In the last two equations, $h^0_a,\;h^0_b$  are understood to be elements of $SO(3)$.\\

\Autoref{closure_h} implies that the three vectors $\{ h^0_a\rhd \vec \ell_\ab\}$ define, at fixed $a$, the sides of a triangle; in turn, \autoref{transp_h} implies that they are moreover coherently identified among them, so to form an (oriented) tetrahedron.\footnote{For example, one possible realization of such a tetrahedron is given by the following positions of its vertices: $\vec x_1=\vec O$ (the origin of $\mathbb R^{3}$), $\vec x_2=\vec\ell_{34}$, $\vec x_3=\vec\ell_{42}$ and $\vec x_4=\vec\ell_{23}$. Here we named the vertices with the index of the face opposite to it, and orientations are chosen as in \autoref{FIGURA!!!}.} The latter identification is due to the combinatorics of the spin foam vertex (see \autoref{tetrahedra_id}). \\

\bFIG[width=11cm]{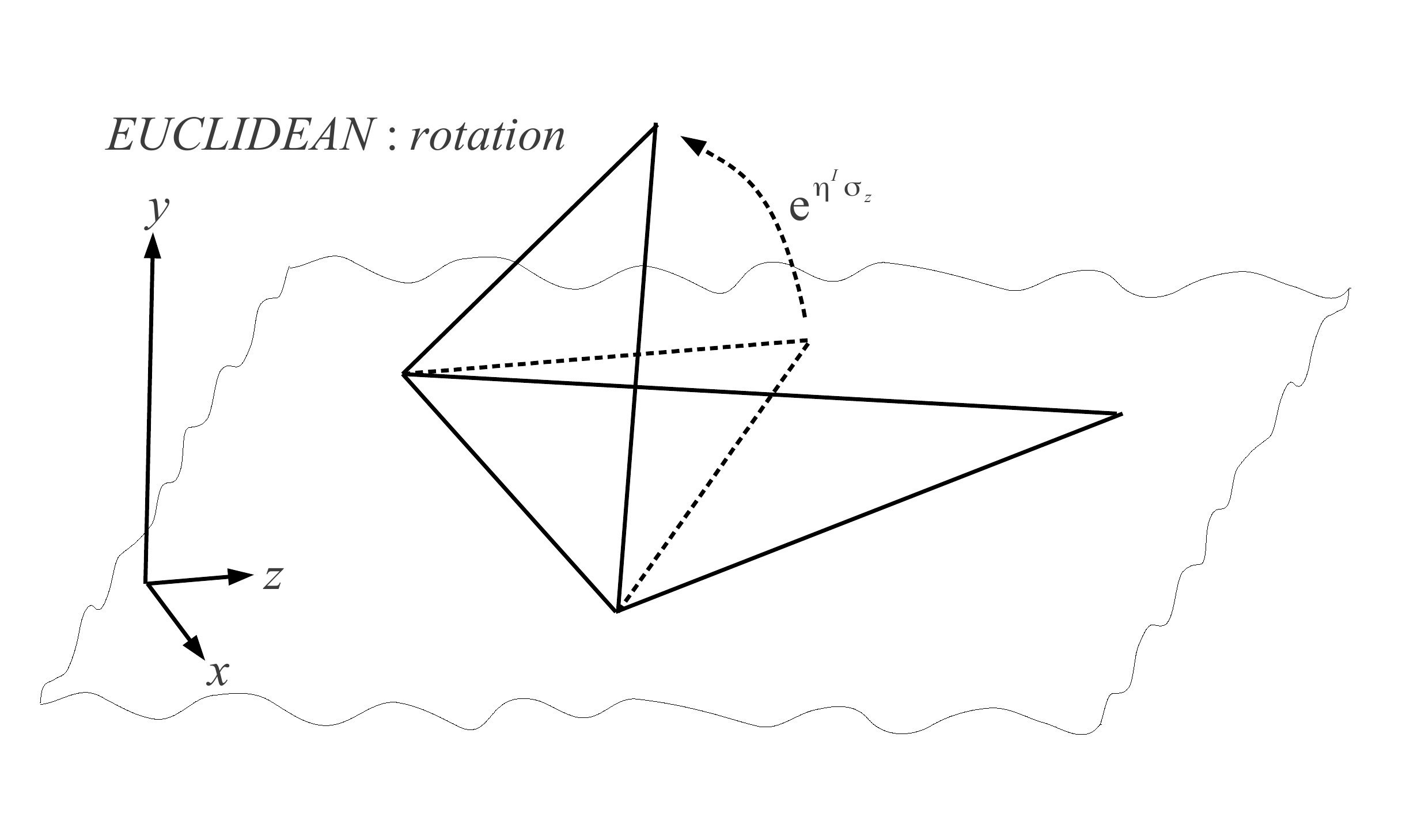}
\caption{The figure represents the action of a $\SU$ element $h=\exp{\eta^I\sigma_z}$ on the dashed triangle lying in the $xz$ plane. A second (fixed) triangle is also shown.}
\label{eucl_rot}
\eFIG

To extract the geometric content of the parallel transport equations, i.e. in order to calculate the dihedral angles between the faces of the tetrahedron, define the \emph{oriented} normal vectors to the triangles
\be
\vec F_a:=-\frac{1}{2}\epsilon_{abcd}\vec \ell_\ab\times\vec\ell_{ac}\;.
\ee
Here no sum on repeated indices is performed, and $\epsilon_{abcd}$ is the completely antisymmetric symbol used for the gauge fixing of the previous section. Notice, that this gauge fixing, through the previous definition, also fixes an orientation for the tetrahedron and its faces, as it was anticipated in the previous subsection.\footnote{Indeed, three vectors satisfying the closure equations, can be organized in two different ways, each of which provides a different orientation for the triangle. Therefore, asking for $\vec F_a$  to be the oriented normal to the triangle, picks out only one of these two possibilities.} Furthermore, notice that this expression is independent of the choice of the two sides of the triangle $a$ involved, thanks to the closure equation and to the antisymmetry of the $\epsilon_{abcd}$. Now, the cosine of the \emph{internal} dihedral angle $\Theta_\ab^E\in[0,\pi]$ between two faces of the tetrahedron, is given by minus the (normalized) scalar product of the two outwards (or inwards) pointing normals to the faces of the tetrahedron, which are e.g. $g_a\rhd \vec F_a$ and $-g_b\rhd\vec F_b$, 
therefore:
\be
\cos \Theta^E_\ab  :=  \frac{(h^0_a\rhd\vec F_a).(h^0_b\rhd \vec F_b)}{||F_a||\;||F_b||} =  \frac{\vec F_a.(G_\ab\rhd \vec F_b)}{||F_a||\;||F_b||},
\label{ang_diedro}
\ee
where \autoref{7.23} was used.\\

By gauge fixing the triangles $a$ and $b$ to lie in the $xz$ plane, we have
\be
\vec F_a=A_a \hat y,\quad\text{where}\quad A_a:=-\frac{1}{2}\epsilon_{abcd}\hat y.(\vec \ell_\ab\times\vec\ell_{ac})=\frac{1}{2}\epsilon_{abcd}\sin{2\vartheta^a_{bc}}\geq0,
\ee
thanks to the gauge fixing condition. Also (in spin 1 representation)\footnote{Recall the formulas for the Lorentz generators (spin 1 $\sldc$ generators):
$$
\mathcal J_x=  \left(\begin{array}{c|ccc} 0 & 0 & 0 & 0 \\\hline 0 & 0 & 0 & 0 \\ 0 & 0 & 0 & -1 \\ 0 & 0 & 1 & 0 \end{array} \right),\quad
\mathcal J_y=  \left(\begin{array}{c|ccc} 0 & 0 & 0 & 0 \\\hline 0 & 0 & 0 & 1 \\ 0 & 0 & 0 & 0 \\ 0 & -1 & 0 & 0 \end{array} \right),\quad
\mathcal J_z=  \left(\begin{array}{c|ccc} 0 & 0 & 0 & 0 \\\hline 0 & 0 & -1 & 0 \\ 0 & 1 & 0 & 0 \\ 0 & 0 & 0 & 0 \end{array}\right),
$$$$
\mathcal K_x=  \left(\begin{array}{c|ccc} 0 & 1 & 0 & 0 \\\hline 1 & 0  & 0 & 0 \\ 0 & 0 & 0 & 0 \\ 0 & 0 & 0 & 0 \end{array} \right),\quad
\mathcal K_y=  \left(\begin{array}{c|ccc} 0 & 0 & 1 & 0 \\\hline 0 & 0 & 0 & 0 \\ 1 & 0 & 0 & 0 \\ 0 & 0 & 0 & 0 \end{array} \right),\quad\text{and}\quad
\mathcal K_z=  \left(\begin{array}{c|ccc} 0 & 0 & 0 & 1 \\\hline 0 & 0 & 0 & 0 \\ 0 & 0 & 0 & 0 \\ 1 & 0 & 0 & 0 \end{array}\right).
$$}
\be
G_\ab = \E^{-2\theta_\ab \mathcal J_y} \E^{2\eta^I_\ab\mathcal J_z}\E^{2\theta_\ba \mathcal J_y}
\ee
where we used the fact that $\eta_\ab=\Im(\eta_\ab)\equiv\eta_\ab^I$. Hence, a straightforward calculation of \autoref{ang_diedro} gives
\be
\cos\Theta^E_\ab =  \cos 2\eta^I_\ab= \cos2(\varphi_\ab-\phi_\ab+\phi_\ba).
\label{cosThetaE}
\ee

This formula, combined with \autoref{7.21}, makes the connection among the parameters $2\eta_\ab$'s and the tetrahedron internal dihedral angles $\Theta^E_\ab$ in the Euclidean case:
\be
\Theta_\ab(j)=\I\Theta^E_\ab\quad\text{[Eucl]}.
\ee

At this point, the reader may be worried about the appearance of the non-physical phases $\phi_\ab$ in the previous formula. However, we shall notice that such phases make no appearance in any formula with \emph{true} physical significance (e.g. a transition amplitude), but just in those formula which constitute an \emph{interpretation} of the transition amplitudes (see the second step in \autoref{S_crit}). 


\subsection{The Lorentzian Sector}

The structure of this section is exactly the same as that of the previous one. They differ just because of some technicalities. In particular, what here is called $\ell_\ab$ will not represent any more the side of one of the triangular faces, but rather the unique normal to it on the plane defined by the face itself. Since this fact implies minor changes both in the algebra and in its geometrical interpretation, the decision was made to maintain the same symbol for the two of them.\\

In the Lorentzian sector, $G_\ab=g_a^{-1} g_b\in\sldc$. \Autorefs{transpJ} and \ref{transp} implies that the $g_a$'s have the following form:
\be
g_a= g g^0_a \forall a,\;\text{with}\; g^0_a\in\sldc,\;\text{and any}\;g\in\sldc.
\ee
where the $g_a^0$, as much as the $G_\ab$, must be understood to be determined by the $j_\ab$'s.\\

Now, for every couple $(ab)$, define the four-vector\footnote{We allowed ourselves an abuse of notation here, since we are assigning here the symbol $\ell_\ab$ to a different entity with respect to the previous section. However, in doing this, we want to stress on the analogous physical interpretation of the two quantities. }
\be
\ell_\ab:= \frac{1}{2}\epsilon_\ab j_\ab \ast \iota(M_y)\wedge\iota(m_\ab)\wedge\iota(Jm_\ab)\;, \quad\text{where}\quad M_y :=\frac{1}{\sqrt{2}} \left(\begin{tabular}{c} \I \\ 1 \end{tabular}\right)\;\Rightarrow \iota(M_y) = \left(\begin{tabular}{c} 1\\ $\hat y$ \end{tabular}\right),
\label{ell}
\ee
where it shall be understood that the $m_\ab$'s have been gauge fixed as in \autoref{eq8}. Moreover, in the previous equation the wedge product stands for the anti-symmetrized tensor product, and the asterisk for Hodge duality.\footnote{For definiteness: $$v_1\wedge\dots\wedge v_n = \frac{1}{n!}\sum_{\sigma\in\Sigma}\text{sign}(\sigma) v_{\sigma(1)}\wedge\dots\wedge v_{\sigma(n)}\;,$$ and $$ T=T^{i_1\dots i_k} \Rightarrow (\ast T)^{i_1\dots i_{d-k}} = \epsilon^{i_1\dots i_{d-k}}_{\phantom{i_1\dots i_{d-k}}i_{d-k+1}\dots i_d} T^{i_{d-k+1}\dots i_d}\;.$$} In coordinates, this can be written as
\be
\ell_\ab^I=
 \epsilon_\ab j_\ab \sum_{J,K,L} \epsilon^I_{\phantom{I}JKL} t^J y^K m_\ab^L,
\label{ell_coord}
\ee
where the following symbols have been introduced: 
\be
t^I= \left(\begin{tabular}{c} 1\\ $\vec 0$ \end{tabular}\right), \quad y^I= \left(\begin{tabular}{c} 0\\ $\hat y$ \end{tabular}\right) \quad \text{and}\quad m_\ab^I= \left(\begin{tabular}{c} 0\\ $\hat m_\ab$ \end{tabular}\right).
\ee
Therefore, it is clear that $\ell_\ab$ is a spacelike four-vector, and that it lies in the three dimensional hyperplane orthogonal to $t^I$. Here it forms an orthonormal triad together with $\hat y$ and $\hat m_\ab$. 
Also, its length is $||\ell_\ab||\equiv\sqrt{\ell_\ab^I{ \ell_\ab}_I}=j_\ab$.\\

Then, we wedge the spin one representation of the stationary point equations in \autorefs{transpJ} and \ref{transp}, with each other and with the four vector $\iota(M_y)$. This simply gives, in terms of the $\ell_\ab$:
\be
g_a^0\rhd\ell_\ab=-g_b^0\rhd\ell_\ba,
\label{ell_transp}
\ee
where we used the shorthand notation $g\rhd  \iota(z)\wedge\iota(z')\wedge\iota(z''):=[g\otimes g\otimes g] \rhd \iota(z)\wedge\iota(z')\wedge\iota(z'')=  \iota(gz)\wedge\iota(gz')\wedge\iota(gz'')$.\\

Furthermore, from the closure equation (\autoref{closure}) for the $\hat m_\ab$ can be immediately deduced a closure equation for the $\ell_\ab$ 
 (this can be most easily done via \autoref{ell_coord}). On this equation one can then act with the group elements $g^0_a$, to obtain
\be
\sum_{b,b\neq a} g^0_a\rhd\ell_\ab=0.
\label{ell_closure}
\ee

The natural geometrical interpretation of \autoref{ell_closure} is that the $\{g^0_a\rhd\ell_\ab\}_{b,b\neq a}$ define in $\mathbb R^{1,3}$ (up to translations) an oriented triangle $a$ with sides $g^0_a\rhd\ell_\ab$. In the same terms, \autoref{ell_transp} shows that the sides of these triangles are coherently identified in such a way to form an oriented tetrahedron.\\

\bFIG[width=11cm]{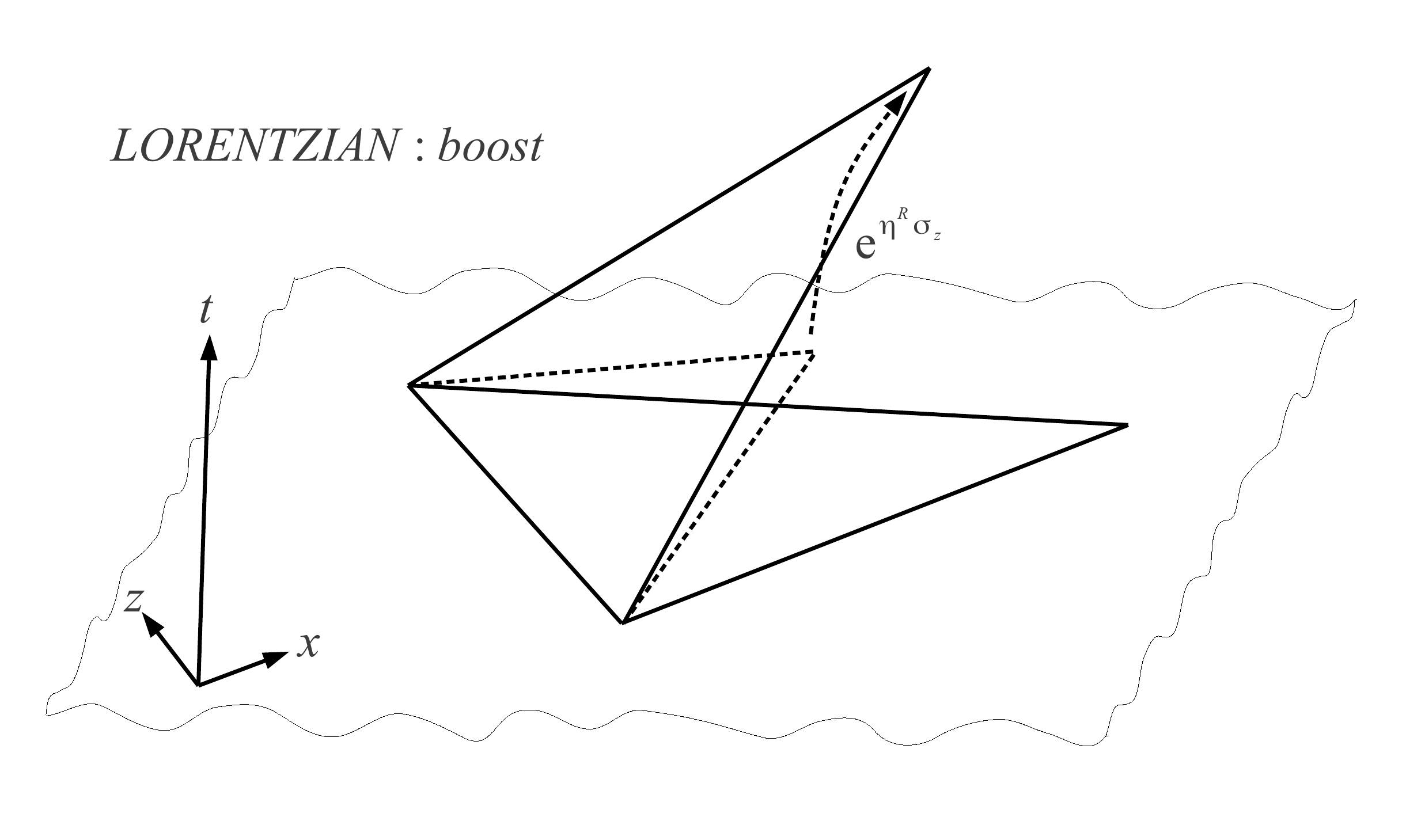}
\caption{The figure represents the action of an $\sldc$ element $g=\exp{\eta^R\sigma_z}$ on the dashed triangle lying in the $xz$ plane. A second (fixed) triangle is also shown. Remark that with respect to \autoref{eucl_rot} the axis on the plane are rotated.}
\label{eucl_rot}
\eFIG

As in the previous section, in order to extract the geometrical content of the parallel transport equations, we define the face bivectors as\footnote{As before for the $\ell_\ab$'s, here we commit an abuse of notation.}
\be
F_a := -\frac{1}{2}\epsilon_{abcd}\ast \ell_{ab}\wedge\ell_{ac}. 
\label{face_biv}
\ee
Using \autoref{ell_coord}, this can be recast in the form
\be
F_a = 2A_a\; t\wedge y, \quad \text{where} \quad A_a :=-\frac{1}{2}\epsilon_{abcd}\hat y.(\vec\ell_{ab}\times\vec\ell_{ac})\geq0.
\label{F_lor}
\ee
Or equivalently, in covariant form
\be
F_a = A_a \;\iota(JM_y)\wedge\iota(M_y).
\ee

In turn, in spin one representation:
\be
G_\ab = \E^{-2\theta_\ab\mathcal J_y}\E^{2\eta_\ba^R\mathcal K_z}\E^{2\theta_\ba\mathcal J_y} ,
\ee
where $\eta^R$ and $\eta^I$ are the real and imaginary parts of $\eta\in\mathbb C$, respectively, and where we used the fact that in the Lorentzian sector $\eta^I_\ab\in\{0,\pi\}$.\\

Thus, the scalar product between two face bivectors is:
\be
\frac{(g_a\rhd F_a)^{IJ}(g_b\rhd F_b)_{IJ}}{||F_a||\;||F_b||} = \frac{F_a^{IJ}(G_\ab\rhd F_b)_{IJ}}{||F_a||\;||F_b||} = \ch 2\eta_\ab^R.
\ee
Analogously to the Euclidean case, this formula gives the dihedral hyperbolic angle between two faces of the Lorentzian tetrahedron:\footnote{Since all the face bivectors are future pointing (\autoref{F_lor}), we are always in the thick wedge case of \cite{BarrettHellmann:LorAsympt}.}
\be
\ch \Theta^L_\ab = \ch 2\eta_\ab^R.
\ee
Therefore if $\Theta^L_\ab$ is the absolute value of the boost parameter among the faces $a$ and $b$ (internal hyperbolic dihedral angle), using \autoref{7.21} we get
\be
\Theta^R_\ab(j)=\Theta^L_\ab\quad\text{[Lor]}.
\ee

\section{Graph Face Orientation\label{section_face}}

Before estimating the radiative corrections to the EPRL-FK propagator, we think a brief comment about face orientation is in order at this point.\\

The EPRL-FK model is invariant under face orientation reversal of the internal faces (and in some sense ``covariant'' with respect to the face reversal of the external faces). This fact had already been emphasized in \cite{MagliaroPerini:LocalSF}, nevertheless we shall give a brief demonstration of this fact in \autoref{appendix_face}.\\

Furthermore, if the coherent states are eliminated from \autoref{EPRL_spin}, using the formula for the resolution of the identity,\footnote{$ \mathbb{I}_{j} = \int_{S_2} \uD_j \hat m \ket{m}_j\bra{m}_j$.} then even the integrand of each face amplitude is invariant under face reversal (see \autoref{appendix_face}):
\be
\Tr_{j_f}\left[\overleftarrow{\prod_{v\in\partial f}} \Y^\dag g_{ve'}^{-1}g_{ve} \Y \right]\text{ is invariant under } f\mapsto f^{-1}\;.
\ee
However, as soon as the coherent states are introduced, a face orientation is automatically picked out, and the integrand is not any more face-reversal invariant.\\

Once the real part of the face action $\Re( S_f)$ is set to zero by one of the stationary point equations, and the $\mathbb{CP}^1$ spinors associated to each face-vertex couple ($z_{vf}$ in the notation of \cite{HanZhang:LorAsympt}) can be expressed as functions of the coherent states $m_{ef}$, it is immediate to see that
\be
\left. S_{f^{-1}}[z_{vf},m_{ef},g_{ve}]\right|_{Z_{vf}\propto_\mathbb{C} m_{ef}} = -\left. S_{f}[z_{vf},m_{ef},g_{ve}]\right|_{Z_{vf}\propto_\mathbb{C} m_{ef}}.
\ee

At the same time, the closure equations themselves are not invariant under face reversal, since $\epsilon_{ef}(v)\mapsto-\epsilon_{ef}(v) \;|\forall v,e\in\partial f$ when $f\mapsto f^{-1}$. As a consequence the $g^{-1}_{ve'}g_{ve}\equiv G_{e'e}^f(v)$ solving the parallel transport equations depend on the face orientation as well. As it shall become clear from the explicit example of the melon graph, this two changes compensate each other, giving - as expected - a face-reversal invariant phase at the stationary points. Indeed, the way we wrote the melon graph amplitude in \autoref{Mel_Gr_ampl} automatically implies that the faces are oriented counter-clockwise and that, once a global choice of $\nu\in\{\pm1\}$ has been made:
\be
\epsilon_\ab = -\tilde\epsilon_\ab = \left\{
\begin{array}{ll}
\nu & \text{if } a<b\\
-\nu & \text{if } a>b
\end{array}
\right. \;.
\label{eq_orientation}
\ee


\section{Radiative Correction\label{section_radiativecorrections}}

In this section, we shall use the analysis of the saddle point equations previously performed, to evaluate the melon graph at its dominant order. In particular, we shall evaluate the partial amplitudes at fixed (large) spins in \autoref{Mel_Gr_ampl} at their stationary points, as well as estimate the rank of the action Hessian at these points so to understand the partial amplitude scaling at large spins. Finally, we shall argue how the sum over all the possible internal spins affects the final result.\\

Formally, the argument goes as follows. First, we rewrite the truncated amplitude as a sum of partial amplitude at fixed spins:
\be
W_{\mathcal M}^\Lambda(j_a,n_a,\tilde n_a) = : \sum_{\{j_a<\Lambda\}} w_{\mathcal M}(j_a,n_a,\tilde n_a; j_\ab)\;.
\label{partial}
\ee
Then, we claim that the procedure studied in \autoref{section_BF} in the context of BF theory applies also to the EPRL-FK amplitude. Summarizing, since the face weight $\mu(j_\ab)$ scales with some positive power $\mu\geq1$ of the spin, we claim that the relevant contributions to the sum in \autoref{partial} come from the configurations where the internal spins are all large. The formal way to achieve this, is to scale the internal spins $j_\ab$ by a factor $\lambda\rightarrow\infty$, therefore ``picking'' the stationary points of $w_\mathcal M$ as the relevant contributions.\\

In the next subsections we shall evaluate the partial amplitudes $w_\mathcal M$ at their (geometrically non-degenerate) stationary points. As it was shown in the previous section, the geometry type of the stationary points (Euclidean or Lorentzian) depends on the value of the internal spins $j_\ab$ only. Therefore, at the two vertices of the melon graph, the two geometries are of the same type.

\subsection{Action at the Stationary Points}
In order to calculate the action at its stationary points, it is enough to combine together the stationary point equations, their solution (\autoref{7.22}), as well as the result of the previous section about the graph face orientation.\\

We first substitute into the total action formula (\autorefs{face_action} and \ref{eq_action}) the solutions to the stationary point equations \autorefs{z_stationary} and \ref{transp} (essentially the parallel transport equations), and their parametrization in terms of the complex parameters $\eta_\ab\in\mathbb C$ introduced in \autoref{eta}:
\begin{eqnarray}
S_\text{stat}  &=& \I\sum_{ab,a<b} j_\ab\left[-2(\varphi_\ab-\tilde\varphi_\ab) + 2\gamma \left(\ln\frac{||Z_\ab||}{||Z_\ba||}-\ln\frac{||\tilde Z_\ba||}{||\tilde Z_\ab||}\right)\right]\notag\\
&=&\I\sum_{ab,a<b} 2j_\ab\left[-\Im(\eta_\ab-\tilde\eta_\ab) + \gamma\Re(\eta_\ab-\tilde\eta_\ab)\right].
\label{S_crit}
\end{eqnarray}
Notice that the formula was used $\Im(\eta_\ab-\tilde\eta_\ab) = \varphi_\ab - \tilde \varphi_\ab$, which stems from the fact that the \emph{same} phase $\phi_\ba-\phi_\ab$ appears in both $\eta_\ab$ and $\tilde\eta_\ab$: therefore the arbitrary phases $\phi_\ab$'s play no role in the relevant formulas.\\

At this point, one just need to combine \autoref{7.22} and the result about the face orientations of \autoref{eq_orientation} and to insert them in the previous formula, to first obtain
\be
\eta_\ab=\frac{1}{2}\nu_v\nu\;\Theta_\ab+[\I\pi]_\ab\quad\text{and}\quad\tilde\eta_\ab=-\frac{1}{2}\nu_{\tilde v}\nu\;\Theta_\ab+\tilde{[\I\pi]}_\ab\quad \text{when $a<b$},
\ee
and then
\be
S_\text{stat}= -\nu(\nu_v+\nu_{\tilde v})\I\sum_{ab,a<b} j_\ab\left[\Im (\Theta_\ab )+ \gamma \Re( \Theta_\ab) \right]-\nu\sum_{ab,a<b} 2j_\ab [\I\pi]^0_\ab.
\label{faceaction1}
\ee
In the last equation we made use of the fact that $2j_\ab\in\mathbb{N}$ to introduced the new symbol\footnote{Recall that the action $S$ appears in the physically meaningful formula only as $\exp S$. Therefore, it is defined only modulo $2\I\pi$.}
\be
[\I\pi]^0_\ab := [\I\pi]_\ab + \tilde{[\I\pi]}_\ab\mod 2\I\pi.
\ee

Finally, remark that the choice of the sign $-\nu$ is irrelevant, since it can be absorbed in that of the signs $\nu_v$ and $\nu_{\tilde v}$. 
%
%
Therefore:
\be
S_\text{stat}= \I(\nu_v+\nu_{\tilde v})\sum_{ab,a<b} j_\ab\left[\Im (\Theta_\ab )+ \gamma \Re( \Theta_\ab) \right]+\sum_{ab,a<b} 2j_\ab[\I\pi]^0_\ab.
\label{faceaction}
\ee
Notice that this expression is manifestly invariant under face-orientation reversal. \\

\subsection{Euclidean Stationary Points\label{section_eucl_st_pts}}

In the Euclidean case, $\eta_\ab\in\I\mathbb R$, and $\Theta_\ab=\I\Theta_\ab^E$. There are essentially three possible phases for the stationary points, depending on the choice of $\nu_v$ and $\nu_{\tilde v}$:
\begin{itemize}
\item[1.] $\nu_v=-\nu_{\tilde v}=\pm 1\Longrightarrow S_{\text{stat}} =0+\sum_{ab,a<b} 2j_\ab [\I\pi]^0_\ab.$;
\item[2\&3.] $\nu_v=\nu_{\tilde v} = \pm1 \Longrightarrow S_{\text{stat}} \equiv\pm \sum_{ab,a<b} 2j_\ab \left(\I\Theta^E_\ab+[\I\pi]^0_\ab\right) = \pm2\I S_\text{Regge}^E[j]+\sum_{ab,a<b} 2j_\ab [\I\pi]^0_\ab.$
\end{itemize}

Here $S^E_\text{Regge}[j_\ab]$ stands for the (classical) Regge action for an Euclidean tetrahedron of sides $\{j_\ab\}$.\\

For the stationary points of cases 2 and 3
, the sum over the spins which is needed to compute the total amplitude can be viewed as a discrete path integral weighted by the Regge action. In the large spin limit where all the $j_\ab$'s scale as $\lambda\rightarrow\infty$, the exponential of the action is rapidly oscillating, and the path integral will be dominated by its stationary points, i.e. by the classical motions of the Regge action for a tetrahedron. In this case, the well-known equations of motions of the Regge action read $\Theta^E_\ab(j)=0\;\forall ab$ \cite{Regge}, which have clearly no solution in the present context. Therefore, we shall consider the contribution to the total amplitude coming from these stationary points as sub-dominant. This argument is not very rigorous, because in this case the spins are not just a scale, but are actually \emph{summed} over (and doubts could even arise about the behaviour of the sum as the cut-off is sent to infinity\footnote{See for example \cite{ChristLangvik:PRspike} for the discussion of a similar issue.}). However, if one think to perform the sum first over the spins at fixed scale and then over the different scales,\footnote{This would correspond to the change from Cartesian to spherical coordinates in the continuous version of the sum over the spins.} one realizes that the first of these sums should be suppressed at high scales because of the usual reason, i.e. because of the absence of stationary points of the action. Nevertheless, the same conclusion is supported by the parallelism with BF-theory, where the divergent sector can be independently calculated and therefore it is possible to verify \emph{a posteriori} the absence of the contributions stemming from the terms 2 and 3  (see \autoref{section_BF}).\\

When $S_\text{stat}=0+\sum_{ab,a<b} 2j_\ab [\I\pi]^0_\ab$, it means that $\tilde\eta_\ab= \eta_\ab+[\I\pi]^0_\ab$, i.e.
\be
G_\ab = [\pm]_\ab \tilde G_\ab\in\SU. 
\label{G=pmtildeG}
\ee
Here the fact was used that $\E^{\I\pi\sigma_z}=-1$
. Remark that following the discussion at the end of \autoref{section7.1}, the choices of the signs in the previous equations are not independent at every face. Rewriting the previous equation in terms of the edge holonomies $H_a:=g_a\tilde g_a^{-1}$ gives
\be
H_a=[\pm]_\ab H_b.
\label{edgeH}
\ee
With \autoref{G=pmtildeG} written in this form, it is easy to understand the meaning of the plus or minus signs. Indeed, renaming $H_1\equiv H$, the previous equation just reads
\be
H_a=\pm_a H,
\ee
where it is now emphasized (by the absence of the square brackets) that the $\pm$ signs are now independent and actually characterize the stationary point. Notice that the holonomy around the face $(ab)$ is given by $H_\ab := H_a^{-1} H_b$, and \autoref{edgeH} imposes it to be
\be
H_\ab = \pm_a\pm_b \mathbb I.
\ee
The solution $H_\ab = \mathbb I$ is exactly the solution one would have expected from this sector, which is equivalent to the Ponzano Regge model, i.e. three dimensional $\SU$-BF theory. Nevertheless, with some surprise, it is found that the stationary points treat equivalently $\pm\mathbb I_\SU$, and are hence only sensitive to the SO(3) structure of the model, \emph{viz.} to its vectorial representation.\\

In the rest of this section we shall calculate the contribution of this family of stationary points to the total amplitude. In a first time we shall ignore the presence of possible $[\I\pi]$ terms in the expressions for  the $\eta_\ab$'s, which shall later be shown to be completely uninfluential. 

\Autoref{G=pmtildeG} (with the plus sign), implies that 
\be
g_a = K g^0_a\quad\text{and}\quad \tilde g_a = \tilde K g_a^0\;,
\ee
where $K,\tilde K$ are arbitrary elements of $\sldc$, and the $g_a^0\in\SU$ are functions of the $j_\ab$'s only. In this notation $H=K\tilde K^{-1}$. Therefore, inserting this result in \autoref{Mel_Gr_ampl}, and following the discussion at the end of \autoref{Sect_sym}, we evaluate the partial amplitudes related to this sector to be
\begin{align}
w^E_{\mathcal M}\sim& \lambda^{6\mu + 12+12 -\frac{1}{2}\rank(\delta^2 S^{E,1}_\text{stat})}\int_{\sldc^{\otimes2}} \hspace{-1.5mm}\D K \D \tilde K \prod_a\int_{S_2}\uD \hat m_a\; \bra{n_a}K g_a^0\ket{m_a}_{\Y,j_a}\bra{m_a}( g_a^0)^{-1}\tilde K^{-1}\ket{\tilde n_a}_{\Y,j_a}\;\notag\\
&=\lambda^{6\mu + 24 -\frac{1}{2}\rank(\delta^2 S^{E,1}_\text{stat})}\int_{\sldc^{\otimes2}} \hspace{-1.5mm}\D K \D \tilde K \prod_a\int_{S_2}\uD \hat m_a\;\bra{n_a}K \ket{m_a}_{\Y,j_a}\bra{m_a}\tilde K\ket{\tilde n_a}_{\Y,j_a}\;,
\end{align}
 where we used the resolution of the identity in terms of the coherent states and the fact that the $\SU$ group elements commute with the EPRL-FK $\Y$-map. Notice that the term $6\mu$ in the exponent of $\lambda$ comes from the face weights, the first 12 comes from the measures $\uD_{j_\ab}\hat m_\ab$, and the second 12 comes from the factors $(2j_\ab+1)^2$ normalizing the measures $\Omega'_\ab\tilde\Omega'_\ab$. \\

This equation shows that at leading order, in the Euclidean sector, the dependence on the internal face spins is confined to a scale factor. This fact allows to estimate the divergence degree of the full amplitude and to explicitly know its dependence on the boundary data. Before writing the result, it is important to estimate $\rank(\delta^2 S^{E,1}_\text{stat})$:
\begin{align}
\rank(\delta^2 S^E_\text{eq}) &= 8\times\dim(\sldc)+12\times\dim(S_2)+12\times\dim(\mathbb{CP}^1)+\notag\\
&\hspace{5.4cm}-4\times\dim(\SU)-2\times\dim(\sldc) \notag\\
&= 72,
\end{align}
where we calculated as in the BF case of \autoref{section_BFevaluation} the dimension of the space of integration minus that of the gauge orbits. Once more, this is only an \emph{upper} bound, which sets a \emph{lower} bound on the total degree of divergence of this sector. We shall ignore this issue, assuming that the bound is saturated as it is in the $\SU$-BF case.\\

Finally, taking into account the six sums over the internal spins, one finds
\be
W^{\Lambda,E}_\mathcal{M}\sim\Lambda^{6(\mu-1)}\int_{\sldc^{\otimes2}} \D K \D \tilde K\; \prod_a\int_{S_2}\uD \hat m_a\;\bra{n_a}K \ket{m_a}_{\Y,j_a}\bra{m_a}\tilde K\ket{\tilde n_a}_{\Y,j_a}\;.
\label{W^E_M}
\ee

The possible asymmetry between the solutions of the stationary point equations at the two vertices coming from the presence of the $[\I\pi]$'s, does not alter this result. Indeed, at the light of the previous discussion, it is clear that one can bring back up the presence of those factors to the change in sign of some of the $g_a$'s (or, equivalently, $\tilde g_a$'s):\footnote{The reader can check that this interpretation of the origin of the terms $[\I\pi]_\ab$ appearing in \autoref{7.22} is coherent with the analysis made at the very end of \autoref{section7.1}. Briefly, if $\eta_{12}\mapsto\eta_{12}+I\pi$, this must be traced back to either $(g_1,g_2)\mapsto(-g_1,g_2)$ or $(g_1,g_2)\mapsto(g_1,-g_2)$. Depending on the possibility that $g_3$ and $g_4$ might also have been mapped to $-g_3$ and $-g_4$ respectively, one finds the different cases listed at the end of \autoref{section7.1}.}
\be
g_a \mapsto -g_a.
\label{gin-g}
\ee
Consequently, the action at the stationary point acquires a term
\be
S_\text{stat}^{E,1} \mapsto S_\text{stat}^{E,1} + 2\I\pi\sum_{b,b\neq a} j_\ab,
\ee
and the amplitudes is modified by an additional factor
\be
\bra{n_a}g_a\ket{m_a}_{\Y,j_a}=\bra{n_a}\Y^\dag g_a\Y\ket{m_a}^{2j_a} \mapsto (-1)^{2j_a}\bra{n_a}g_a\ket{m_a}_{\Y,j_a}\;.
\ee
Therefore, the total effect of the change of sign of a $g_a$ on the partial amplitude is
\be
w_\mathcal{M}^E\mapsto(-1)^{2\sum_{b=1}^4 j_\ab }w_\mathcal{M}^E\;,
\label{9.14}
\ee
where the convention was used that $j_{aa}\equiv j_a$. However, at each edge of a non-null EPRL-FK spin foam graph, the sum of spins must be an integer, since any (formal)\footnote{See e.g. \cite{BarrettHellmann:LorAsympt} and references therein for a discussion about the $\sldc$ intertwiner normalization and $\sldc$-invariance.} integration of the type
\be
\int_\sldc \D g \;g\rhd\bigotimes_i\Y\ket{m_i}_{j_i}
\ee
subsumes an analogous  (well-defined) integration over $\SU$:
\be
\int_\SU \D h \;h\rhd\bigotimes_i\ket{m_i}_{j_i}
\ee
which gives as a result an $\SU$ intertwiner, and therefore can be non-zero only if $\sum_i j_i \in \mathbb N$. This fact, combined with \autoref{9.14}, concludes our proof. \\

Therefore, we conclude that the contribution from the Euclidean sector of the sum of the internal spins is dominated by a term of order $\Lambda^{6(\mu-1)}$ of the form \autoref{W^E_M}. This formula can be further simplified by the use of the Dupuis-Livine map $\K$ (see \autoref{appendix_face}). Indeed, recall:
\be
\K(h,g) := \int_\SU \D k\; \sum_j (2j+1)^2 \chi^j(hk) \chi^{\gamma j,j}(kg)
\ee
from which it follows
\begin{align}
\int_\sldc \D g\;\K(h,g) &= \int_\sldc \D g\int_\SU \D k\; \sum_j (2j+1)^2 \chi^j(hk) \chi^{\gamma j,j}(kg)\notag\\
& = \int_\SU \D k \; \sum_j (2j+1)^2 \chi^j(hk) \int_\sldc \D (kg)\; \chi^{\gamma j,j}(kg)\notag\\
& = 0\;,
\end{align}
where we supposed we could exchange the $\sldc$ and $\SU$ integrals as well as the sum over the spins, and where the translation invariance of the $\sldc$ Haar measure was used. The result follows from the the fact
\be
\int_\sldc \D g \;\chi^{\gamma j,j}(g) = 0\;.
\ee
Therefore, applying the previous formula to the Euclidean dominant contribution \autoref{W^E_M}, one gets:
\begin{align}
\int_{\sldc^{\otimes2}} \D g\D\tilde g &\prod_a \int_{S_2}\uD \hat m_a\; \bra{n_a} g\ket{m_a}_{\Y,j_a}\bra{m_a}\tilde g\ket{\tilde n_a}_{\Y,j_a} =\notag\\
&= \int_{\sldc^{\otimes2}} \D g\D\tilde g\int_\SU \D h\D\tilde h\;  \K(h,g)\K(\tilde h,\tilde g)\prod_a \int_{S_2}\uD \hat m_a\bra{n_a} h\ket{m_a}_{j_a}\bra{m_a}\tilde h\ket{\tilde n_a}_{j_a} \notag\\
&= 0.
\end{align}

\subsection{Lorentzian Stationary Points\label{section_degeneratesector}}
The analysis of the contribution of the Lorentzian stationary points to the full amplitude closely parallels that given for the Euclidean stationary points. The only relevant difference\footnote{Another possible difference could arise from the equations of motions for the ``Lorentzian Regge action'', which may not be the same as for the Euclidean one. However, the expression of the dihedral angles in terms of the spins in the two cases is the same, up to an analytical continuation, so the equations of motions read the same in the two cases.} comes from the fact that in this case the $G_\ab$'s are genuine elements of $\sldc$ (i.e. $\notin \SU$). This means that a crucial step is missing in order to find an equation of the type of \autoref{W^E_M}, i.e. the fact that the $g_a^0$'s do not commute any more with the EPRL-FK $\Y$-map. Therefore, the best that can be done is to write
\be
W^{\Lambda,L}_\mathcal{M} \sim\sum_{\{j_\ab<\Lambda\}}\lambda^{6(\mu-2)}\int_{\sldc^{\otimes2}} \D K\D\tilde K\prod_a\int_{S_2}\uD\hat m_a\;\bra{n_a}Kg_a^0\ket{m_a}_{\Y,j_a}\bra{m_a}(g_a^0)^{-1}\tilde K^{-1}\ket{\tilde n_a}_{\Y,j_a}\;,
\label{lor_contrib}
\ee
where the $g_a^0\in\sldc$ are highly non-trivial functions of the internal face spins $\{j_\ab\}$.

\newpage
\section{The Total Amplitude\label{section_totalamplitude}}

At the light of the previous calculations, we can conclude that the amplitude of the melon graph in the EPRL-FK model, at fixed external coherent states $\{\ket{n_a}_{j_a},\ket{\tilde n_a}_{j_a}\}$, with arbitrary spins $\{j_a\}$, is dominated by a divergent contribution
\begin{align}
W^{\Lambda}_\mathcal{M}\sim&\Lambda^{6(\mu-1)} \int_{\sldc^{\otimes2}} \D K\D \tilde K\prod_a \int_{S_2} \uD \hat m_a \;\bra{n_a}K\ket{m_a}_{\Y,j_a}\bra{m_a}\tilde{K}\ket{\tilde n_a}_{\Y,j_a} +\notag\\
&\quad+\sum_{\{j_\ab<\Lambda\}}\lambda^{6(\mu-2)}\int_{\sldc^{\otimes2}} \D K\D \tilde K\prod_a\int_{S_2}\uD\hat m_a\;\bra{n_a}Kg_a^0\ket{m_a}_{\Y,j_a}\bra{m_a}(g_a^0)^{-1}\tilde K^{-1}\ket{\tilde n_a}_{\Y,j_a}\;, +\notag\\
&\quad + o(\Lambda^{6(\mu-1)}).
\label{final_melon_ampl}
\end{align}
\section{About the Approximations\label{section_approx}}

In this section we briefly review the hypothesis and approximations needed to obtain the result presented in the previous section (\autoref{final_melon_ampl}).\\

First, we started with the hypothesis that the relevant regime to study the convergence properties of the melon graph full amplitude is that in which all the internal spins scale \emph{uniformly} to infinity. Even if we showed that this hypothesis (when combined to other claims we are going to discuss in a moment) leads to the correct result in the context of $\SU$-BF theory, it will continue to be an hypothesis on which all the calculation crucially relies.\\

Then, in order to use this hypothesis, we used the stationary phase approximation for the integrals. Rigorously speaking this is a viable approximation only if the integration domain is compact; therefore in the context of the EPRL-FK model, where multiple integrals are performed over the non-compact group $\sldc$, a further hypothesis is needed, i.e. some kind of well-behavedness of the integrand functions which suppresses their contribution for large group elements.\footnote{Some preliminary numerical results seem to show that this is indeed the case.}\\

Furthermore, (at least in the context of the non-degenerate sector) we neglected the possibility that the Hessian of the action could be degenerate at the stationary points (obviously, except for those directions related to the symmetries of the action); to support this hypothesis, we showed that it is consistent with our analysis of the self-energy of $\SU$-BF theory.\\

Finally, we claimed that the terms associated to two tetrahedra with the same parity are suppressed with respect to those appearing in \autoref{final_melon_ampl}, because of the strong interferences they undergo once the sum over the internal spins is performed; this hypothesis also parallels one in the $\SU$-BF theory, which, in this case, is compatible with to the correct result.

\newpage
\section{The Spike Interpretation\label{section_spike}}

Up to now, we analysed the diverging part of the amplitude in terms of a \emph{three} dimensional ``classical'' geometry, as our (closure) equations were indicating. In the standard interpretation of the divergences, however, they are associated to \emph{four} dimensional spikes: keeping the boundary of the graph fixed (in our cases a topological three-sphere), an additional point is added at the ``centre'' of the triangulation dual to the boundary and is let roam the four dimensional space. To this roaming is then associated a possible divergence (in the case of BF theory, all such configurations are gauge equivalent, and therefore integrating upon them gives a divergent factor associated to the gauge orbit\\

\bFIG[width=11cm]{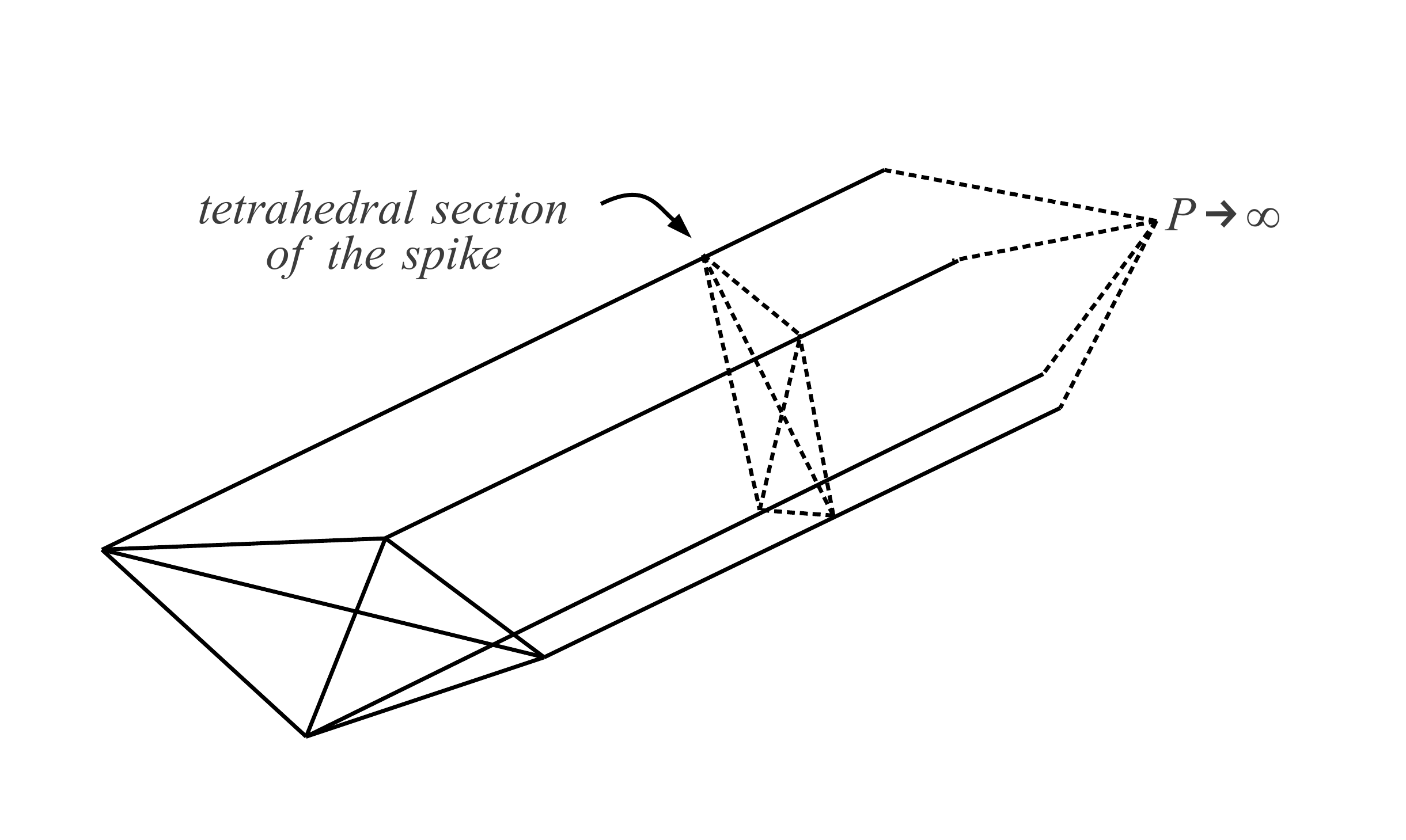}
\caption{A graphical representation of the ``spike'' interpretation in four space dimensions.}
\label{spike}
\eFIG

In order to recover this point of view - at least to some extent - we shall consider the ``full'' closure equation at each internal edge:
\be
\epsilon_{aa} j_a \hat m_a + \sum_{b,b\neq a} \epsilon_\ab j_\ab \hat m_\ab = \vec 0 ,
\label{full_cl}
\ee
where $\epsilon_{aa}=\pm1$ according to the choice of orientation of the external face $a$. This equation holds once an $\SU$ integration is performed at each edge (such an integration is subsumed by the $\sldc$ integrations: $\int_\sldc \D g_a$). This equations holds true if and only if the vectors $\{\vec\ell_\ab\}_{b, b\neq a}$ and $\vec\ell_a := \epsilon_{aa}j_a\hat m_a$ are the oriented normals to the faces of an Euclidean tetrahedron, with length proportional to the face area.\\

If the internal spins $\{j_\ab\}$ are now uniformly rescaled by a factor $\lambda\rightarrow\infty$, while the external spins $\{j_a\}$ are kept fixed, the tetrahedra associated to the full closure equations become more and more ``spiky''. Moreover, the normals to the rescaled faces tend to lie on one single plane and fulfill \autoref{closure}, instead of \autoref{full_cl}, to order $O(\lambda^{-1})$. Therefore the ``reduced'' closure equation is associated to a normal section of the previous spiky tetrahedron. In this sense, we can think we trade a four geometry for a three geometry in our dissertation.\\

However, this in not completely correct. Indeed, in order to have a  fully consistent four geometry, parallel transport equations must be imposed along the external faces as well; but parallel transport equations are an approximate consequence of the largeness of the spins associated to a given face. Therefore, nothing of this kind holds for general external faces.\\

Nevertheless, if we force ourselves to think in terms of spikes, the natural interpretation of the origin of the divergences we found, is the one given in \cite{ChristLangvik:PRspike}, where it was associated to the sum over orientation implicit in the EPRL-FK model (as well in most spin foam models). We shall come back to this point in the conclusions.

\section{Conclusions\label{section_conclusions}}

In this paper we calculated the dominant contributions to the Lorentzian EPRL-FK melon graph, i.e. to its self-energy, at fixed boundary states. This is formally done with the insertion of a rigid cut-off $\Lambda$ on the $\SU$ representations characterizing the internal faces of the graph. Then, in order to study the convergence properties of the sum, the relevant large-spin limit of the partial amplitudes has been taken. This allowed us - under some hypothesis - to use the stationary phase approximation for the integrals. Hence, the equations defining the stationary points have been solved explicitly, and the solution have been interpreted in geometrical terms. The dominant contributions, for those sets of spins which allow their existence, are found to be given\footnote{With the \emph{caveats} thoroughly discussed in \autoref{section_approx}.} by two equal Euclidean, or Lorentzian (depending on values of the spins one is considering), tetrahedra of opposite parity.\\

The result about the amplitude is then twofold. On one side we calculated the self-energy scaling in $\Lambda$, as a function of the spin-foam face-weight $\mu(j)$. On the other side, we worked out the structure of the dependence on the boundary states of the dominant part of the graph amplitude.\\

The scaling of the EPRL-FK self-energy is found to reflect the Euclidean results of \cite{PeriniRovelliSpeziale:SelfEnergy,KrajewskiMagnen:EPRL-GFT}.\footnote{This is done in a different setting from that used in \cite{PeriniRovelliSpeziale:SelfEnergy,KrajewskiMagnen:EPRL-GFT} (though the one of \cite{KrajewskiMagnen:EPRL-GFT} is similar). Nonetheless, our calculation is straightforwardly generalizable to the Euclidean case and, at the light of this fact, it is clear that the two results must coincide. Indeed, our technique essentially calculates the dimension of the integration space used to calculate the amplitude minus the dimension of the symmetries of the amplitude itself. Moreover the only relevant variables for the calculation turn out to be the group variables (this is not explicitly shown in the text, but the reader can rapidly demonstrate it by showing that the contributions coming from the ``auxiliary'' variables $\{\hat m_a\}$ and $\{z_\ab\}$ cancel thanks to their measure scalings). Finally, since $\dim SO(4)=\dim \sldc$, the equality between the Euclidean and Lorentzian results follows.} In particular the dominant contribution diverges as
\be
W^\Lambda_\mathcal{M}\sim O(\Lambda^{6(\mu-1)})
\ee
in the spin cut-off $\Lambda$. Here, $\mu$ is defined by the asymptotic behaviour of the spin-foam face weight: $\mu(\lambda)\sim\lambda^\mu$ as $\lambda\rightarrow\infty$. In particular, this means that if the face weight is the one imposed by the invariance of the spin-foam amplitude under face splitting ($\mu(j)=2j+1$), then the dominant (and consequently only) divergence present in the melon graph is logarithmic in the cut-off:
\be
\mu(j)=2j+1\quad\Longrightarrow\quad W^\Lambda_\mathcal{M}\sim O(\ln\Lambda).
\label{log}
\ee

At this point one could try and interpret the cut-off model as a qualitatively viable and more treatable version of the full $q$-deformed model. In the $q$-deformed model a cut-off on the $\SU$ representations arises naturally, and clues exist which point to the relation of this cut-off with the cosmological constant $\Lambda_{CC}$. In particular, $\Lambda\sim{\Lambda_{CC}/}{l_P^2}\approx10^{120}$, $l_P$ being the Planck length. Then the logarithm appearing in \autoref{log} would not be very large at all! Indeed:
\be
\ln\Lambda\sim\ln(10^{120})\approx280,
\label{logplanck}
\ee
which is a small number when compared to those one would expect when considering representations of the order of $\Lambda$ itself.\\
Moreover, the previous estimation seems to be meaningful only for radiative corrections of phenomena taking place at the Planck scale: if one takes into account the fact that there exist a scale at which our analysis in terms of the stationary point approximation starts to be viable, and that this scale is set by (some large number times) the boundary spins,\footnote{Otherwise the closure equation in the form of \autoref{closure} would not be accurate.} then \autoref{log} should be replaced with something more similar to
\be
W^\Lambda_\mathcal{M}\sim O\left(\ln\frac{\Lambda}{j_\text{ext}}\right).
\ee
Hence, for physics at the galactic scale $j_\text{ext}\sim{(10^5\text{ly})^2}/{l^2_P}\approx10^{110}$. This implies
\be
\ln\frac{\Lambda}{j_\text{ext}}\sim\ln(10^{10})\approx20,
\ee
which is essentially of order 1!\footnote{At our scale $j_\text{ext}\sim{1\text{m}^2}/{l^2_P}\approx10^{70}$, hence $\ln\frac{\Lambda}{j_\text{ext}}\sim\ln(10^{50})\approx115$, which is of the same order as \autoref{logplanck}. }
Therefore, this suggestive setting leads us to speculate that in a theory with cosmological constant, the supposedly very large divergences associated to melon graphs could be, in the end, not large at all (but not small either!). In general, the result would clearly be scale dependent, and the larger the (length) scale, the smaller the corrections. Remark, however, that the framework we used for the calculations breaks down at the cosmological scale, since it is strongly based on the existence of a neat scale hierarchy: $j_\text{ext}\ll\Lambda$.\\

For what concerns the other result, i.e. the dependence of the dominating part of the amplitude on the boundary states, it reflects the very nature of the EPRL-FK model. Indeed, its form strongly relies on the EPRL-FK $\Y$ map.  We report here the final formula:
\begin{align}
W^{\Lambda}_\mathcal{M}\sim&\Lambda^{6(\mu-1)}\int_{\sldc^{\otimes2}} \D K\D \tilde K \prod_a \int_{S_2} \uD \hat m_a \;\bra{n_a}K\ket{m_a}_{\Y,j_a}\bra{m_a}\tilde{K}\ket{\tilde n_a}_{\Y,j_a} +\notag\\
&\quad+\sum_{\{j_\ab<\Lambda\}}\lambda^{6(\mu-2)}\int_{\sldc^{\otimes2}}\D K\D\tilde K\prod_a\int_{S_2}\uD\hat m_a\;\bra{n_a}Kg_a^0\ket{m_a}_{\Y,j_a}\bra{m_a}(g_a^0)^{-1}\tilde K^{-1}\ket{\tilde n_a}_{\Y,j_a}\; +\notag\\
&\quad + o(\Lambda^{6(\mu-1)}).
\label{final_melon_ampl}
\end{align}
The second term in this formula is a new feature of the EPRL-FK model with respect to BF-theory. It stems from the presence of a Lorentzian sub-sector of the solutions to the stationary point equations. In this sector the group elements $g_a^0$ are genuine $\sldc$ elements and do not commute with the EPRL-FK $\Y$ map. The explicit evaluation of this term is obstructed by the implicit (and very intricate) dependence of the $g_a^0$ on the internal spins $\{j_\ab\}$.One can try and guess that because of this dependence, the sum over the internal spins gets suppressed. This would give a result quite intriguing for its simplicity:
\be
W^{\Lambda}_\mathcal{M}\sim\Lambda^{6(\mu-1)}\int_{\sldc^{\otimes2}} \D K\D \tilde K \prod_a \int_{S_2} \uD \hat m_a \;\bra{n_a}K\ket{m_a}_{\Y,j_a}\bra{m_a}\tilde{K}\ket{\tilde n_a}_{\Y,j_a} + o(\Lambda^{6(\mu-1)})\;.
\label{finaleq}
\ee
The structure of this formula reminds the EPRL-FK bare propagator, which coincides with that of $\SU$ BF-theory.\footnote{See e.g. \cite{Krajewski:GFTs}.} Formally,
\be
\mathbb P := \int_\SU\D h \;\bigotimes_a h \;,\quad\bra{n_a}\mathbb P \ket{\tilde n_a}_{j_a}\equiv\mathbb P(j_a,n_a,\tilde n_a) = \int_\SU \D h \prod_a \;\bra{n_a}h\ket{\tilde n_a}_{j_a}.
\ee
To stress the similarity of \autoref{finaleq} with the propagator, and to write it in a more compact form, define the (formal\footnote{Problems could arise from the non-compactness of $\sldc$.}) operator $\mathbb T_\gamma$:
\be
\mathbb T_\gamma :=  \int_\sldc\D K \;\bigotimes_a\Y^\dag K \Y  \;,\quad\bra{n_a}\mathbb T_\gamma \ket{\tilde n_a}_{j_a}\equiv\mathbb T_\gamma(j_a,n_a,\tilde n_a) = \int_\sldc \D K \prod_a \;\bra{n_a}K\ket{\tilde n_a}_{\Y,j_a}.
\ee
Remark that $\mathbb T_\gamma$ is \emph{morally} nothing else than $`` \K\rhd \mathbb P"$ (see \autoref{appendix_face}). Notice, also, that $\mathbb T_\gamma = \mathbb P \mathbb T_\gamma = \mathbb T_\gamma \mathbb P$, since the $\sldc$ integration subsumes an $\SU$ integration. In terms of this new operator, the EPRL-FK self energy at its dominant order (\autoref{finaleq}) reads simply
\be
W^{\Lambda}_\mathcal{M}(j_a,n_a,\tilde n_a)\sim\Lambda^{6(\mu-1)} \bra{ n_a}\mathbb T_\gamma^2\ket{\tilde n_a}_{j_a}\;.
\ee

Remark that $\mathbb T_\gamma$ is \emph{not} a projector (on the contrary $\mathbb P$ is), therefore $\mathbb T_\gamma^2\neq \mathbb T_\gamma$. This is because of the presence of the EPRL-FK $\Y$ map and is therefore intrinsic to the nature of the model. For this very same reason, $\mathbb T_\gamma$ is an operator which appears in many practical applications of the EPRL-FK model (e.g. in Spin Foam Cosmology \cite{sfcosmology}) and therefore deserves a study of its general properties. For example, its semi-classical behaviour at large spins $\{j_a\}$ was recently unravelled in \cite{jacek}.\\

Finally, remark also that the most diverging part of the (non-degenerate sector of the) melon graph amplitude is \emph{not} proportional to the bare propagator of the model. The consequences of this fact on the renormalization group of the EPRL-FK model and on the physical implication of the model itself rest to be studied. Nonetheless it seems clear that the dynamic of the renormalized model would result modified with respect to the tree level one. Whether this modification goes in the direction of invalidating the results about the EPRL-FK semiclassical limit, or rather in the direction of enriching them (e.g. by modifying the semiclassical limit to an approximation of General Relativity richer than Regge calculus), is still an open question. For the moment, we just remark that a possible mechanism for this is the emergency of non-trivial gluing conditions due to the  dependence of the renormalized propagator on the spins $\{j_a\}$. Indeed, something very similar happens in the $\SU$-BF case (see \cite{BenGelounBonzom:BFren}), where the result of the melon renormalization procedure is  the introduction of a group Laplacian in the renormalized GFT Lagrangian.

\section*{Acknowledgements}
I thank my advisor, Carlo Rovelli, for his guidance, as well as Muxin Han, Hal Haggard, Thomas Krajewski, Mingyi Zhang, Simone Speziale and Daniele Oriti for fruitful discussions at different stages of this work.

\appendix
\section{Euclidean or Lorentzian?\label{appendix_euclideanorlorentzian}}
In this section we shall give a more straightforward characterization in terms of the spins of Euclidean or Lorentzian nature of the stationary points. Indeed, in \autoref{section7.1} a formula is given which allows for a distinction of the Euclidean from the Lorentzian case, in terms of the spins (\autoref{ch2eta}). However, now that we have a correspondence between the spinorial data at the stationary points and geometrical tetrahedra, we can analyse the same issue from a different perspective.\\

In order for the $\ell_\ab$'s to be solutions of the closure equations, the spins $\{j_\ab\}_{b,b\neq a}$ must satisfy triangular inequalities. We shall suppose these inequalities hold strictly, i.e. we shall \emph{not} consider the case of degenerate triangles. One can ask if further conditions are needed to define a tetrahedron in $\mathbb R^{1,3}$. To answer this question, we go through the following geometrical construction. It was shown that all the faces of the considered tetrahedra are spacelike. Therefore, it is always possible to boost the whole tetrahedron in such a way that face 4 lies on the $xz$-plane. Further rotating the tetrahdron around the $\hat y$ axis, allows to align side 13 to the $\hat x$-axis. Finally, translating the tetrahedron one can make coinced vertex 1 with the origin of the Cartesian system (see \autoref{tetra_lati}). In this configuration, the equations for the tetrahedron edge lengths read
\be
\left\{
\begin{array}{l}
j_{13}^2 = x_3^2\\
j_{12}^2 = x_2^2+z_2^2\\
j_{23}^2 = (x_3-x_2)^2+z_2^2\\
j_{14}^2 = x_4^2+z_4^2+y_4^2-t_4^2\\
j_{34}^2 = (x_4-x_3)^2+z_4^2+y_4^2-t_4^2\\
j_{24}^2 = (x_4-x_2)^2+(z_4-z_2)^2+y_4^2-t_4^2\\
\end{array}\right. ,
\ee
where $(x_i,y_i,z_i,t_i)$ are the four coordinates of the vertex $i=1,\dots,4$. 
The first three equations concern the triangle 1 and involve the resolution of a quadratic equation, whose solution is guaranteed to exist by the triangular inequalities. Subtracting equation four from five gives a linear equation for $x_4$, and so does the subtraction of equation five from six for $z_4$. Thanks to their linearity, these equations are readily solved in terms of the $j$'s and of the solution to the first three equations. Equation four becomes then an equation for $(z_4^2-t_4^2)$:
\be
y_4^2-t_4^2 = j_{14}^2-x_4^2-z_4^2 \gtrless 0\;,
\ee
where the right hand side of this equation can be either positive or negative. Provided that the $zt$-plane is appropriately boosted, in the first case the tetrahedron lies in the hypersurface orthogonal to $t^I$ and is said to be Euclidean; while in the latter case, it lies in the hypersurface orthogonal to $y^I$ and is said to be Lorentzian.\\

\bFIG[width=9cm]{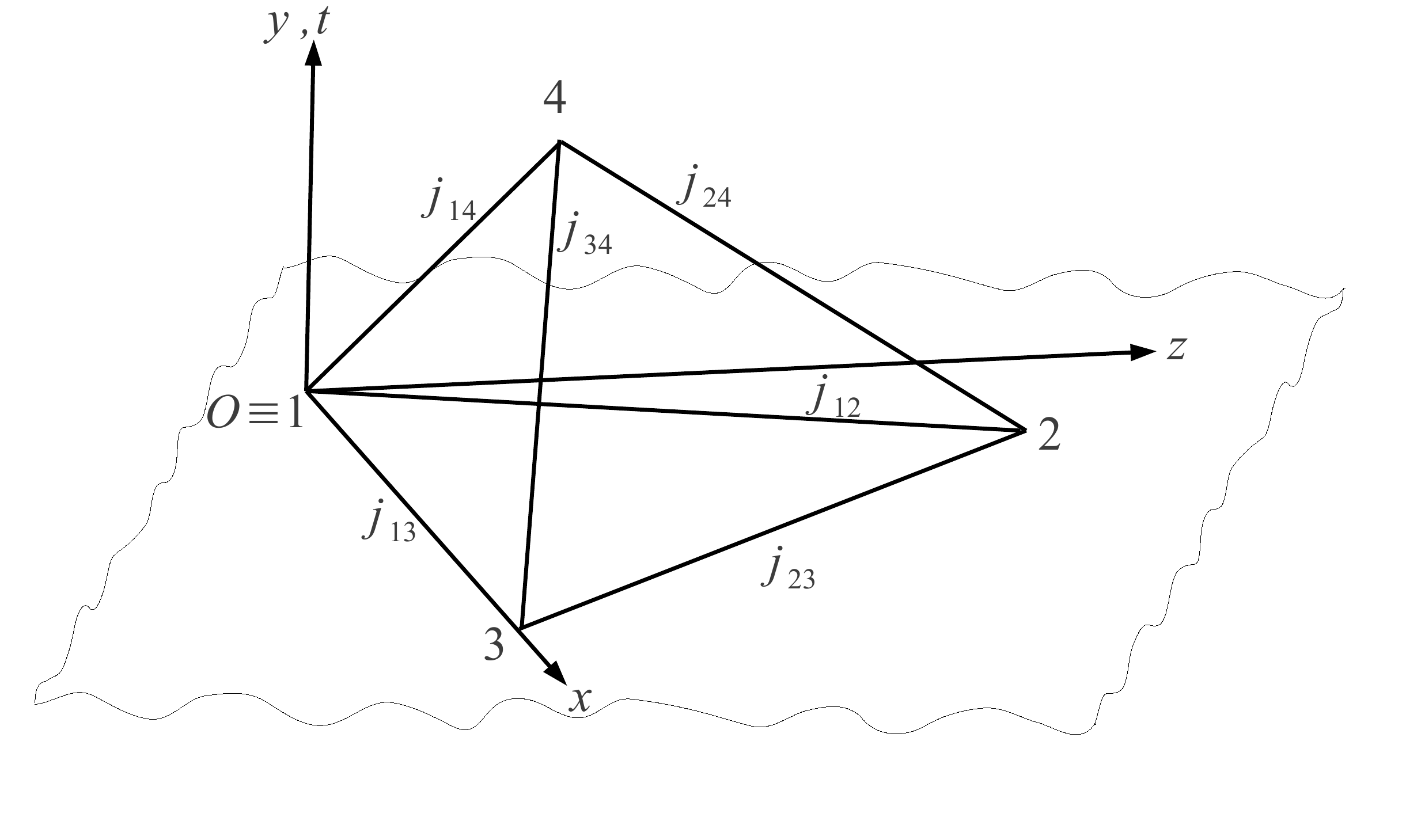}
\caption{The six edge lengths $\{j_{12},j_{13},j_{23},j_{14},j_{24},j_{34}\}$ closing into a tetrahedron in $\mathbb R^4$, with the axis chosen as in the text.}
\label{tetra_lati}
\eFIG

Remark that during the previous procedure, two quadratic equations had to be solved. This implies that two sign choices had to be made. The first of these choices corresponds to the choice of orientation of the tetrahedron basis (the face lying in the $xz$-plane). The second one correspond to the sign of $y_4$ or $t_4$ in the Euclidean or Lorentzian cases, respectively.%
Therefore any set of six edge lengths $\{j_\ab\}$ satisfying triangular inequalities for any subset $\{j_\ab\}_{b,b\neq a}$, defines up to global rotations and boosts, and up to space and time inversion, a tetrahedron in $\mathbb R^{1,3}$. The tetrahedron can be either Euclidean or Lorentzian. In the first case the edge lengths satisfy also the condition of positivity of the so-called Caley-Menger determinant \cite{CaleyMenger,ChristLangvik:PRspike}.

\section{The Dupuis-Livine Map and Face Reversal\label{appendix_face}}
In this appendix the following equation is shown to hold, from which it is straightforward to deduce the face reversal symmetries of the EPRL-FK model:
\be
\Tr_j\left[ \overleftarrow{\prod_e}\Y^\dag G_e \Y \right] = \Tr_j\left[ \overrightarrow{\prod_e}\Y^\dag G_e^{-1} \Y \right]\;.
\ee

In order to to this, it is useful to introduce the Dupuis-Livine map \cite{DupuisLivine:ProjSpinNet} $\K(h,g)$:
\begin{align}
\begin{array}{rccl}
\K :& \mathcal{L}_2[\SU] & \longrightarrow&  \mathcal{C}[\sldc] \\
&\phi(h) & \longmapsto&  \hspace{.5cm}\Phi(g):=\int_\SU \D h\; \K(h,g)\phi(g)
\end{array}\;,
\end{align}
whose kernel, with little abuse of notation, is defined by
\begin{align}
\K(h,g) := \int_\SU \D k \sum_j (2j+1)^2 \chi^j(hk)\chi^{\gamma j, j}(kg)\;.
\label{app_K}
\end{align}
Here $\chi^j$ and $\chi^{p,k}$ are characters of $\SU$ and $\sldc$ in their $j$ and $(p,k)$ representation respectively.
\\

From this appendix's point of view, the crucial property of Dupuis-Livine map $\K$ is that
\be
\K(h,g) = \K(h^{-1},g^{-1}).
\label{app_K_inv}
\ee
To show this, beside the fact that $\int_\SU \D k =\int_\SU \D k^{-1}$, it is enough to recall that $\chi^j(h)=\chi^j(h^{-1})$ and $\chi^{p,k}(g) = \chi^{p,k}(g^{-1})$. The first of these identities follows from $h={h^{-1}}^\dag$ and the fact that $\SU$ characters are real,\footnote{Explicitely, using the fact that $\forall h\in\SU \;\exists u\in\SU,\theta\in[0,2\pi]$ such that $h=u\E^{\I\theta\sigma_z}u^{-1}$: $$\chi^j(h)=\chi^j\left(u\E^{\I\theta\sigma_z}u^{-1}\right)=2 \sum_{m=-j}^{j} \cos(m\theta)=2\frac{\sin\left(\left(j+\frac{1}{2}\right)\theta\right)}{\sin\left(\frac{1}{2}\theta\right)}.$$} while the second one can be shown by mean of the map \cite{Ruhl:sl2c,BarrettHellmann:LorAsympt,Krajewski:GFTs} $\mathcal J:\mathcal H^{p,k}\mapsto\mathcal H^{p,k}$. Indeed, it has the following properties:
\begin{align}
(\mathcal J g\psi, g \psi')_{p,k} = (\mathcal J \psi,\psi')_{p,k} \quad\text{and}\quad (\psi,\psi')_{p,k} = (\mathcal J \psi',\mathcal J \psi)_{p,k}\quad\forall \psi,\psi'\in\mathcal H^{p,k}\;\forall g\in\sldc
\end{align}
where $(\cdot,\cdot)_{p,k}$ stands for the Hermitian scalar product in $\mathcal H^{p,k}$. From these, it follows that
\begin{align}
(\mathcal J \psi, g\mathcal J \psi)_{p,k}  = (\mathcal J g^{-1} \psi,\mathcal \psi)_{p,k} = (\psi,g^{-1} \psi)_{p,k}\quad\forall \psi\in\mathcal H^{p,k}.
\end{align}
Therefore, since if $\{\psi_i\}_i$ is an orthonormal basis then $\{\mathcal J\psi_i\}_i$ is an orthonormal basis as well, by summing the previous equation over such a basis, one immediately finds the sought conclusion: $\chi^{p,k}(g) = \chi^{p,k}(g^{-1})$.\\

Using Schur's orthogonality relations, \autoref{app_K} can be rewritten as
\be
\K(h,g) =\sum_j (2j+1) D^j(h^{-1})^n_{\phantom{n}m}D^{\gamma j,j}(g)^{jm}_{\phantom{jm}jn}\;,
\ee
where no sum is taken over the indices $m$ and $n$, and where $D^j(h)$ and $D^{p,k}(g)$ are the matrices of the elements $h\in\SU$ and $g\in\sldc$ in representation $j$ and $(p,k)$, respectively. Therefore, using Peter-Weyl theorem, one finds
\be
\phi(h)=\sum_j\sum_{m,n} \phi^{(j)\;n}_{\phantom{(j)\;n}m} D^j(h)^m_{\phantom{m}n} \;\;\stackrel{\K}{\longmapsto}\;\;\Phi(g) = \sum_j \phi^{(j)\;n}_{\phantom{(j)\;n}m} D^{\gamma j, j}(g)^{jm}_{\phantom{jm}jn}\;.
\ee
Formally, being $D^{\gamma j, j}(g)^m_{\phantom{m}n}=\bra{j,m}\Y^\dag D^{\gamma j, j}(g)\Y\ket{j,n}$, this can be thought as
\be
\Phi(g):=(\K\rhd\phi)(g)\approx\phi(\Y^\dag g \Y).
\ee

Using this notation, it is immediate to see that a face (partial) amplitude in the EPRL model is the transformed through $\K$ of an $\SU$-BF one:
\be
\Tr_j\left[ \overleftarrow{\prod_e}\Y^\dag G_e \Y \right] = \prod_e \int_\SU \D H_e \;\K(H_e,G_e)\; \chi^j\left( \overleftarrow{\prod_e} H_e \right) \;.
\ee

Now, the partial amplitude of the same face of the previous equation, taken with opposite orientation is
\begin{align}
\Tr_j\left[ \overrightarrow{\prod_e}\Y^\dag G_e^{-1} \Y \right]&= \prod_e \int_\SU \D H_e \;\K(H_e,G^{-1}_e)\; \chi^j\left( \overrightarrow{\prod_e} H_e \right)\notag\\
&= \prod_e \int_\SU \D H_e \;\K(H_e^{-1},G_e)\;\chi^j\left( \overleftarrow{\prod_e} H_e^{-1} \right)\notag\\
&=\Tr_j\left[ \overleftarrow{\prod_e}\Y^\dag G_e \Y \right]\;,
\end{align}
where \autoref{app_K_inv} was used, as well as the fact that $\int_\SU \D h = \int_\SU \D h^{-1}$ and $\chi^j(h)=\chi^j(h^{-1})$ for $h\in\SU$.

\section{The Degenerate Sector\label{appendix_degenerate}}

In Appendix C of \cite{KrajewskiMagnen:EPRL-GFT}, it was shown that the geometrically degenerate sector of the melon graph in the Euclidean EPRL-FK model contributed to the graph amplitude \emph{more} than the non-degenerate one. This result was achieved via explicit estimation of the asymptotics of the $15j$ and $9j$ symbols building the Euclidean version of the EPRL-FK vertex amplitude. Moreover, this was done by fixing the external spins (the $\{j_a\}$ in our notation) exactly to zero.\\

Here, we try to translate their result to our language, which better suits the Lorentzian version of the model. This shall also clarify the geometrical origin of the dominance of this sector in the EPRL-FK model, while it is \emph{sub}-dominant in the $\SU$-BF theory (as was indirectly shown in \autoref{section_BF}).
Nevertheless, we shall not go into the details of explicitly evaluating the contribution of this sector.\\

We start by classifying the degenerate sectors of the melon graph. In particular, there exists three subsectors of degenerate configurations, depending on how many triangular inequalities are saturated:
\begin{itemize}
\item[1.] the triangular inequalities are strictly satisfied, but the Caley-Menger determinant of the spins is null;
\item[2.] at one (and only one) edge the triangular inequalities are saturated;
\item[3.] at more than one edge the triangular inequalities are saturated .
\end{itemize} 
This sectors correspond to geometrical configurations in which the tetrahedron is flattened to lie on a plane in three different ways:
\begin{itemize}
\item[1.] all the four vertices lie on a plane, but no numerical coincidence exists among the vertex positions (the faces are triangles with non-zero area);
\item[2.] a face is degenerate, therefore three vertices lie on a line, while the fourth one can lie anywhere in space, but on the line spanned by the three other vertices.
\item[3.] the four vertices are aligned. This sector shall be dubbed \emph{maximally degenerate}.
\end{itemize} 

For the previous coincidences to be possible, a certain number of constraints must be satisfied by the $\{j_\ab\}$ in such a way that the six independent sums on the $j$'s are effectively reduced by 1,2 and 3 sums, respectively. At the same time, the stationary points acquire more symmetries; this fact tends to enhance the divergence related to this sector. The balance between these two phenomena shall determine the relevance of the degenerate sector over the non-degenerate one.\\

For simplicity in this section we will only analyse the maximally degenerate case, i.e. case 3. here above. Moreover, we will also avoid to go into the algebra of the stationary point equations, even if this would be almost straightforward in the maximally degenerate case. Therefore, the presented arguments will be essentially of a geometrical nature.\\

For a better understanding, we start from the simple case of $\SU$-BF theory. At each edge, the degenerate configuration is represented by three points aligned on a segment. It is intuitively clear that a new symmetry must associated to this configuration. Indeed, this is the case, and the symmetry is a $U(1)$ symmetry around the direction of the edge itself. Therefore, the four parallel transports within a tetrahedron $h_a$ (and $\tilde h_a$) are all determined up to one of these $U(1)$, except for the fact that a constraint among these symmetries exists. One can realise that this is the case by observing that in the spinor representation a $U(1)$ rotation along the spinor axis introduces a phase, and the four phases so introduced within a tetrahedron must compensate each other. Finally, four of these symmetries can just be reabsorbed in the $\SU$ rotational freedom we have at each edge (i.e. can be reabsorbed in the definition of the edge reference frames). As a result the total rank of the Hessian is (at most) reduced of
\be
\rank(\delta^2 S^\text{BF}_\text{deg})-\rank(\delta^2 S^\text{BF}_\text{non-deg}) = -(8-2-4)\times\dim(U(1)) =  -2.
\ee
Therefore, taking into account that there are 3 less free spins on which to sum, one can estimate the degree of divergence of the degenerate sector in the $\SU$-BF theory:
\be
W^\text{BF}_{\mathcal{M},\text{deg}}\sim \Lambda^{-3}\times\Lambda^{\frac{1}{2}\left[\rank(\delta^2 S^\text{BF}_\text{deg})-\rank(\delta^2 S^\text{BF}_\text{non-deg})\right]}\times W^\text{BF}_{\mathcal{M},\text{non-deg}}
\sim \Lambda^{-2}\times W^\text{BF}_{\mathcal{M},\text{non-deg}},
\ee
which means that the degenerate sector is subdominant with respect to the non-degenerate one.\\

Unluckily, things might different for the EPRL-FK model. The reason is that in this model the vertex boundary states are the same as in $\SU$-BF theory, however the parallel transports double the copies of $\SU$ present in the bulk.\footnote{This is exactly true in the Euclidean EPRL-FK model, and morally true in the Lorentzian one, where the holonomies are in $\sldc$, whose algebra, in the fundamental representation, is essentially a complex, instead of real, $\SU$ algebra.} As a consequence, the symmetries on the degenerate boundary states induce a larger freedom in the bulk. For clarity we shall treat the Euclidean and Lorentzian EPRL-FK models independently, even if they lead to the same result.\\

In the Euclidean model, each $Spin(4)$ holonomy is made of a self-dual and an antiself-dual part $g=(g^+,g^-)$, each in $\SU$, and whose combined action on the boundary states is dictated by the fusion coefficients (see \cite{EPRL:EPRL}). Therefore the counting we did in the $\SU$-BF theory applies almost independently on the self-dual and antiself-dual, the only exception being the fact that the freedom to redefine the edge reference frame is not doubled (the edge boundary state being the same as in the previous case!). Therefore,
\be
\rank(\delta^2 S^\text{Eucl}_\text{deg})-\rank(\delta^2 S^\text{Eucl}_\text{non-deg}) = -[2\times(8-2)-4]\times\dim(U(1)) =  -8.
\ee
Hence,
\be
W^\text{Eucl}_{\mathcal{M},\text{deg}}\sim \Lambda^{-3}\times\Lambda^{\frac{1}{2}\left[\rank(\delta^2 S^\text{Eucl}_\text{deg})-\rank(\delta^2 S^\text{Eucl}_\text{non-deg})\right]}\times W^\text{Eucl}_{\mathcal{M},\text{non-deg}}
\sim \Lambda^{1}\times W^\text{Eucl}_{\mathcal{M},\text{non-deg}}.
\ee
This result is in complete accord with the calculation performed in Appendix C of \cite{KrajewskiMagnen:EPRL-GFT}, with the only \emph{caveat} that they are using a different face weight\footnote{Remark that the GFT construction is less flexible in the choice of the edge and face weights, than that of Spin Foams.} (essentially the BF theory one, which has $\mu=2$).\\

In the Lorentzian case, being the parallel transports in $\sldc$, thanks to the symmetries of the degenerate boundary state, they are defined only up to $U(1)_\mathbb{C}$ element, which has dimension 2.\footnote{The physical interpretation of this is related to the possibility of boosting the faces along their degeneracy direction.} On the other side, only 4 \emph{real} $U(1)$ can be reabsorbed in the redefinition of the edge reference frames. Therefore,
\be
\rank(\delta^2 S^\text{Eucl}_\text{deg})-\rank(\delta^2 S^\text{Eucl}_\text{non-deg}) = -(8-2)\times\dim(U(1)_\mathbb{C})-4\times\dim(U(1)) =  -8.
\ee
And finally,
\be
W_{\mathcal{M},\text{deg}}\sim \Lambda^{-3}\times\Lambda^{\frac{1}{2}\left[\rank(\delta^2 S_\text{deg})-\rank(\delta^2 S_\text{non-deg})\right]}\times W_{\mathcal{M},\text{non-deg}}
\sim \Lambda^{1}\times W_{\mathcal{M},\text{non-deg}}.
\ee

In conclusion, this estimation hints to the fact that the geometrically degenerate sector contributes \emph{more} than the non-degenerate one to the EPRL-FK self-energy. Nevertheless, one has to be cautious. Indeed, the degenerate sector happen to be close to the boundary of the region of validity of some of our assumptions and the treatment we gave to the problem could break down and must be carefully checked. On the other side, this sector might be simple enough to allow an explicit analysis of the Hessian of the action.\\

\newpage

\bibliographystyle{ieeetr}
\bibliography{BIBmelon}

\end{document}